\documentclass[aps,floats,floatfix,twocolumn,prb,showpacs,10pt,superscriptaddress]{revtex4-1}
\usepackage[T1]{fontenc}
\usepackage{graphicx}
\usepackage{color}
\usepackage{upgreek}
\usepackage{setspace}
\usepackage{amssymb}
\usepackage{amsmath}
\usepackage{pstricks}
\usepackage{dsfont}
\usepackage{fancyhdr}
\usepackage{array}
\usepackage{multirow}
\usepackage{booktabs}
\usepackage{diagbox}
\usepackage{hhline}
\usepackage{footmisc}
\usepackage{float}
\usepackage{soul}
\usepackage{hyperref}

\allowdisplaybreaks[4]

\newcolumntype{M}[1]{>{\centering\arraybackslash}m{#1}}

\begin{document}
\title{Dual parquet scheme for the two-dimensional Hubbard model:\\ Modelling low-energy physics of high-$T_c$ cuprates with high momentum resolution}
\author{Grigory V. Astretsov}
\affiliation{Russian Quantum Center, Skolkovo innovation city, 121205 Moscow, Russia}
\affiliation{Department of Physics, Lomonosov Moscow State University, Leninskie gory 1, 119991 Moscow, Russia}
\author{Georg Rohringer} 
\affiliation{Russian Quantum Center, Skolkovo innovation city, 121205 Moscow, Russia}
\affiliation{Institute of Theoretical Physics, University of Hamburg, 20355 Hamburg, Germany}
\author{Alexey N. Rubtsov}
\email{ar@rqc.ru}
\affiliation{Russian Quantum Center, Skolkovo innovation city, 121205 Moscow, Russia}
\affiliation{Department of Physics, Lomonosov Moscow State University, Leninskie gory 1, 119991 Moscow, Russia}
\date{\today}

\begin{abstract}
We present a new method to treat the two-dimensional (2D) Hubbard model for parameter regimes which are relevant for the physics of the high-$T_c$ superconducting cuprates. Unlike previous attempts to attack this problem, our new approach takes into account all fluctuations in different channels on equal footing and is able to treat reasonable large lattice sizes up to 32x32. This is achieved by the following three-step procedure: (i) We transform the original problem to a new representation (dual fermions) in which all purely local correlation effects from the dynamical mean field theory are already considered in the bare propagator and bare interaction of the new problem. (ii) The strong $1/(i\nu)^2$ decay of the bare propagator allows us to integrate out all higher Matsubara frequencies besides the lowest using low order diagrams. The new effective action depends only on the two lowest Matsubara frequencies which allows us to, (iii) apply the two-particle self-consistent parquet formalism, which takes into account the competition between different low-energy bosonic modes in an unbiased way, on much finer momentum grids than usual. In this way, we were able to map out the phase diagram of the 2D Hubbard model as a function of temperature and doping. Consistently with the experimental evidence for hole-doped cuprates and previous dynamical cluster approximation calculations, we find an antiferromagnetic region at low-doping and a superconducting dome at higher doping. Our results also support the role of the van Hove singularity as an important ingredient for the high value of $T_c$ at optimal doping. At small doping, the destruction of antiferromagnetism is accompanied by an increase of charge fluctuations supporting the scenario of a phase separated state driven by quantum critical fluctuations.

\end{abstract}

\maketitle

\let\n=\nu \let\o =\omega \let\s=\sigma


\section{Introduction}
\label{sec:intro}

Some of the features that make cuprates distinctly different from other known families of superconductors are the record high critical temperature $T_c$ at ambient pressure and the simultaneous presence of different collective modes in the phase diagram (for a review see, e.g., Refs.~\onlinecite{Damascelli2003,Vishik2010,Vishik2018}). At half-filling (i.e., in the undoped parent compound) and at low doping ($\delta\lesssim0.03$), an antiferromagnetic (AFM) phase is observed. Strong spin fluctuations prevail in a much broader doping region ($\delta\sim0.2\ldots0.4$)\cite{Dagotto94}, surrounding the narrow AFM state. This is not surprising taking into account the layered structure of cuprate compounds: in a purely planar system the AFM state would be fully suppressed at finite temperature\cite{Mermin1966} while the fluctuations would still be present\cite{Schaefer2015-2}. However, it can be disputed how the AFM state breaks down at the microscopic level. Several scenarios have been proposed including the formation of spin and charge stripes\cite{Vojta2009, Fujita2012}, a phase separated state where extra charges form droplets\cite{Stepanov2018} in the AFM medium and even glassy phases\cite{Seibold2014}. On the single-electron level, a destruction of the AFM ordering is accompanied by the formation of the famous pseudogap state \cite{Yoshida2006}, in which the Fermi surface near the anti-nodal direction is destroyed, likely because of the strong collective fluctuations.\cite{Gunnarsson2015}

A further increase of doping away from half-filling results in the formation of the superconducting  (SC) state with d-wave symmetry.\cite{Wollman1993,Shen1993} In the (temperature vs. doping) phase diagram, it forms a dome peaked at optimal doping $\delta_{opt}\approx 15\%$. While this feature is virtually common to all hole-doped cuprate compounds, the nature of the superconducting phase is, however, still highly debated. The absence of an isotope effect\cite{Pringle2000,Greco2015} of the form $T_c\sim1/\sqrt{M}$ suggests that the pairing glue for the electrons is not (exclusively) due to phonons. Instead, collective excitations of the electrons themselves, such as the above mentioned antiferromagnetic spin fluctuations, may generate\cite{Schmalian1998,Scalapino2012} an effective attractive interaction. As an additional ingredient which is potentially responsible for the high value of $T_c$ at optimal doping the presence of a van Hove singularity has been discussed.\cite{Markiewicz1997,Piriou2011} 



The above mentioned formation of a pseudgap as well as the anomalous non-Fermi-liquid-like behavior of certain transport properties with doping and temperature\cite{Legros2019,Jenkins2010} indicate that the physics of the cuprates originates from --or is at least substantially affected by-- strong correlation effects between the electrons\cite{Lee2006}. This is consistent with the widely accepted assumption that the physical properties of cuprates are dominated by the electrons in the partially filled $3d_{x^2-y^2}$ orbital of the copper atoms in the CuO$_2$ planes. Therefore, the planar single-band Hubbard model\cite{Hubbard64} on a square lattice, which incorporates correlation effects via a purely local on-site Coulomb repulsion $U$, is commonly used for the theoretical description of these compounds. One should mention, that this model neglects several possibly important degrees of freedom in realistic cuprate crystals: The tunnel coupling between different CuO$_2$ planes, phonon degrees of freedom which may be responsible for the experimentally observed charge density waves\cite{Reznik2006,Greco2015,Miao2017}, or oxygen $p$-orbitals which can give rise to a metal-to-insulator transition of charge-transfer type.\cite{Avella2013,Hansmann2014} Nevertheless, the Hubbard model is usually considered as a minimal model which incorporates the important correlations effects in the Cu $d_{x^2-y^2}$ orbitals.\cite{Anderson2002}

Thus, mapping out the phase diagram of the Hubbard model is vitally important for our understanding of the physics of cuprates. Calculations for the Hubbard model are, however, extremely difficult in the parameter regimes relevant to cuprate systems. Although, the d-wave superconductivity arises in both the strong and the weak coupling limit of the Hubbard Hamiltonian, in the cuprates the value of the Coulomb interaction $U$ is comparable to the bandwidth $W$ ($U\!\sim\!W$). This prevents any perturbative treatment in both $U$ and $W$ starting from the weakly interacting Fermi gas or isolated atoms, respectively. The sign problem\cite{Loh1990}, on the other hand, imposes severe difficulties for lattice Quantum Monte Carlo calculations away from half-filling. Embedding approaches such as the density matrix embedding theory\cite{Knizia2012} or the site occupation embedding theory\cite{Senjean2019} are restricted to zero (or very low) temperature and/or to one-dimensional systems. Similar problems arise for the density matrix renormalization group technique\cite{Hallberg2006} which provides almost exact results in one dimension but is very hard to extend to higher dimensions. A recently suggested numerical renormalization group method\cite{Huang2018} is confined to zero temperature and small lattice sizes ($4\times N$ stripes). Functional renormalization group (fRG) approaches\cite{Halboth2000,Metzner2012} are a powerful tool to take into account the mutual screening between competing bosonic modes. However, they can provide an accurate quantitative description only at weak coupling.

To overcome the above-mentioned difficulties, a comprehensive theory for the solution of the 2D Hubbard model should be able to correctly capture at least the two most important features of this system: (i) It should incorporate nonperturbatively strong local correlation effects leading to the spectral weight-transfer and renormalization of the electronic density of states, and (ii) it has to take into account long-range correlation effects due to competing collective bosonic modes in different scattering channels. The situation becomes even more complicated because these phenomena occur at very different energy scales ranging from $10$meV [for (ii)] to several eV [for (i)]. 

As for local correlations, the dynamical mean field theory (DMFT) \cite{Metzner1989, Georges1992, Georges1996} has become the standard tool for the description of correlated model systems and materials by replacing the actual lattice of interacting sites by a single interacting site embedded in a self-consistent noninteracting bath. In this way, DMFT takes into account all purely local correlations but captures long-range bosonic collective modes only on a mean-field level. In particular, the mutual interaction and screening effects between bosonic fluctuations in competing channels are not sufficiently considered.


The most straightforward way to overcome this problem and include spatial nonlocality in the framework of DMFT is to use cluster methods instead of the single-site scheme. This has lead to the development of the cellular DMFT (CDMFT) and the dynamical cluster approximation (DCA) which consider a cluster of interacting site in real or momentum space, respectively. CDMFT calculations with a properly periodized 2$\times$2 cluster indeed yield a SC phase\cite{Lichtenstein2000, Maier2005}. DCA approaches \cite{,Yang2011,Chen2013,Gull2013,Chen2015} applied to larger clusters have been able to capture antiferromagnetic spin fluctuations and a dome-like SC phase with $T_c$ quite close to the experiment. Several important physical conclusions have been drawn from the DCA results. In particular, they are consistent\cite{Khatami2010,Yang2011} with the scenario of a quantum critical point (QCP) \cite{Varma1999,Broun2008, Sachdev2010} underlying the SC area in phase diagram and moreover predict various phenomena such as phase separation\cite{Khatami2010}, a Lifshitz transition\cite{Chen2012}, a momentum-sector-selective metal-insulator transition\cite{Gull2009} or pseudogap behavior\cite{Gunnarsson2015,Gunnarsson2016,Gunnarsson2017} .


Unfortunately, cluster approaches can take into account nonlocal correlations only within the cluster size. This is limited by the exponential growth of the Hilbert space to about $4\!\times\!4$ sites 
even within a single orbital model. Phenomena such as a QCP or spin fluctuations are, however, intrinsically long-ranged. Convergence of the results with the cluster size can be therefore be questioned. 
To circumvent such problems, diagrammatic extensions\cite{Rohringer2018RMP} of DMFT have been suggested in the last decade. They allow, at least in principle, to handle correlations at all length and energy scales on equal footing. These approaches construct a perturbation theory around DMFT using the Green's function and the local two-particle vertices of DMFT as building block for the diagrammatic expansions. These schemes benefit from a natural separation of high-energy local physics, which is accounted for by DMFT, and the low-energy bosonic collectve modes, which are treated diagrammatically. Various flavors of diagrammatic extensions of DMFT have been developed, such as the dynamical vertex approximation (D$\Gamma$A)\cite{Toschi2007}, the dual fermion (DF)\cite{Rubtsov2008} theory, the dual boson (DB)\cite{Rubtsov12} scheme, the one-particle irreducible approach (1PI)\cite{Rohringer2013}, the TRILEX\cite{Ayral2015,Ayral2016a}, and the QUADRILEX\cite{Ayral2016} method, as well as mergers of (extended) DMFT and fRG such as the DMF$^2$RG\cite{Taranto2014} and the 2PI-fRG\cite{Katanin2019} approaches. They differ mainly in the choice of diagrams which are constructed on top of DMFT. In most cases, random phase approximation (RPA)\cite{Mahan2000}- or fluctuation exchange (FLEX)\cite{Bickers1989}-like diagrams in {\em one} scattering channel have been considered. This, however, requires an {\sl a priori} knowledge of the dominating fluctuations and cannot describe the interplay between different channels which is necessary to obtain $d$-wave superconductivty from the repulsive Hubbard model. Multichannel FLEX diagrams\cite{Bickers2004}, on the other hand, indeed predict d-wave superconductivity. However, different flavors of diagrammatic extension of DMFT that use such diagrams yield very different phase diagrams, showing rather poor coincidence with the experimentally observed one. For example, TRILEX produces the superconducting dome, but not the AF peak. FLEX-like diagrams with DF\cite{Otsuki2014}, which consider the particle-hole and particle-particle channel, can describe antiferromagnetism well and give rise to the superconductivity, but they do not show the SC-dome structure. FLEX+DMFT calculations\cite{Kitatani2015}, on the other hand, produce a far too broad dome. The  D$\Gamma$A-ladder calculation of the pairing vertex does indicate a dome which is, however, located at very small doping when a realistic electron dispersion is considered.\cite{Kitatani2019} Overall, multichannel FLEX-like diagrams suffer from the insufficient treatment of the mutual screening between competing fluctuations.



A quantitatively accurate theory which takes into account all mutual screening effects between the different bosonic modes and is self-consistent at the one- and the two-particle level can be built from the parquet equations\cite{Diatlov1957,Bickers2004}. They construct all one- and two-particle correlation functions from the fully irreducible vertex\cite{Rohringer2012} of the system. Approximating this vertex with the bare interaction leads to the so-called parquet approximation\cite{Yang2009,Tam2013}. In the framework of diagrammatic extension of DMFT, the D$\Gamma$A replaces the fully irreducible vertex by the corresponding local one of DMFT\cite{Li2016}. However, although the method formally obeys a polynomial complexity, in practice it is numerically very expensive as it requires the full two-particle vertex functions which depend on three frequencies and three momenta. This restricts practical calculation to very small momentum grids of about 6x6 sites, even if one uses highly elaborated parametrizations of the frequency\cite{Karrasch2008,Wentzell2016,Tagliavini2017} and the momentum\cite{Eckhardt2018} grids or solves the parquet equations by means of the multiloop fRG\cite{Kugler2018,Kugler2018a} technique.

To mitigate these limitations, we propose a parquet method in which the Matsubara frequency grid is reduced to the lowest Matsubara frequencies. 
Our method can be presented as a three-step procedure: First, we solve the local DMFT impurity problem which provides us the basic elements for our diagrammatic expansion, i.e., the DMFT Green's function as well as the local vertex function. Secondly, we integrate out the higher Matsubara frequencies using the dual-fermion theory which provides an optimal framework for a diagrammatic expansion around DMFT. Third, we solve our low-frequency effective model by means of the parquet equations which is possible due to the reduction to only a few Matsubara frequencies for lattices up to 32x32 sites. 


The plan of the paper is the following: In Sec.~\ref{sec:formalism} we present the basic formalism and derive our method. Our results for the 2D Hubbard model are discussed and compared to other approaches in Sec.~\ref{sec:results}. Specifically we plot the phase diagram, calculate the fluctuations in spin, charge and superconducting channels for different points of the high-temperature phase, and analyze the pairing glue and the origin of the dome structure observed in our study.  Sec.~\ref{sec:conclusions} is devoted to conclusions and an outlook.


\section{Model and method}
\label{sec:formalism}

\subsection{Definition of the model and DMFT}

We consider the Hubbard Hamlitonian on a 2D square lattice:
\begin{equation}
\label{equ:Hubbard}
\hat{H} = \sum_{\mathbf{k}} (\epsilon_\mathbf{k} -\mu) \hat{c}^\dagger_{\mathbf{k}\sigma}\hat{c}_{\mathbf{k}\sigma} + U\sum_i \hat{n}_{i\uparrow}\hat{n}_{i\downarrow},
\end{equation}
where $\hat{c}^{(\dagger)}_{\mathbf{k}k(i)\sigma}$ is an annihilation (creation) operator for an electron with momentum $\mathbf{k}$ (or at lattice site $i$) and spin $\sigma\!=\!\uparrow\!,\!\downarrow$. $\hat{n}_{i\sigma}\!=\!\hat{c}^\dagger_{i\sigma}\hat{c}_{i\sigma}$. The dispersion relation is given by $\epsilon_\mathbf{k}\!=\!-2t(\cos k_x\!+\! \cos k_y)\!-\!4t'\cos k_x \cos k_y\!-\!2t'' (\cos 2k_x\!+\! \cos 2k_y)$, $\mu$ is the chemical potential, and $U$ denotes the Coulomb repulsion between two particles at the same lattice site. 

Within the DMFT approximation, one replaces the actual lattice of interacting sites by a single interacting site (=impurity) which hybridizes with a noninteracting bath. This corresponds to an Anderson impurity model (AIM) which can be represented by the action $S_{\text{imp}}$
\begin{align}
    \label{equ:impurityaction}
    S_{\text{imp}}[c^\dagger,c]&=\sum_{\nu\sigma}\left[-i\nu+\Delta_\nu-\mu\right]c^\dagger_{\nu\sigma}c_{\nu\sigma}\nonumber\\&+U\int_0^{\beta}d\tau\;c^\dagger_{\uparrow}(\tau)c_{\uparrow}(\tau)c^\dagger_{\downarrow}(\tau)c_{\downarrow}(\tau),
\end{align}
where $c^{(\dagger)}$ are the Grassmann fields corresponding to the operators $\hat{c}^{(\dagger)}$, $\tau\!\in\![0,\beta]$ is an imaginary time, $\nu\!=\!\frac{\pi}{\beta}(2n\!+\!1)$, $n\!\in\!\mathds{Z0}$, the corresponding fermionic Matsubara frequency, and $\beta\!=\!1\!/\!T$ the inverse temperature. Summations over Matsubara frequencies include the normalization factor $\beta^{-1}$. The hybridization function $\Delta(\nu)$ between the impurity and the bath is determined by the DMFT self-consistency condition which requires the local part of the lattice Green's function in the DMFT approximation (i.e., with the self-energy replaced by the local one of the impurity problem) to be identical to the corresponding impurity Green's function:
\begin{equation}
    \label{equ:DMFTcondition}
    \sum_\mathbf{k} G^{\text{DMFT}}_{\nu \mathbf{k}}\equiv\sum_\mathbf{k} [i\nu-\epsilon_\mathbf{k}+\mu-\Sigma_\nu]^{-1}=g(\nu),
\end{equation}
where $\Sigma_\nu$ is the local impurity self-energy and $g_\nu\!=\![\nu\!-\!\Delta_\nu\!+\!\mu\!-\!\Sigma_\nu]^{-1}$ the impurity Green's function. The $\sum_\mathbf{k}\!\equiv\!\frac{1}{V_\text{BZ}}\int_{\text{BZ}}d^2k$ is the normalized momentum integral over the first Brilluoin zone (BZ) with the volume $V_{\text{BZ}}$. The impurity self-energy $\Sigma_\nu$ captures all purely local correlations of the system while nonlocal correlations are neglected. In the next section, we will outline our new approach which constructs nonlocal correlations from the DMFT starting point.

\subsection{Low-frequency model and parquet equations}
\label{sec:lowfreqmodelparquet}

Our new approach to find an (approximate) solution of the Hubbard model can be divided into three steps which we outline in the following.

\subsubsection{Local correlations and DF transformation}
\label{sec:loccorrDFtransform}

Since an exact calculation of the one- and two-particle correlation functions for the Hamiltonian in Eq.~(\ref{equ:Hubbard}) is not possible so far we have to apply perturbation theory. However, a perturbative expansion in the bare Green's function and the bare interaction {\em cannot} capture the important local correlations and the related Mott\cite{Mott1968} physics. Hence, a reformulation of perturbation theory in terms of the DMFT Green's function and the local DMFT two-particle vertex [i.e., the vertex of the AIM Eq.~(\ref{equ:impurityaction})] is highly desirable. In this way, all purely local correlations are included already in the building blocks of a Feynman diagrammatic expansion while the latter itself will add nonlocal correlation effects which are absent in DMFT.  

The DF theory\cite{Rubtsov2009} provides a convenient formal framework for the construction of a diagrammatic perturbation theory around DMFT. In this approach, the action of the Hubbard Hamiltonian Eq.~(\ref{equ:Hubbard}) is separated into a local impurity part and a remainder
\begin{equation}
\label{equ:splitaction}
S_\text{Hubbard}[c^\dagger_i,c_i] = \sum_i S_\text{imp}[c^\dagger_i,c_i] + \sum_{\nu \mathbf{k}\sigma} (\epsilon_\mathbf{k} - \Delta_\nu)c_{\nu \mathbf{k}\sigma}^\dagger c_{\nu \mathbf{k}\sigma}.
\end{equation}
To obtain an effective perturbation theory around the local DMFT physics, we separate local and nonlocal degrees of freedom by decoupling the second term on the right-hand side of Eq.~(\ref{equ:splitaction}) via a Hubbard-Stratonovich transformation\cite{Hubbard1959,Stratonovich1957} (for the explicit procedure see, e.g., Ref.~\onlinecite{Rohringer2018RMP}). The corresponding Hubbard Stratonovich fields $f^\dagger$ and $f$ are typically labelled ``dual fermions''. One can now integrate out the original fields $c^\dagger$ and $c$ to obtain the action $S_{\text{Hubbard}}$ of the Hubbard model in terms of the dual particles

\begin{equation}
\label{equ:dualaction}
S[f,f^\dagger]=-\sum_{\nu \mathbf{k} \sigma}\widetilde{G}^{-1}_{0,\nu \mathbf{k}}f^\dagger_{\nu \mathbf{k} \sigma}f_{\nu \mathbf{k} \sigma} +\sum_i V[f_i^\dagger,f_i],
\end{equation}
where $\widetilde{G}_{0,\nu \mathbf{k}}=G^{\text{DMFT}}_{\nu \mathbf{k}}\!-\!g_\nu$ is the bare dual propagator which is given by the difference between the full momentum dependent and the local DMFT Green's function and, hence, accounts for the nonlocal degrees of freedom. The effective interaction $V[f^\dagger,f]$ between the dual fermions is given by local two-, three-, $\ldots$ particle vertices of the AIM. While the role of three- and more-particle terms has not been fully clarified so far\cite{Rohringer2013, Ribic2017b}, a truncation at the two-particle level is a reasonable approximation considering that the original Hubbard interaction is of two-body type. With this approximation the effective interaction between the dual fermions becomes
\begin{equation}
    \label{equ:dualinteraction}
    V[f^\dagger_i,f_i]\approx-\sum_{\nu\nu'\omega}\sum_{\sigma\sigma'}\gamma^{(2)}_{\nu\nu'\omega,\sigma\sigma'}f^\dagger_{\nu i\sigma}f_{(\nu+\omega) i\sigma}f^\dagger_{(\nu'+\omega) i\sigma'}f_{\nu' i\sigma'},
\end{equation}
where $\gamma^{(2)}$ denotes the local two-particle vertex of the AIM and $\omega\!=\!\frac{\pi}{\beta}2m$, $m\!\in\!\mathds{Z}$, is a bosonic Matsubara frequency.

\subsubsection{Effective low-frequency model}
\label{sec:lowfreqmodel}

The complex frequency dependence of the effective interaction $V[f^\dagger,f]$ makes a diagrammatic expansion for the action in Eq.~(\ref{equ:dualaction}) very difficult. In particular, the dependence of the vertex $\gamma^{(2)}$ on three frequencies typically restricts the choice of Feynman diagrams to rather simple topologies such as ladders in a single scattering channel. Since this is not sufficient for the description of competing fluctuations, a simplification of the frequency and/or momentum dependence of the one- and two-particle correlation functions is highly desirable.

The second step, and at the same time central idea, of our new approach is a reduction of complexity in the frequency domain in the dual action Eq.~(\ref{equ:dualaction}). To achieve this, we split the $f$-variable in the spirit of Wilson's renormalization group\cite{Wilson1975} into a low- and a high-energy part

\begin{eqnarray}
f^{(\dagger)}_< = \begin{cases}
f^{(\dagger)}_{\nu \mathbf{k} \sigma}, & \text{if $|\nu| \leqslant \nu_\text{max}$}\\
0, & \text{if $|\nu| > \nu_\text{max}$}
\end{cases} \quad
f^{(\dagger)}_> = \begin{cases}
0, & \text{if $|\nu| \leqslant \nu_\text{max}$}\\
f^{(\dagger)}_{\nu \mathbf{k} \sigma}, & \text{if $|\nu| >\nu_\text{max}$,}
\end{cases} \qquad
\end{eqnarray}
where $\nu_\text{max}$ is the cutoff frequency.

This allows us to separate the total action in Eq.~(\ref{equ:dualaction}) into a lesser part $S_0[f_<]$ which depends only on $f^{(\dagger)}_<$ and a greater part which depends on both $f_<^{(\dagger)}$ and $f_>^{(\dagger)}$:
\begin{equation}
    \label{equ:separateaction}
    S[f^\dagger,f]=S_0[f^\dagger_<,f_<]+S_0[f^\dagger_>,f_>]+S_>[f^\dagger_>,f_>,f^\dagger_<,f_<].
\end{equation}

\begin{figure}[t!]
\begin{center}
\includegraphics[width=\linewidth]{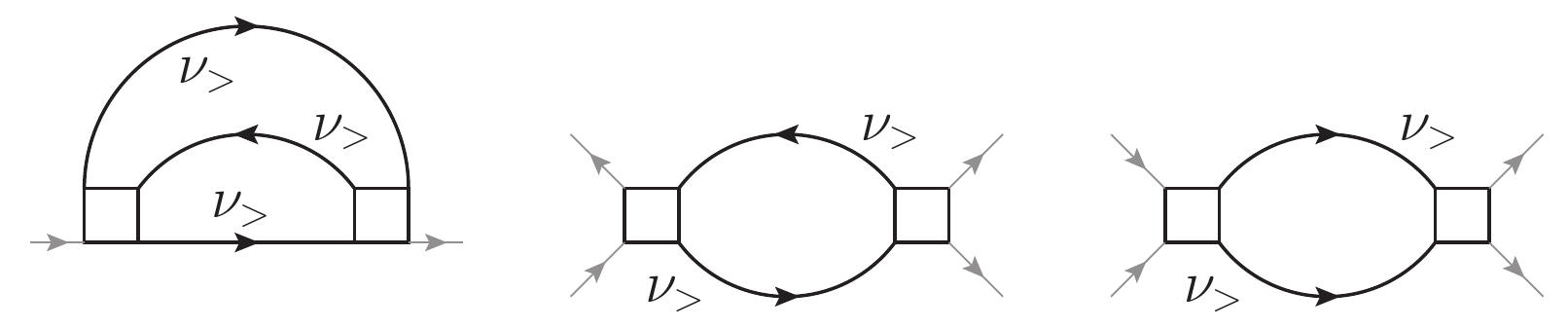}
\caption{Corrections for the renormalized self-energy and vertex functions. A white box represents the local interaction $\gamma^{(2)}$.}
\label{fig:renormalization}
\end{center}
\end{figure}

Then, we integrate out high-frequency fields in the functional integral representation of the partition function $Z$ considering the diagrams depicted in Fig.~\ref{fig:renormalization} for the perturbative expansion of $e^{-S_>}$: 

\begin{widetext}
\begin{equation}
\begin{split}
Z = \int \mathcal{D}f\mathcal{D}f^\dagger e^{-S[f,f^\dagger]} = \int \mathcal{D}f_<\mathcal{D}f_<^\dagger {D}f_>\mathcal{D}f_>^\dagger  e^{-S_0[f_<,f_<^\dagger] -S_0[f^\dagger_>,f_>] - S_>[f_<,f_<^\dagger,f_>,f_>^\dagger] } = \\
\int \mathcal{D}f_<\mathcal{D}f_<^\dagger e^{-S_0[f_<,f_<^\dagger]} \int \mathcal{D}f_>\mathcal{D} f_>^\dagger e^{-S_0[f^\dagger_>,f_>]-S_>[f^\dagger_<,f_<,f^\dagger_>,f_>]} = Z_{0>} \int \mathcal{D}f_<\mathcal{D}f_<^\dagger e^{-S_\text{eff}[f_<,f_<^\dagger]}.
\end{split}
\end{equation}
\end{widetext}

which gives rise to the effective low-frequency action
\begin{align}
\label{equ:effectivelowfreqaction}
S_\text{eff}[f^\dagger_<,f_<] &= S_0[f^\dagger_<,f_<] + \ln\langle e^{-S_>[f^\dagger_<,f_<,f^\dagger_>,f_>]} \rangle_>\nonumber\\&\approx S_0[f^\dagger_<,f_<] - \frac{1}{2}\left(\langle S_>^2\rangle_>-\langle S_>\rangle_>^2\right),
\end{align}
where $\langle\ldots\rangle_>$ denotes an expectation value with respect to the high-frequency fields, i.e., taking the path integral over all fields $f^\dagger_\nu$ and $f_\nu$ with $\lvert\nu\rvert\!>\!\pi/\beta$. The resulting action $S_{\text{eff}}$ depends only on the lowest fermionic Matsubara frequencies $\nu\!=\!\pm\frac{\pi}{\beta}$. To present the explicit expression, we use a simplified notation adopting a multi-index $1\!\widehat{=}\!(\nu,k,\sigma)$ which includes frequency, momentum, and spin degrees of freedom. In this notation, the effective action reads

\begin{equation}
S_\text{eff} = -\sum \mathcal{G}_{12}^{-1} f^\dagger_{1<} f_{2<} + \frac{1}{4}\sum \mathcal{V}_{1234} f^\dagger_{1<} f_{2<} f^\dagger_{3<} f_{4<},
\end{equation}
where the effective bare propagator $\mathcal{G}_{12}$ is given by
\begin{equation}
\mathcal{G}_{12<}^{-1} = \widetilde{G}^{-1}_{0,12<}- \frac{1}{2}\sum\gamma^{(2)}_{1345}\gamma^{(2)}_{6782}\widetilde{G}_{0,36>}\widetilde{G}_{0,47>}\widetilde{G}_{0,85>},
\end{equation}
and the effective bare interaction $\mathcal{V}_{1234}$ can be represented as
\begin{equation}
\begin{split}
    \mathcal{V}_{1234} = \gamma^{(2)}_{1234} + \frac{1}{4}\sum\left(2\gamma^{(2)}_{1278}\gamma^{(2)}_{5634}-\frac{1}{2}\gamma^{(2)}_{1836}\gamma^{(2)}_{5274}\right)\times\\\times\widetilde{G}_{0,58>}\widetilde{G}_{0,76>},
\end{split}
\end{equation}
where a summation/integration over repeated indices is assumed.

Let us stress, that similar simplifications of the frequency dependence of the vertex function has been used also for fRG studies of the 2D Hubbard model\cite{Honerkamp2003,Metzner2012}. However, these approaches differ in two crucial aspects from our approach. 

(i) First, and probably most important, in the above-mentioned fRG studies the simplification in the frequency domain has been applied directly to the Hubbard model in Eqs.~(\ref{equ:Hubbard}). In this case, the bare propagator for the construction of a diagrammatic perturbation theory decays as $1/\nu$ for $\nu\!\rightarrow\!\infty$. On the contrary, starting from the dual action in Eq.~(\ref{equ:dualaction}), the bare propagator decays much faster, i.e., as $1/\nu^2$, which supports the general picture of the DF approach as a theory, which constructs low-energy (or low-frequency) corrections around the DMFT solution of the Hubbard model.

(ii) The second difference between our new technique and the above-mentioned fRG schemes is that in the latter approaches the frequency space has just been truncated to the lowest Matsubara frequency, losing information about the effect of the high-energy onto the low-energy physics. In our approach, this information from higher frequencies is partially taken into account by our downfolding procedure. This may be particularly important in the presence of strong correlations where the formation of Hubbard subbands occurs at such high frequencies. The overall importance of the downfolding procedure and the question whether the diagrams in Fig.~\ref{fig:renormalization} are indeed sufficient for a comprehensive description of the effective model requires, however, further inverstigation.


\subsubsection{Parquet equations}
\label{sec:parquet}

The simplifications leading to $S_{\text{eff}}$ in the previous section allow us now to apply highly advanced diagrammatic techniques for the treatment of our effective problem. The parquet formalism is an approach, which is able to take into account fluctuations in all scattering channels (i.e., spin, charge, and particle-particle or pairing channel) as well as their mutual interaction and screening effects. It constructs the full two-particle vertex function $F$ from a single input quantity, i.e., the fully irreducible vertex function $\Lambda$. Moreover, the one-particle self-energy $\Sigma$ is derived from the full vertex $F\!\equiv\!\gamma^{(2)}$ via the equation of motion which guarantees the consistency between the one- and the two-particle correlation functions.

\begin{figure}[t!] 
\begin{center}
\includegraphics[width=\linewidth]{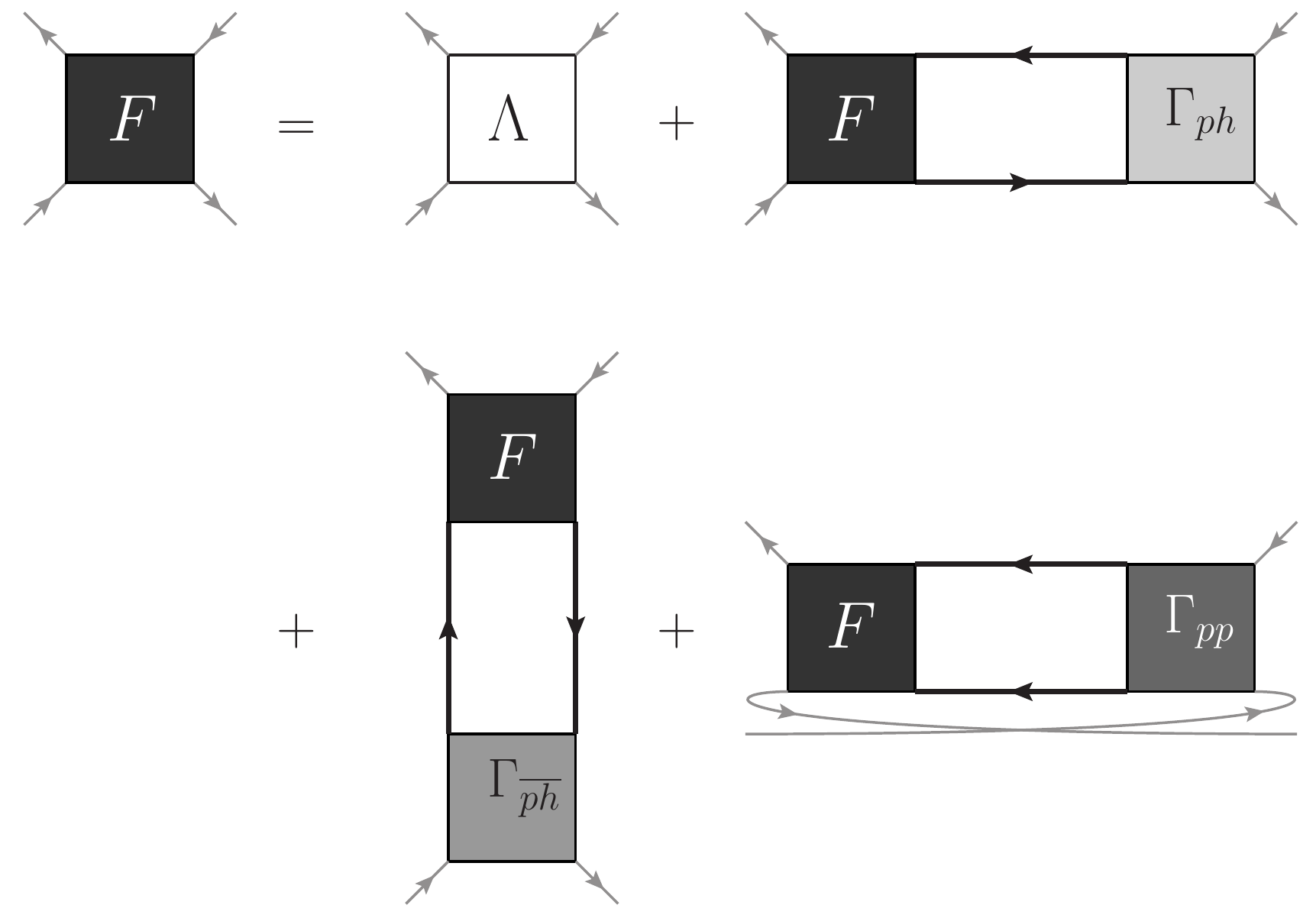}
\caption{Parquet equation.}
\label{fig:parquet}
\end{center}
\end{figure}

\begin{figure}[t!]
\begin{center}
\includegraphics[width=\linewidth]{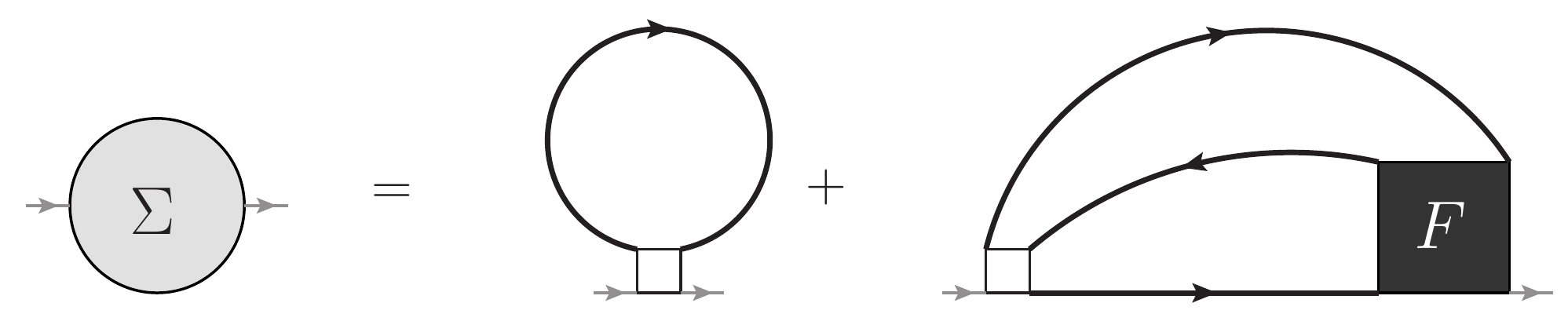}
\caption{Schwinger-Dyson equation. A white box represents the bare interaction $\mathcal{V}$ of the model.}
\label{fig:schwinger-dyson}
\end{center}
\end{figure}

For completeness, let us briefly recapitulate the parquet formalism (for details we refer to Appendix~\ref{app:details} and Refs.~\onlinecite{Rohringer2012,Rohringer2018RMP}): The full vertex $F$ can be decomposed into a fully irreducible vertex $\Lambda$ and vertices $\Phi_r$ reducible in particle-hole, particle-hole transverse and particle-particle channels ($r=ph,\overline{ph},pp$).  This decomposition is expressed by the purely algebraic {\em Parquet equation} 
\begin{equation}
\label{equ:parquet}
F = \Lambda + \Phi_{ph} + \Phi_{\,\overline{ph}} + \Phi_{pp},
\end{equation}
which is represented diagrammatically in Fig.~\ref{fig:parquet}. The reducible vertices $\Phi_r$, in turn, correspond to ladder diagrams in the given scattering channel $r$ which are constructed from the corresponding irreducible vertex $\Gamma_r\!=\!F-\Phi_r$ as $\Phi_r\!=\!F GG \Gamma_r$ (where $G$ denotes the single-particle Green's function). This gives rise to the so-called Bethe-Salpeter (BS) equations in all three scattering channels $r=ph$, $\overline{ph}$, and $pp$. 

\begin{subequations}
\label{equs:BS}
\begin{align}
\label{equ:bs}
    F_{1234} &= \Gamma^{ph}_{1234} + F_{1256} \beta G_{57}G_{86} \Gamma^{ph}_{7834}, \\
    F_{1234} &= \Gamma^{\overline{ph}}_{1234} + F_{4256} \beta G_{57}G_{86} \Gamma^{\overline{ph}}_{1837}, \\
    F_{1234} &= \Gamma^{pp}_{1234} + F_{1256} (-\beta/2) G_{57}G_{68} \Gamma^{pp}_{7834},
\end{align}
\end{subequations}
which are depicted diagrammatically in Fig.~\ref{fig:parquet}. Let us stress that, in contrast to the simple algebraic parquet Eq.~(\ref{equ:parquet}), the BS Eqs.~(\ref{equ:bs}) are complex integral equations since a summation over the repeated indices 5,6,7, and 8 has to be performed. For a fixed input $\Lambda$, Eqs.~(\ref{equ:parquet}) and (\ref{equ:bs}) are iterated for a given one-particle Green's function $G$ (see small inner loop on the right-hand side of the flow diagram in Fig.~(\ref{fig:loop-parquet}). After convergence, a new self-energy is obtained from $F$ via the so-called Schwinger-Dyson equation (or Heisenberg equation of motion) 
\begin{equation}
\label{equ:schwinger_dyson}
    \Sigma_{12} = \mathcal{V}_{12,34}G_{34} + \mathcal{V}_{13,45} \beta^2 G_{36}G_{74}G_{58} F_{67,82},
\end{equation}
which is depicted diagrammatically in Fig.~\ref{fig:schwinger-dyson}. From the updated $\Sigma$, a new Green's function is obtained through the Dyson equation $G\!=\!(G_0^{-1}\!-\!\Sigma)^{-1}$ which is used, in turn, to reiterate the vertex functions (see large outer loop in the flow diagram in Fig.~\ref{fig:loop-parquet}). 

\begin{figure}[t!]
\begin{center}
\includegraphics[width=\linewidth]{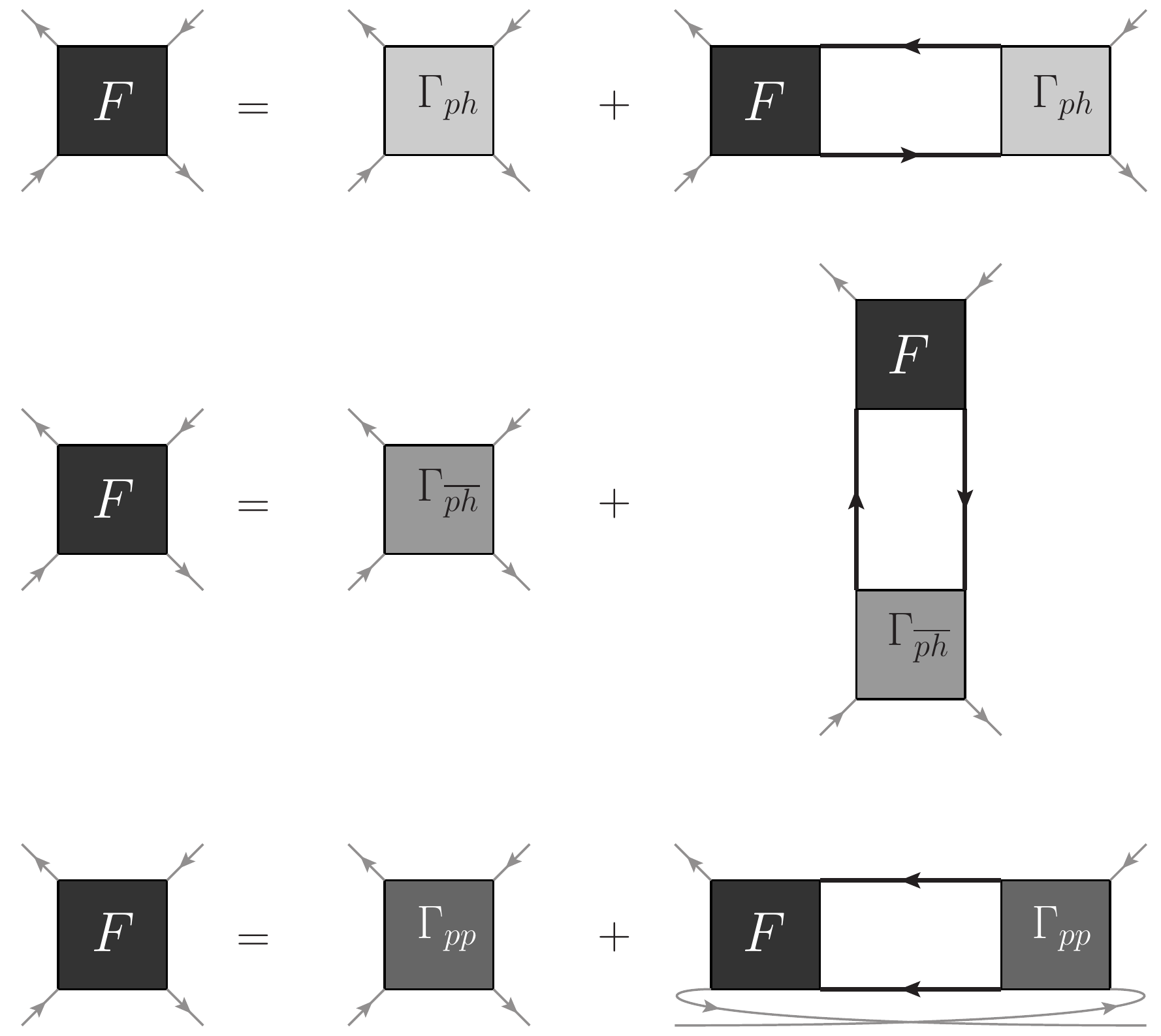}
\caption{Bethe-Salpeter equations in all channels.}
\label{fig:bethe-salpeter}
\end{center}
\end{figure}

Since the fully irreducible vertex $\Lambda$ of a system is in general not known, approximations for this quantity have to be applied. The most simple one, the so-called parquet approximation\cite{Bickers2004}, replaces the fully irreducible vertex simply by the bare interaction. Applying this procedure directly to the Hubbard model in Eq.~\ref{equ:Hubbard}, leads to a theory which is applicable only at weak coupling. Moreover, a truncation to only the lowest Matsubara is less justified for the same reasons as discussed at the end of Sec.~\ref{sec:lowfreqmodel}. On the contrary, the bare interaction of our effective theory $\mathcal{V}$ has inherited the local strong coupling Mott physics from DMFT, so that the high-energy physics is taken into account while changing to the new variables. Hence, the choice $\Lambda = \mathcal{V}$ and working with lowest Matsubaras only should yield reasonable results at strong 
coupling and, at the same time, captures the physics of competing bosonic fluctuations via the ladder and parquet diagrams of the parquet formalism. Let us again stress, that due to the simplified frequency structure of our theory, the parquet equations can be solved on a much finer momentum grid than in previous parquet calculations\cite{Li2017}.

\subsection{Technical details}
\label{sec:technicaldetails}

We solve the impurity problem using an Exact Diagonalization solver and compute local Green's functions and local two-particle vertex functions for a given temperature and chemical potential. The Parquet solver extensively uses CUDA for parallel computations. 

\begin{figure}[t!] 
\begin{center}
\includegraphics[width=\linewidth]{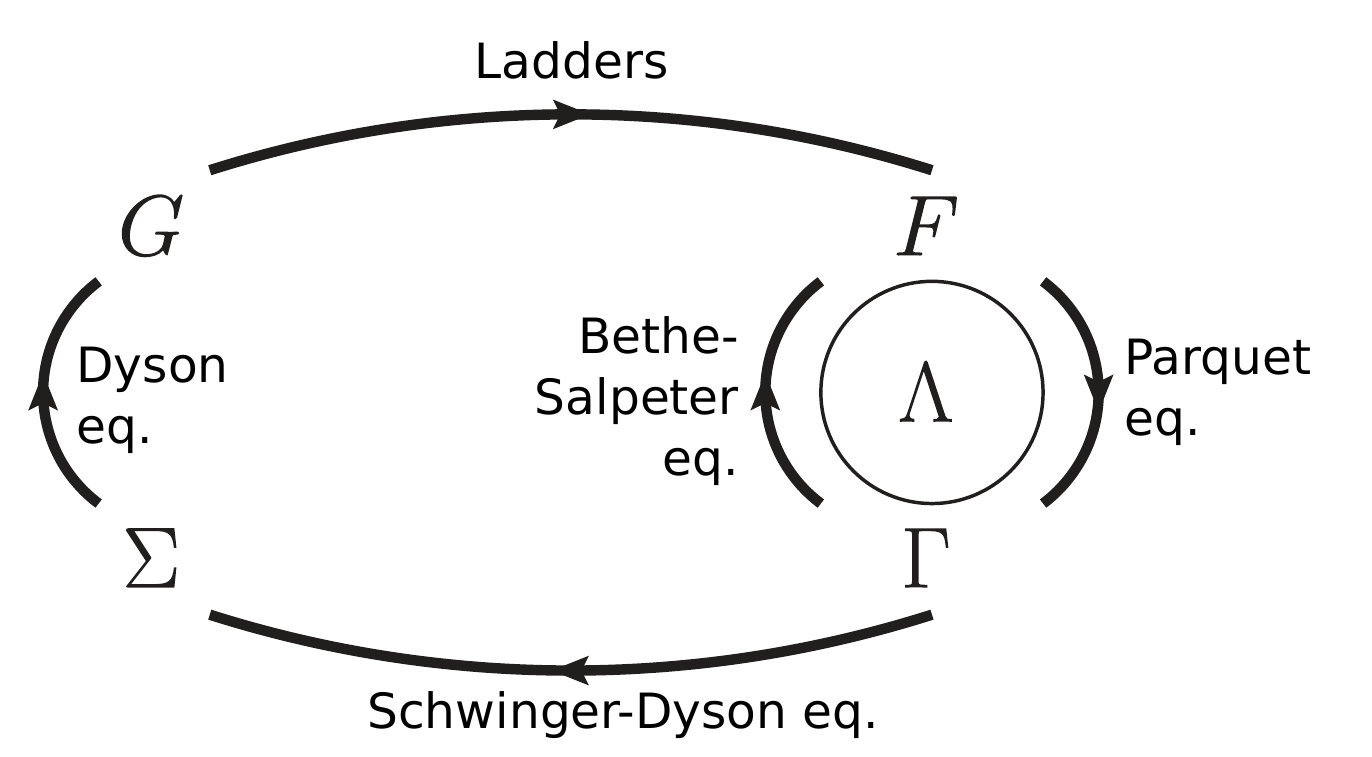}
\caption{Scheme of the algorithm for solving the Parquet equations.}
\label{fig:loop-parquet}
\end{center}
\end{figure}

As discussed in the previous section, the Parquet formalism involves the iteration of an inner and an outer self-consistency loop for the vertex function $F$ and the self-energy $\Sigma$, respectively (see Fig.~\ref{fig:loop-parquet}). The initial guess of $\Sigma$ and $F$ is essential for the convergence of the algorithm to a stable solution. The most convenient way for the initial guess is to choose all irreducible vertices $\Gamma$ and the full vertex $F$ equal to $\Lambda$, and put an imaginary part of the self-energy equal to a very large number (several bandwidths) in order to sufficiently suppress the size of the ladders in the BS equation in the first iteration. To avoid instabilities, damping factors for updates have been introduced:
\begin{subequations}
\begin{align}
\Sigma &= \alpha \Sigma_{n} + (1-\alpha)\Sigma_{n-1},\\
\Gamma &= \alpha \Gamma_{n} + (1-\alpha)\Gamma_{n-1}.
\end{align}
\end{subequations}
We have also enforced conservation of crossing symmetry during the calculations\cite{Tam2013}.

The main advantage of the Parquet formalism with respect to cluster calculations is that it scales algebraically with the size of the system ($O((n_\nu\times n_k\times n_s)^4)$, where $n_s=2$ is a number of spin components) in contrast to the exponential scaling of cluster methods. Let us emphasize that in our case a further reduction of complexity is achieved by $n_\nu\!=\!2$.


\section{Results}
\label{sec:results}

In this section, we present the results for the 2D Hubbard model on a square lattice as obtained by the approach discussed in the previous Sec.~\ref{sec:formalism}. For the nearest-, next-nearest- and next-next-nearest-neighbor hopping parameters we have selected the values $t\!=\!0.25$eV, $t'=-0.2t$ and $t''=0.1t$, respectively, which are relevant for the high-temperature superconducting cuprate compound BSSCO\cite{Nicoletti2010}. The interaction value has been chosen as $U\!=\!8t$ which corresponds to an intermediate-to-strong coupling regime for which at half-filling a Mott metal-to-insulator transition is observed in DMFT\cite{Georges1996}. For the solution of the parquet equations we have discretized the Brillouin zone using $16$ $\mathbf{k}$-points in each direction for most of the calculations and $32$ $\mathbf{k}$-points in specific situations in order to analyze the dependence on the grid size.

\begin{figure}[t!] 
\begin{center}
\includegraphics[width=\linewidth]{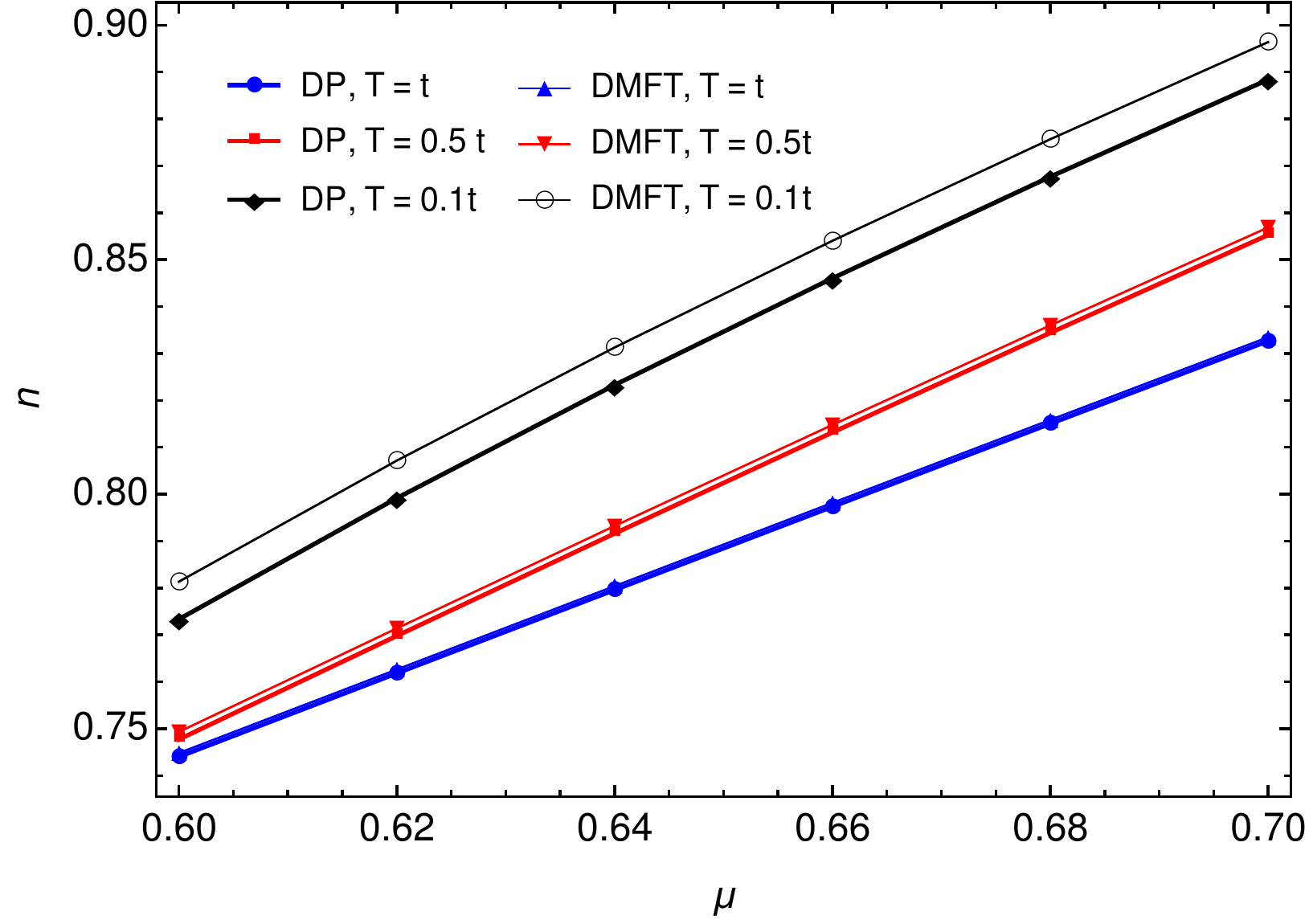}
\caption{Average number of particles per lattice site $n\!=\!\langle \hat{n} \rangle$ vs. chemical potential. Thick and thin lines correspond to dual parquet and DMFT respectively. }
\label{fig:n-mu}
\end{center}
\end{figure}

The calculations have been performed at different values of the chemical potential $\mu$. The corresponding number of particles per lattice site $n\!=\!\langle \hat{n}\rangle\!=\!\langle\hat{n}_{i\uparrow}\rangle\!+\!\langle\hat{n}_{i\downarrow}\rangle$ or, correspondingly, the doping $\delta\!=\!1\!-\!n$ have been calculated in the standard way by summing the one-particle Green's function $G(\nu,\mathbf{k})$ over frequencies and momenta, i.e., $n\!=\!\frac{1}{\beta}\sum_{\nu\mathbf{k}}G(\nu,\mathbf{k})e^{i\nu\delta}=1/2+\frac{1}{\beta}\sum_{\nu\mathbf{k}}\operatorname{Re} G(\nu,\mathbf{k})$ with $\delta\!\rightarrow\!+0$. For the lowest Matsubara frequency $\lvert\nu\rvert\!=\!\pi/\beta$, the one-particle Green's function of our downfolded model has been obtained by means of the parquet equations, while for the larger frequencies the Green's function of DMFT has been used for the calculation. This procedure is justified by the fact, that nonlocal correlations (or correlations in general) affect most strongly the low-energy physics while at higher frequencies the Green's function approaches its non-interacting value.  However, it should be mentioned that a rigorous verification of this assumption requires a systematic extension of the effective low-frequency model to two and more Matsubara frequencies, which will be considered in a future research work (see outlook in Sec.~\ref{sec:conclusions}). In any case, while close to half-filling and at very low temperatures the dependence of $n$ on $\mu$ can exhibit a rather complicated singular behavior\cite{Nourafkan2018}, Fig.\ref{fig:n-mu} shows that for higher values of $T$ considered here, $n(\mu)$ is a regular featureless function which does not differ much from the curve obtained using DMFT. For lower values of temperature however the difference is increased but still stays regular and relatively small near the optimal doping. This has allowed us to straightforwardly recast our results in the following sections, which have been originally obtained for fixed values of $\mu$, in terms of the filling $n$ (or, equivalently, of the doping $\delta\!=\!1\!-\!n$).

\subsection{Phase diagram}
\label{sec:phasediagram}

\begin{figure}[t!]
\begin{center}
\includegraphics[width=\linewidth]{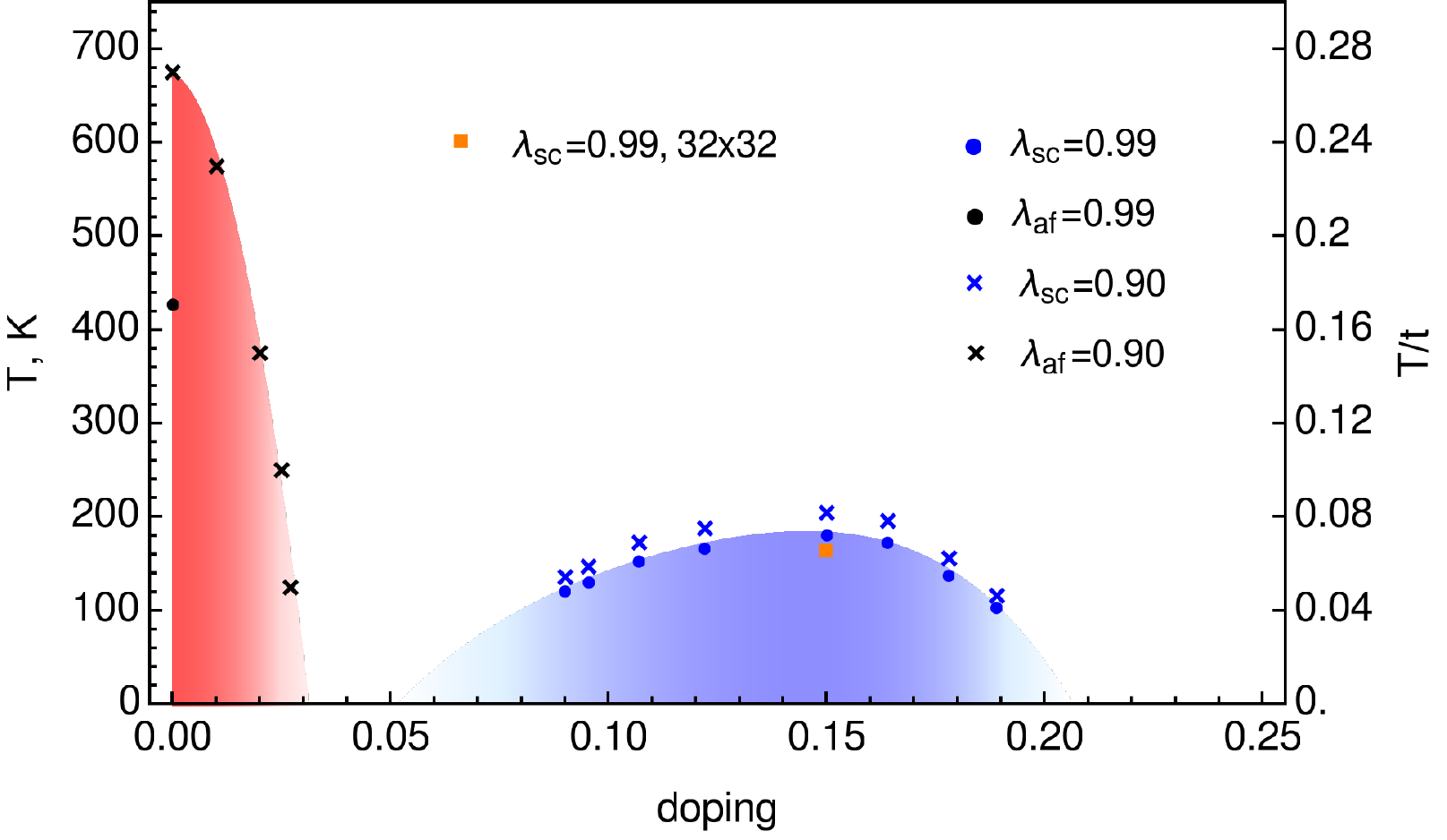}
\caption{Phase diagram of the hole-doped Hubbard model describing a CuO$_2$ monolayer of BSSCO at $U=8t$. Dots and crosses correspond to eigenvalues equal to 0.99 and 0.9 in the magnetic (black) and pairing (blue) channels, respectively, for a 16x16 lattice. The blue region indicates the superconducting and the red region the antiferromagnetic phase. The orange square indicates $\lambda_\text{SC}=0.99$ for a 32x32 lattice. For clarity, the temperature is measured both in units of hoping (right scale) and in Kelvins (left scale).}
\label{fig:pdcolor}
\end{center}
\end{figure}

Fig.~\ref{fig:pdcolor} shows the phase diagram of the Hubbard model as a function of temperature and doping. Close to half-filling ($\delta\!=\!0$), we observe an antiferromagnetic phase with a quasi-long-range order while at larger values of the doping a superconducting phase emerges. 

The notion of ``phase'' in the present context should be clarified. Let us stress that here it does {\sl not} refer to a state with a real long-range order in the thermodynamics limit which would be signaled by a diverging spin or pairing susceptibility at the transition point. In fact, in 2D a long-range magnetic order is restricted to $T\!=\!0$ according to the Mermin Wagner theorem\cite{Mermin1966}. On the other hand, a superconducting state of Kosterlitz-Thouless type can exist in 2D at finite temperatures, characterized by a diverging correlation length but a finite susceptibility\cite{Kosterlitz1973} which can, however, not be accessed by our method. 

Instead, the colored regions in Fig.~\ref{fig:pdcolor}, which mark the different phases, indicate the areas where the corresponding antiferromagnetic or superconducting fluctuations become very large which is reflected in a sizable magnitude of the corresponding susceptibilities $\chi_r$ or vertex functions $F_r$. These correlation functions are obtained from the Bethe Salpeter equation 
\begin{equation}
    \label{equ:BSexplizit}
    \sum_{P_1}(\delta_{PP_1}-\Gamma_{r}^{PP_1Q}G_{P_1}G_{P_1+Q})F_{r}^{P_1P'Q}=\Gamma_{r}^{PP'Q}
\end{equation}
in the spin ($r\!=\!m$) and pairing ($r=pp$) [and, in the next section, charge ($r\!=\!d)$] channels, respectively. Here, we have used a condensed notation where $P\!=\!(\omega,\mathbf{k})$ [$Q\!=\!(\nu,\mathbf{q})$] corresponds to a composite index for a fermionic [bosonic] Matsubara frequency and a momentum vector (for the exact definition of all vertex functions and channel indices we refer to the Appendix). For a given irreducible vertex $\Gamma_r$, Eq.~(\ref{equ:BSexplizit}) represents a matrix equation in the $PP'$ space for the calculations of $F_r$. Since the evaluation of $F_r$ requires the inversion of the operator $\delta_{PP'}-\Gamma_r^{PP'Q}G_PG_{P+Q}$, it is obvious that $F_r$ strongly increases when the largest (leading) eigenvalue $\lambda_r^Q$ of the kernel $\Gamma_r^{PP'Q}G_PG_{P+Q}$ approaches $1$:
\begin{equation}
    \label{equ:eigenvalue}
    \sum_{P_1}\Gamma_r^{PP_1Q}G_{P_1}G_{P_1+Q}\Phi_r^{P_1Q}=\lambda_r^Q\Phi_r^{PQ},
\end{equation}
where $\Phi_r^{PQ}$ denotes the corresponding eigenvector. In this section, we are interested in antiferromagnetic spin and pairing instabilities which are related to the eigenvalues $\lambda_{\text{AF}}\!\equiv\!\lambda_m^{\omega\!=\!0,\mathbf{q}\!=\!(\pi,\pi)}$ and $\lambda_{\text{SC}}\!\equiv\!\lambda_{pp}^{\omega\!=\!0,\mathbf{q}=(0,0)}$. The values of $\lambda_{\text{AF}}$ and $\lambda_{\text{SC}}$ have been determined for various fillings $\delta$ starting from the high-temperature unordered state. We have then gradually decreased the temperature until one of the eigenvalues reached a value close to one indicating the proximity of the corresponding instability. Let us note that for each value of the temperature, we had to perform a series of calculations with fixed chemical potential $\mu$ until the desired filling $n$ (or doping $\delta$) was found.

Following the described strategy, we can identify  an antiferromagnetic region (red shaded area) at small doping in Fig.~\ref{fig:pdcolor} whose border (black crosses in Fig. ~\ref{fig:pdcolor}) is defined by the corresponding antiferromagnetic eigenvalue $\lambda_{\text{AF}}$ of $GG\Gamma_{\text{m}}$ approaching a value of $0.9$ upon lowering the temperature. A further decrease in temperature (for a given doping) allows us to find an eigenvalue of $\lambda_{\text{AF}}\!\sim\!0.99$ even closer to one (black dot in Fig.~\ref{fig:pdcolor}). For larger values of $\delta$, we observe an increase of the superconducting eigenvalue $\lambda_{\text{SC}}$ defining the border of the superconducting region of the phase diagram (blue shaded area). Interestingly, the difference between the temperatures at which $\lambda_{\text{SC}}\!=\!0.9$ (blue crosses) and $\lambda_{\text{SC}}\!=\!0.99$ (blue dots) is much smaller than for the antiferromagnetic case where $\Delta T\!\sim\!300$K at $\delta\!\sim\!0$. This indicates that strong superconducting fluctuations are restricted to a small area of the phase diagram around the superconducting dome while the AF fluctuations seem to be sizable in a larger region of the phase diagram. This finding is also consistent with the Mermin Wagner theorem which predicts the ordered state only at $T\!=\!0$ while rather large AF fluctuations extend to a wide region of the phase diagram. 

\subsection{Leading eigenvalues}
\label{sec:eigenvalues}

\begin{figure}[t!]
\begin{center}
\includegraphics[width=\linewidth]{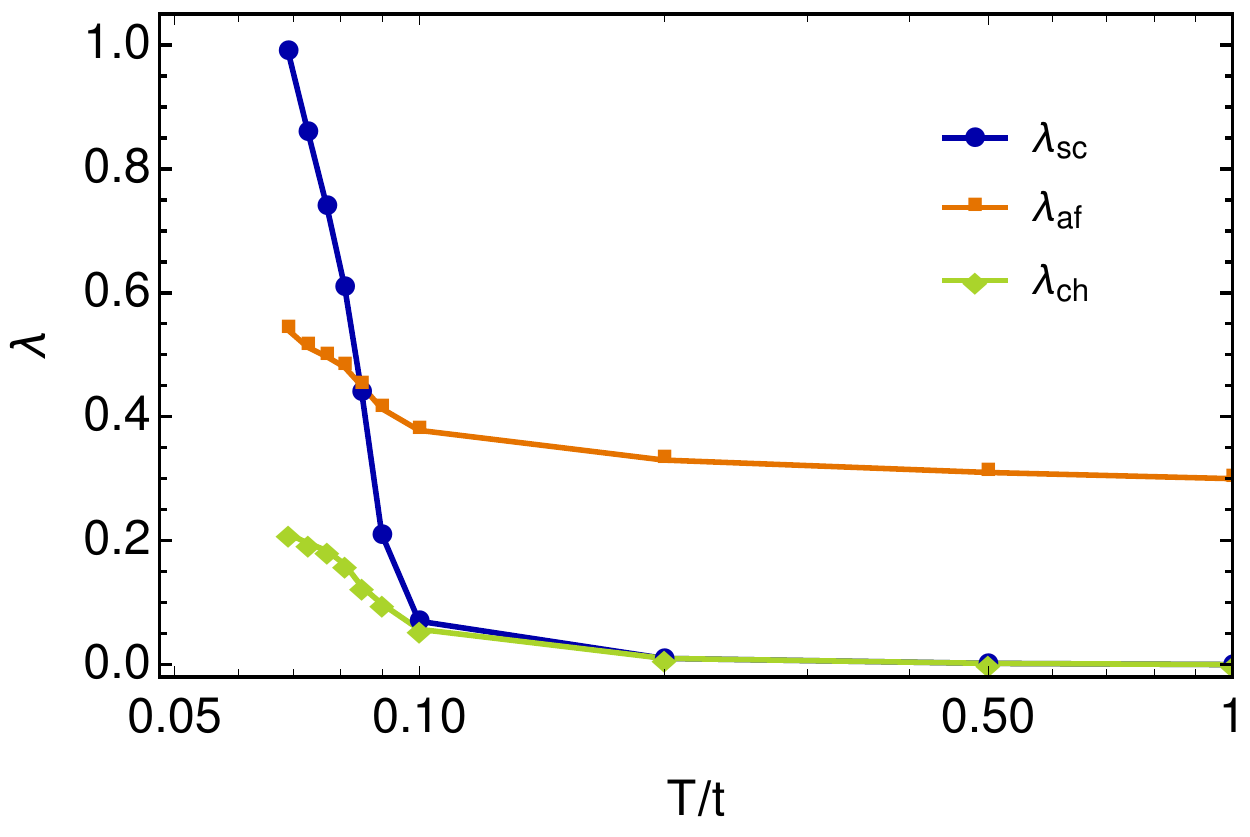}
\caption{Dependence of leading superconducting, spin and charge eigenvalues on temperature at the optimal doping $\delta=0.15$.}
\label{fig:leadeigen1}
\end{center}
\end{figure}

\begin{figure}[t!]
\begin{center}
\includegraphics[width=\linewidth]{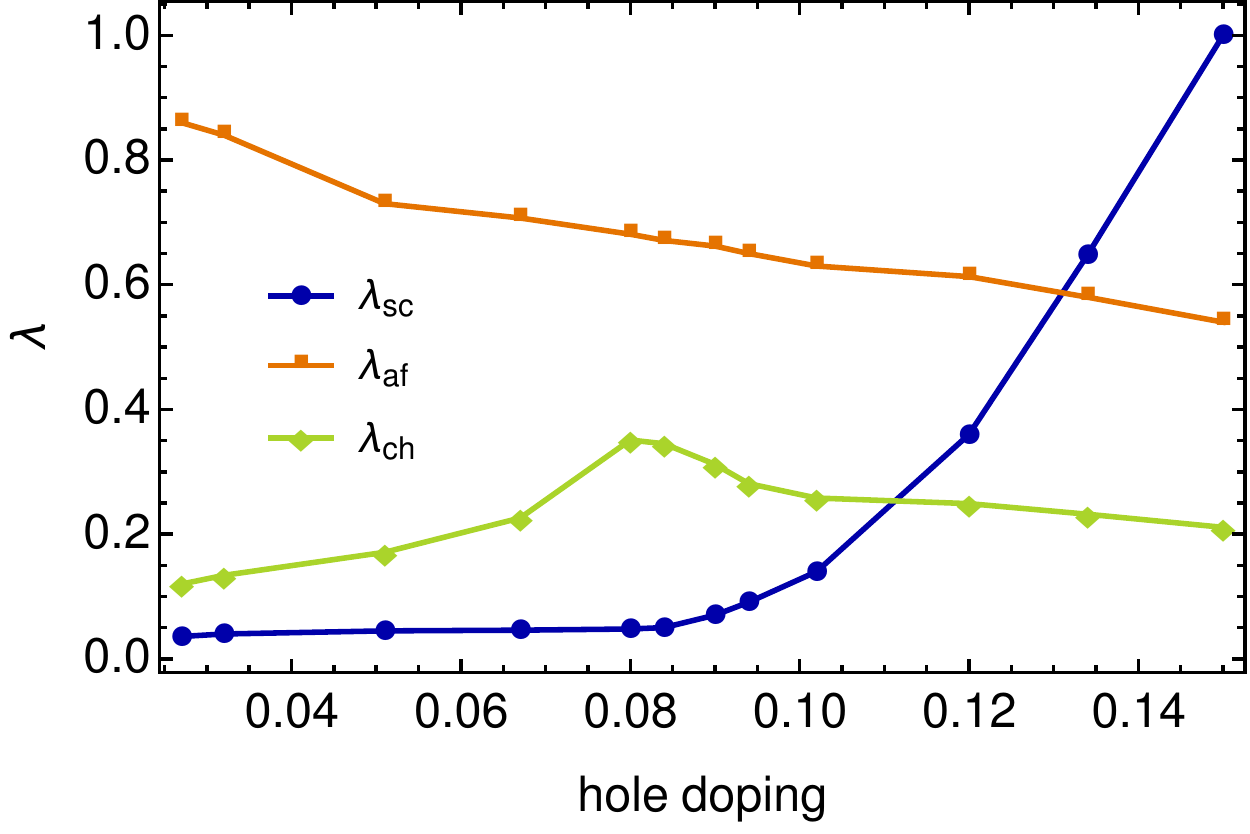}
\caption{Dependence of leading superconducting, spin and charge eigenvalues on temperature at the doping $\delta=0.025$.}
\label{fig:leadeigen3}
\end{center}
\end{figure}

\begin{figure}[t!]
\begin{center}
\includegraphics[width=\linewidth]{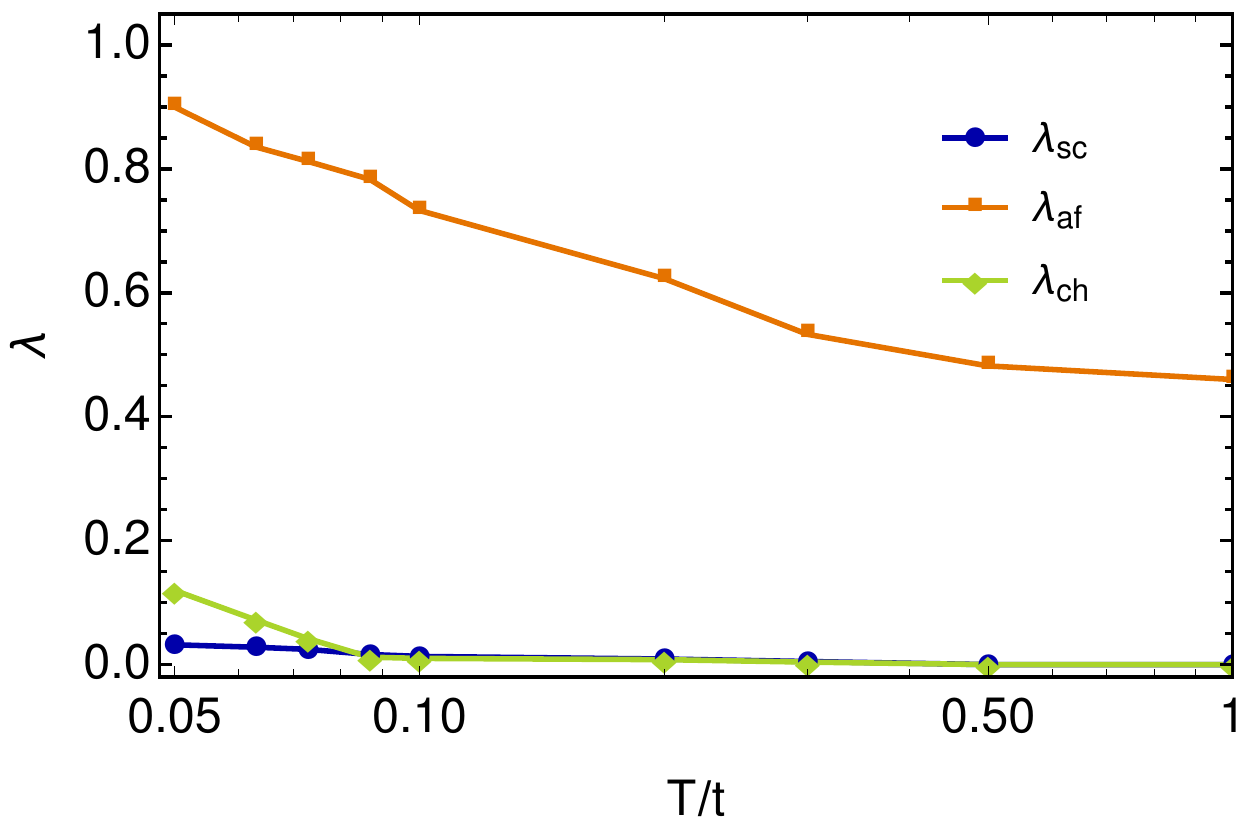}
\caption{Dependence of leading superconducting, spin and charge eigenvalues on doping at the temperature corresponding to the top of the superconducting dome $T=0.075t$.}
\label{fig:leadeigen2}
\end{center}
\end{figure}

In order to obtain further insights into the nature of the different phases and their related fluctuations, we present here a more detailed analysis of the leading eigenvalues for the antiferromagnetic spin ($\lambda_{\text{AF}}$), the superconducting ($\lambda_{\text{SC}}$) and the uniform charge ($\lambda_{\text{CH}}\!\equiv\!\lambda_d^{\nu\!=\!0,\mathbf{q}\!=\!(0,0)}$) channel in Figs.~\ref{fig:leadeigen1}-\ref{fig:leadeigen2}. 

Figure~\ref{fig:leadeigen1} shows $\lambda_{\text{AF}}$ and $\lambda_{\text{SC}}$ as a function of the temperature at optimal doping $\delta\!=\!0.15$ where the superconducting dome reaches its maximum. The antiferromagnetic eigenvalue $\lambda_{\text{AF}}\!\sim\!0.4$ is considerable in a broad temperature range and increases slightly only at the lowest temperature. The superconducting eigenvalue $\lambda_{\text{SC}}$, on the other hand, is almost $0$ at higher temperatures and increases rapidly only at temperatures very close to the superconducting dome where it crosses and eventually becomes larger than $\lambda_{\text{AF}}$. At the lower value of the doping ($\delta\!=\!0.025$), the superconducting fluctuations are strongly suppressed over the entire accessible temperature range as it can be seen in Fig.~\ref{fig:leadeigen3}. As expected, the antiferromagnetic fluctuations are larger close to half-filling and increase upon lowering the temperature. Finally, let us analyze the leading eigenvalues as a function of doping at the maximum temperature of the superconducting dome (see Fig.~\ref{fig:leadeigen2}). As expected, for lower values of $\delta$ antiferromagnetic fluctuations dominate while $\lambda_{\text{SC}}$ is almost zero. Upon increasing doping one approaches the superconducting dome which is reflected in a strong increase of $\lambda_{\text{SC}}$ which eventually crosses the leading antiferromagnetic eigenvalue. The latter exhibits a much weaker doping dependence and is sizable also in the region where $\lambda_{\text{SC}}$ approaches $1$. We can, hence, conclude that antiferromagnetic fluctuations are relevant in the entire phase diagram with an expected maximum at low values of doping and temperatures while superconducting fluctuation are sharply restricted to a region very close to the superconducting dome. This also agrees with the fact that antiferromagnetic fluctuations provide the effective ``pairing glue'' for the electrons in order to form a superconducting state and, hence, should not be small close to the superconducting area of the phase diagram.

\begin{figure*}[t!]
\begin{center}
\includegraphics[width=\linewidth]{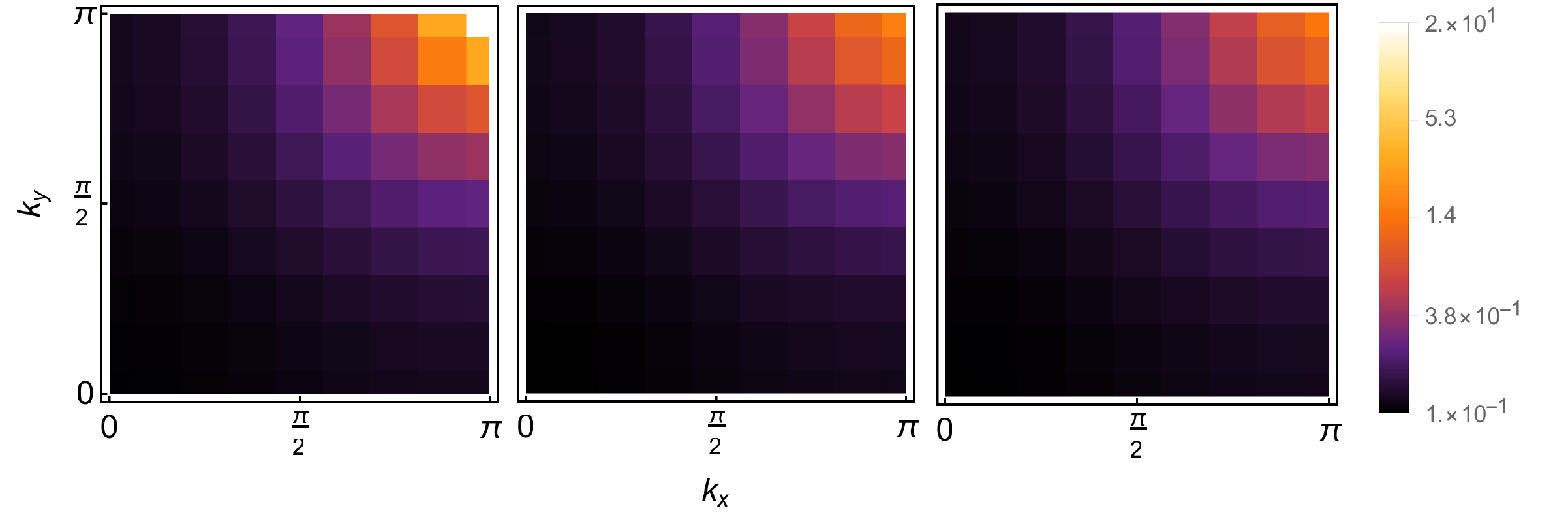}
\caption{Spin susceptibilities calculated for a 16x16 lattice at three different values of the doping $\delta=0.03$ (left), $0.15$ (center) and $0.19$ (right) at $T=0.07t$ where the superconducting dome reaches its maximum.}
\label{fig:afchi}
\end{center}
\end{figure*}

Let us discuss the role of uniform [$\mathbf{q}=(0,0)$] charge fluctuations which are reflected by the behavior of the corresponding eigenvalue $\lambda_{\text{CH}}$. As expected for a system with a repulsion between the particles, charge fluctuations are suppressed in a large region of the phase diagram. In particular, close to half-filling ($\delta=0.025$ in Fig.\ref{fig:leadeigen3}) $\lambda_{\text{CH}}$ is almost 0 for all temperatures with a small increase for $T\rightarrow0$. Interestingly, at optimal doping (Fig.~\ref{fig:leadeigen1}) this increase upon lowering $T$ is much more pronounced which indicates that in the region where $d$-wave superconductivity prevails also charge fluctuations become non-negligible. Turning our attention to the doping dependence of $\lambda_{\text{CH}}$ in Fig.~\ref{fig:leadeigen2}, we observe a very interesting feature: For the maximal superconducting temperature $T=0.075$, $\lambda_{\text{CH}}$ exhibits a maximum at $\delta\!\sim\!0.08$. One can speculate that this remarkable behavior might be a signature of a quantum critical point triggered by charge fluctuations which is responsible for the physics observed in a wide range of the phase diagram although it has to be said that such features are typically found at larger values of doping in DCA calculations.\cite{Yang2011}. Moreover, enhanced charge fluctuations support the picture that AF order breaks down due to the formation of a phase separated state\cite{Khatami2010} in which the holes form (maybe virtually) droplets in the AF background as predicted in Ref.~\onlinecite{Stepanov2018}. While in the latter paper only nonlocal fluctuations in the spin channel were taken into account, our dual parquet calculations allow to access the charge channel also, and the emergence of an increased charge susceptibility at $\mathbf{q}=(0,0)$ indeed provides a new and solid argument in favor of the phase separation picture. We note however that our results cannot exclude other scenarios for the breakdown of antiferromagnetism, such as striped phase (see Sec.~\ref{sec:typefluct}) for a different choice of the model parameters, in particular the value of $t'$.


Let us finally comment on the dependence of the results on the number of $\mathbf{k}$-points in the Brillouin zone used for the calculations. While the AF fluctuations exhibit a rather strong dependence on the momentum grid\cite{Schaefer2015-2} due to the exponentially large correlation length at low values of $T$, the difference between $16$ and $32$ $\mathbf{k}$-points (see orange dot in Fig.~\ref{fig:pdcolor}) for the definition of the superconducting region at optimal doping is very small, which indicates the stability of the superconducting dome w.r.t. to larger momentum grids.

\subsection{Type of superconducting and spin fluctuations}
\label{sec:typefluct}

\begin{figure}[t!]
\begin{center}
\includegraphics[width=7.5 cm]{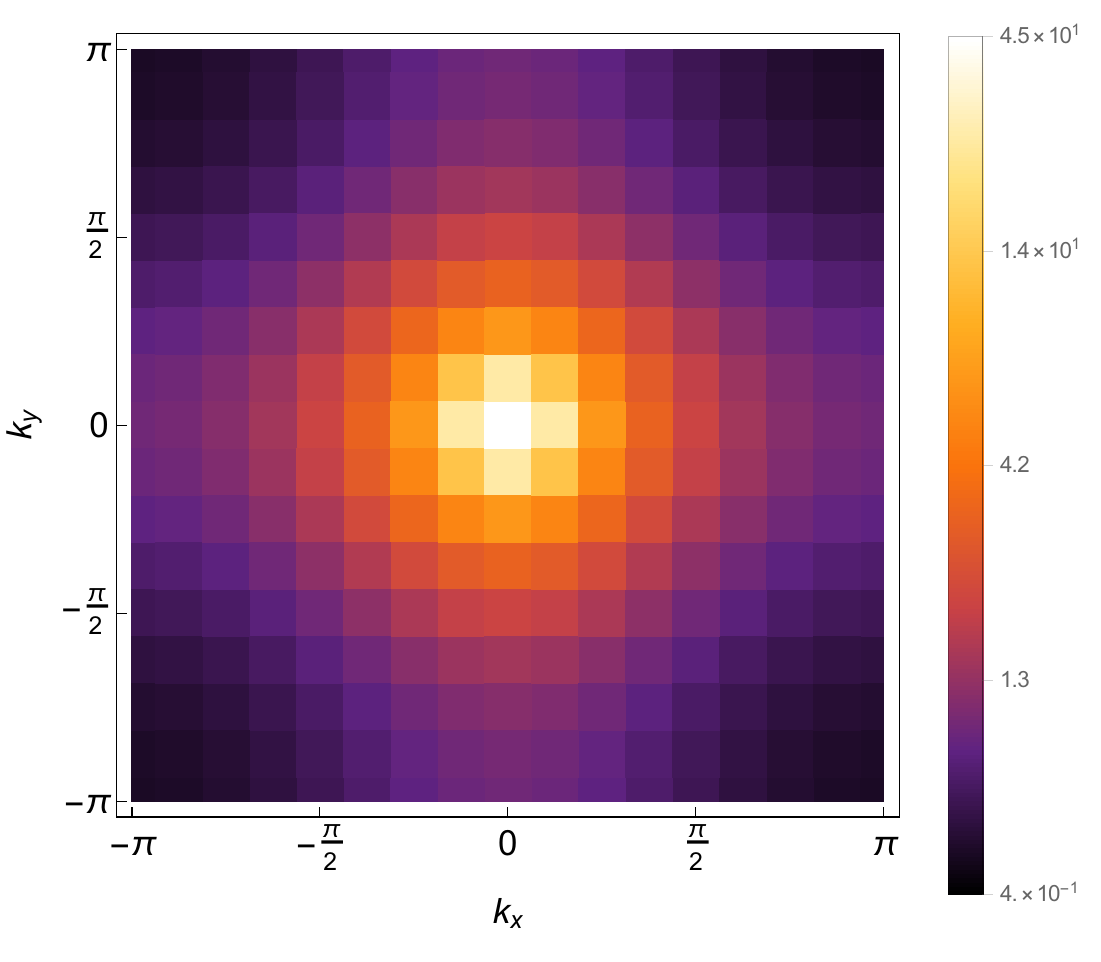}
\caption{$d$-wave superconducting susceptibility at optimal doping for a 16x16 lattice.}
\label{fig:scchi}
\end{center}
\end{figure}

\begin{figure*}[t!]
\begin{center}
\includegraphics[width=\linewidth]{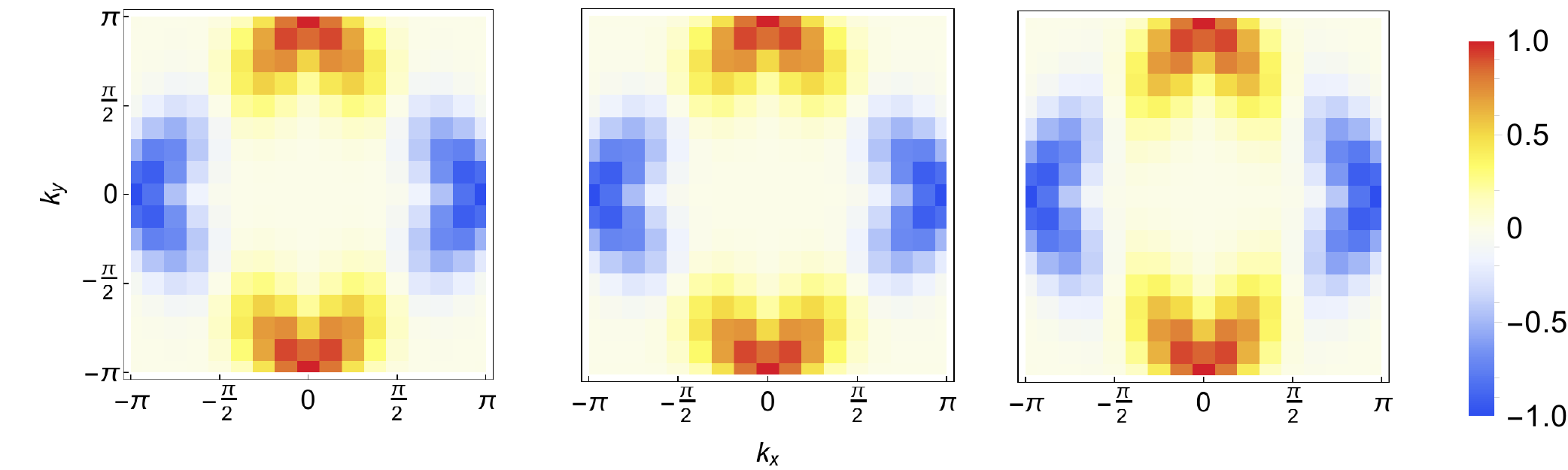}
\caption{Eigenfunctions of Bethe-Salpeter equations in the particle-particle channel for a 16x16 lattice at the dopings $\delta=0.09$ (left), $0.15$ (center), and $0.19$ along the superconducting dome.}
\label{fig:scchieigen}
\end{center}
\end{figure*}

\begin{figure}[t!]
\begin{center}
\includegraphics[width=7.5 cm]{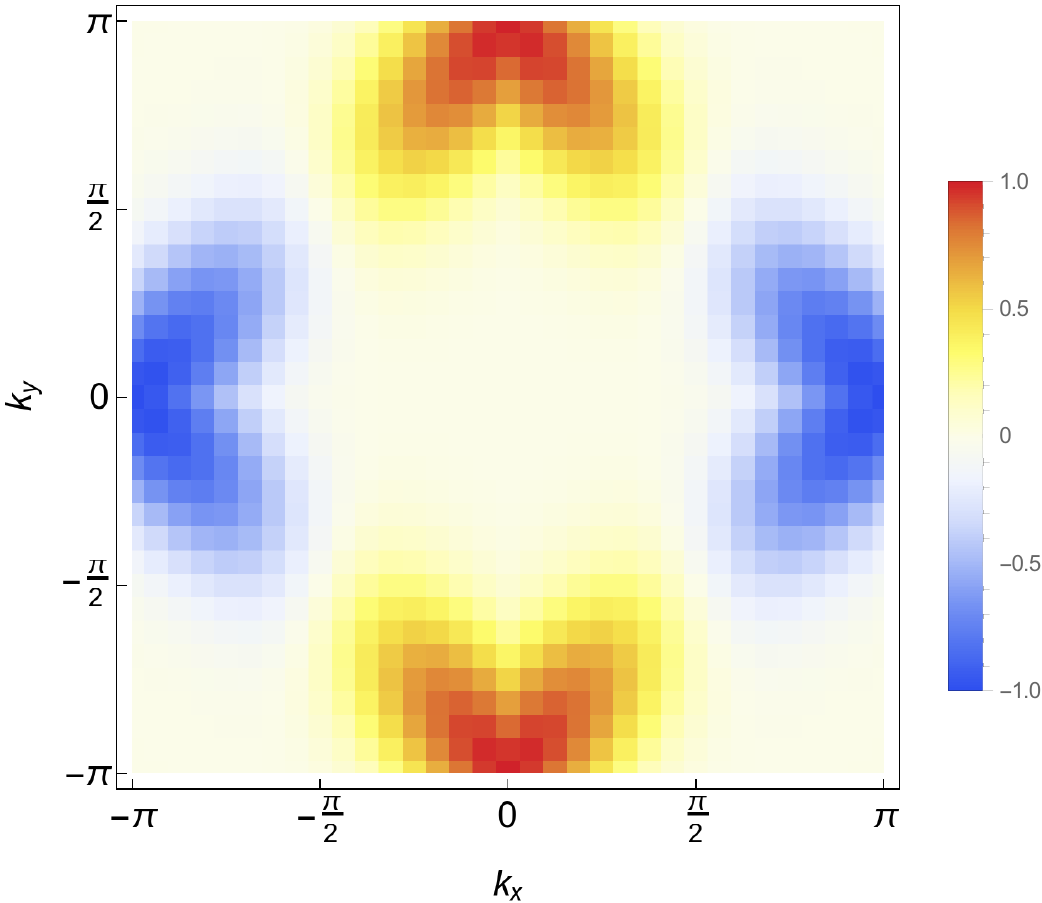}
\caption{An eigenfunction of Bethe-Salpeter equations in the particle-particle channel at optimal doping for a 32x32 lattice.}
\label{fig:eigenvectorpp32}
\end{center}
\end{figure}

In this section, we investigate the type of the magnetic and superconducting fluctuations which we have identified in the previous sections. To this end we analyze the momentum dependence of the corresponding susceptibilities and eigenvectors of the BS kernels which are related to the leading eigenvalues [see Eq.~(\ref{equ:eigenvalue})].

An evaluation of the physical susceptibility $\chi_r^Q$ requires the summation of the generalized susceptibility $\chi_r^{PP'Q}$ 
\begin{align}
    \label{equ:constructgenchi}
    \chi_r^{PP'Q}&=-\beta G_PG_{P+Q}\delta_{PP'}\nonumber\\&-G_{P}G_{P+Q}F_r^{PP'Q}G_{P'}G_{P'+Q},
\end{align}
over the fermionic indices $P$ and $P'$ (i.e., over the fermionic Matsubara frequencies $\nu$ and $\nu'$ as well as over the momenta $\mathbf{k}$ and $\mathbf{k'}$):
\begin{equation}
    \label{equ:sucsphys}
    \chi_r^Q=\sum_{PP'}\chi_r^{PP'Q}.
\end{equation}
For the evaluation of this expression we consider only the two lowest Fermionic Matsubara frequencies for the summation over the fermionic indices, because only $\nu\!=\!\pm\frac{\pi}{\beta}$ are present in our downfolded effective action.

One could think about a more sophisticated solution in analogy to the above calculation of $n(\mu)$, where we have adopted the DMFT Green's function for the summation over the higher frequencies $\lvert\nu\rvert\!>\!\frac{\pi}{\beta}$ (see the discussion at the beginning of Sec.~\ref{sec:results}). However, using the DMFT results for the {\it two-particle} quantity $\chi_r^{PP'Q}$ for $\lvert\nu\rvert,\lvert\nu'\rvert\!>\!\frac{\pi}{\beta}$ in Eq.~(\ref{equ:sucsphys}) is questionable as the DMFT susceptibilities are obtained from a single Bethe-Salpeter equation in the given channel $r$ neglecting mutual screening effects between the channels. Hence, such a construction might add a bias to the (channel-unbiased) parquet results for the low-frequency model and, hence, worsens the results. Moreover, it is currently unclear how to treat the situation where $\nu\!=\!\pm\frac{\pi}{\beta}$ and $\lvert\nu'\rvert\!>\!\frac{\pi}{\beta}$ (or vice versa) which would require a deeper analysis of this problem by tracing the calculation of the generalized susceptibilities within the downfolding procedure. In general we expect, that the above described calculation of susceptibilities, using only the lowest fermionic Matsubara frequencies, yields reasonable results because long-range bosonic fluctuations are typically assiciated with low energies. Nevertheless, the problem of a consistent evaluation of $\chi_r^Q$ within our low-frequency model requires further investigation.

Figure~\ref{fig:afchi} shows the spin susceptibility $\chi_m^{\omega,\mathbf{q}}$ for the bosonic frequency $\omega\!=\!0$ as a function of $q_x$ and $q_y$. Consistent with the analysis of the eigenvalues, at low doping $\delta\!=\!0.03$ antiferromagnetic spin fluctuations dominate which is indicated by a strong peak at $\mathbf{q}\!=\!(\pi,\pi)$. For larger dopings, this maximum of $\chi_s^{\omega\!=\!0,\mathbf{q}}$ is reduced but, nevertheless, remains at the wave vector $\mathbf{q}\!=\!(\pi,\pi)$. This proves that the spin fluctuations, which have been shown to be sizable in the entire phase diagram, are indeed of antiferromagnetic nature in the whole parameter regime under investigation. This is also consistent with the results obtained earlier by DCA in Ref.~\onlinecite{Chen2013}. In particular, our data preclude the emergence of spin stripes which have been discussed controversially on both the experimental\cite{Fujita2012,Tranquada2012} and the theoretical\cite{Vojta2009} side. This further strengthens the assumption that antiferromagnetism break down due to the formation of a phase separated state rather than due to a stripe ordered state as discussed at the end of Sec.~\ref{sec:eigenvalues}.

Let us now turn our attention to the pairing fluctuations. As expected, close to a superconducting instability, $\chi_{pp}^{\omega\!=\!0,\mathbf{q}}$ is strongly peaked at $\mathbf{q}\!=\!0$ close to the superconducting dome. This feature is well reproduced by our numerical data in Fig.~\ref{fig:scchi} which also gives further support for the validity of our approximation for the calculation of $\chi_r^Q$. The physically more interesting question, however, concerns the nature of the superconducting fluctuations. Such an information is encoded in the momentum dependence of the eigenvectors $\Phi_{pp}^{P(Q=0)}$ [see Eq.~(\ref{equ:eigenvalue})] of the BS equation which corresponds to the leading superconducting eigenvalue. In Fig.~\ref{fig:scchieigen} these eigenfunctions are shown for three different values of the doping at temperatures close to the superconducting dome (i.e., where the corresponding leading eigenvalue is $0.99$). One observes a clear $d$-wave structure corresponding to an eigenfunction $\Phi(\mathbf{k})\!\sim\!f_d(\mathbf{k})\!=\!\cos(k_x)\!-\!\cos(k_y)$ for all values of the doping. In order to quantify this qualitative finding we have projected $\Phi(\mathbf{k})$ onto the $s$-, $p$- and $d$-wave form factors [$f_s(\mathbf{q})\!=\!1$, $f_p(\mathbf{q})\!=\!\cos(k_x)$ or $\cos(k_y)$ and $f_d(\mathbf{k})\!=\!\cos(k_x)\!-\!\cos(k_y)$], respectively. This results in a fraction of 0.86, 0.92 and 0.85 for the $d$-wave, 0.04, 0.03 and 0.05 for the $p$-wave and 0.1, 0.05 and 0.1 for the $s$-wave contribution for $\delta\!=\!0.09$, 0.15 and 0.19, respectively. These numerical results, hence, clearly reflect the $d$-wave nature of the superconducting fluctuations. Let us stress that our results for the susceptibilities are stable with respect to the size of the momentum grid as it is illustrated in Fig.~\ref{fig:eigenvectorpp32} for a 32x32 $\mathbf{k}$-lattice.

\subsection{Pairing glue and origin of the dome structure}
\label{sec:analysisdome}

\begin{figure}[t!]
\begin{center}
\includegraphics[width=\linewidth]{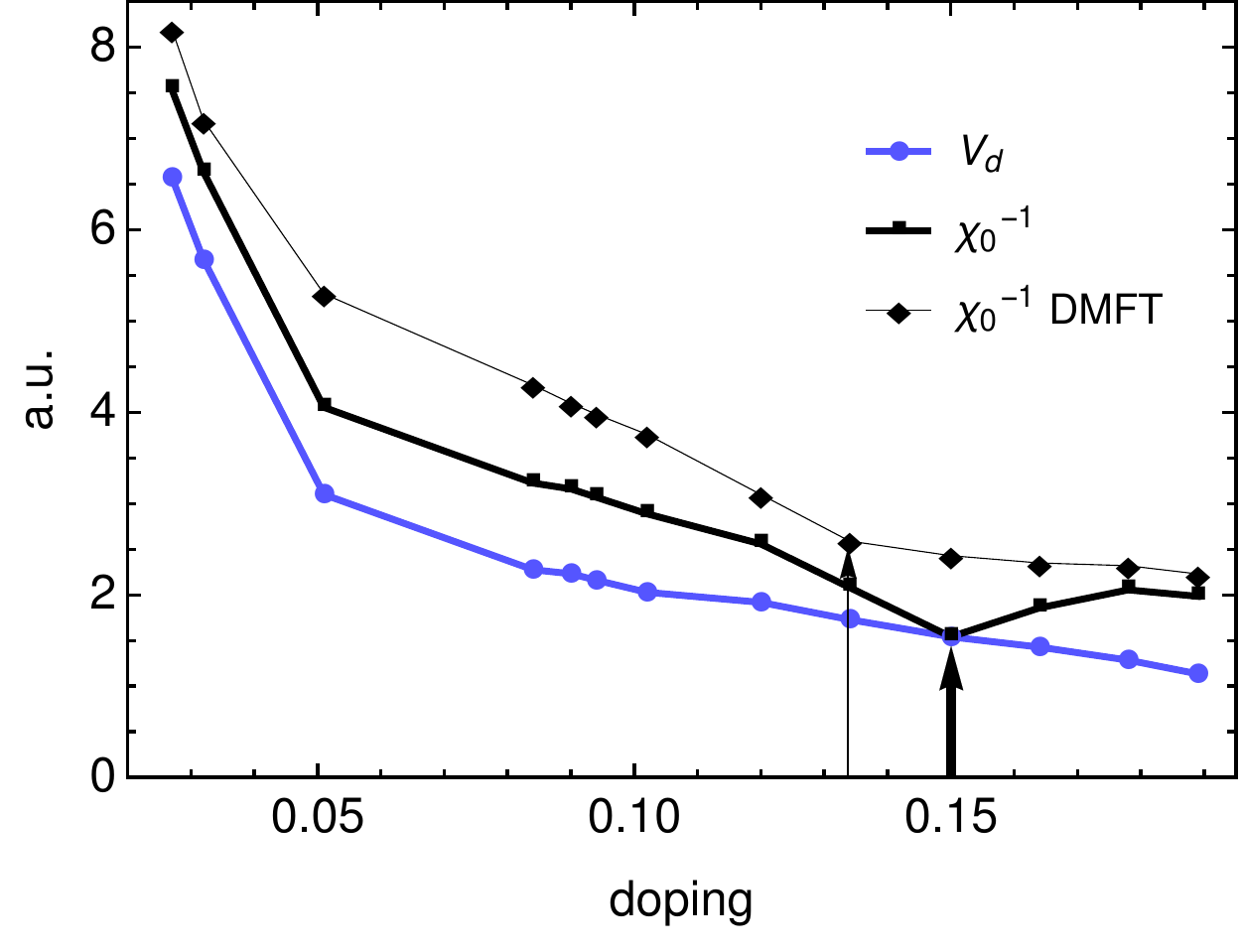}
\caption{Doping dependence of the effective $d$-wave attraction and the inverse bubble at the temperature corresponding to the top of the superconducting dome, $T=0.075t$. Thin curve depicts the DMFT result for the inverse bubble. Black arrows point out at the peculiarities in the behavior of $\chi_0^{-1}$.}
\label{fig:Vd}
\end{center}
\end{figure}

\begin{figure}[t!]
\begin{center}
\includegraphics[width=\linewidth]{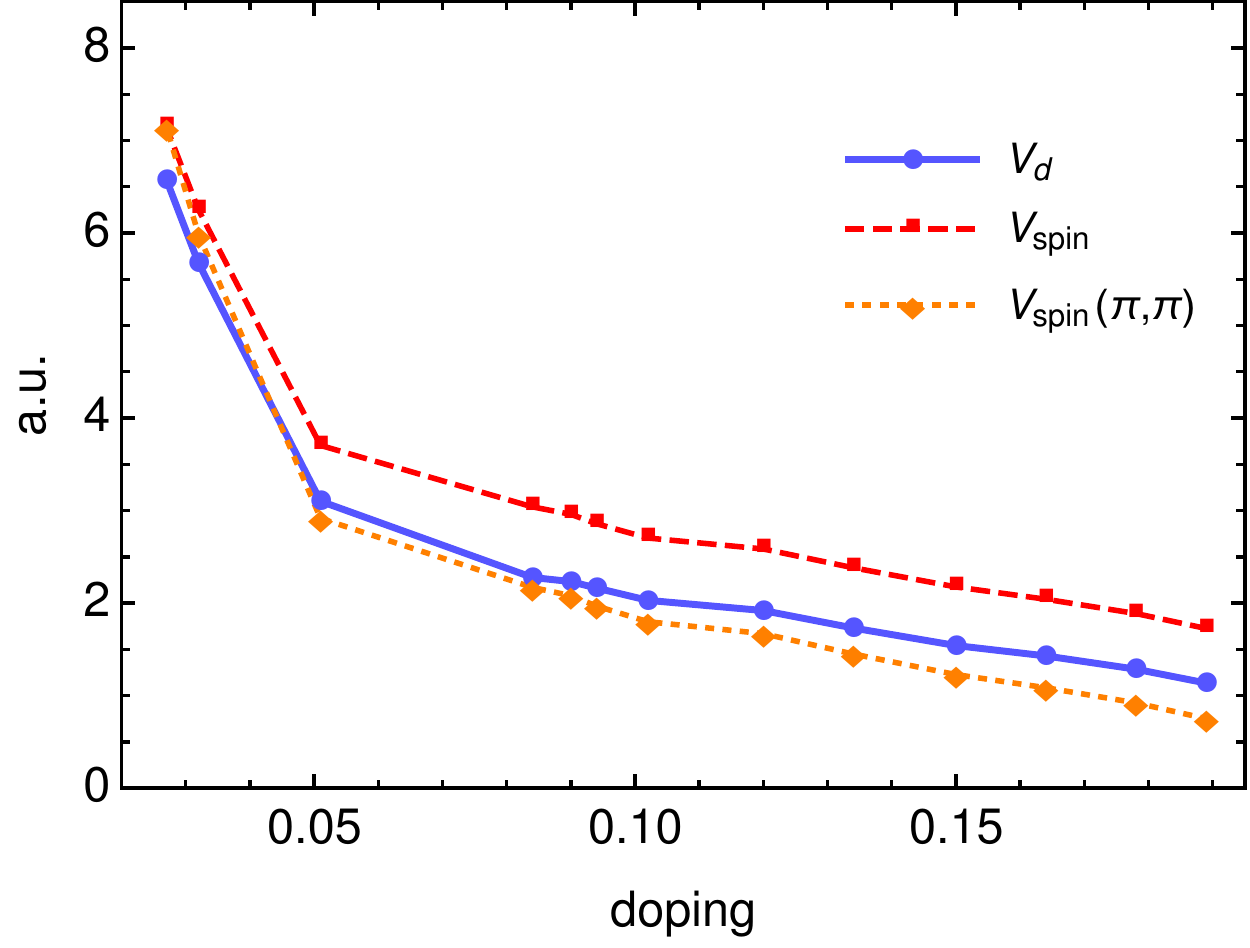}
\caption{Doping dependence of the effective $d$-wave attraction and its magnetic contribution at the temperature corresponding to the top of the superconducting dome, $T=0.075t$. The value of $V_{spin}(\pi,\pi)$ is as a contribution from the corresponding grid-point of the $16\times 16$ lattice.}
\label{fig:Vdmagnetic}
\end{center}
\end{figure}

An important question regarding the phase diagram of the 2D Hubbard model concerns the origin of the pairing glue and -related to this problem- the reason for the dome-like structure of the superconducting region. To this end, we analyze the BS equation for the generalized susceptibility $\chi_{s(\text{inglet})}^{PP'Q}$ in the (singlet) pairing channel
\begin{equation}
    \label{equ:bschisinglet}
    \left[\chi_s^{PP'Q}\right]^{-1}=\left[\chi_0^{PP'Q}\right]^{-1}+\Gamma_s^{PP'Q},
\end{equation}
where $\chi_0^{PP'Q}\!=\!\frac{1}{2\beta}G_P  G_{-P-Q}\delta_{PP'}$ is the bare pairing bubble and $\Gamma_s^{PP'Q}$ two-particle irreducible vertex in the singlet pairing channel. Considering the relation between $\chi_r$ and $F_r$ in Eq.~(\ref{equ:constructgenchi}), Eq.~(\ref{equ:bschisinglet}) corresponds one-to-one to the BS equation for $F_{pp}$ discussed in Sec.~\ref{sec:phasediagram} and, hence, an eigenvalue $\lambda_{SC}\!=\!1$ indicates a divergence of $\chi_s^{PP'Q}$ [i.e., a vanishing of the l.h.s. of Eq.~(\ref{equ:bschisinglet})]. Obviously, a zero on the l.h.s. of Eq.~(\ref{equ:bschisinglet}) has to be generated by the interplay between the bare pairing susceptibility $\chi_0^{PP'Q}$ and the irreducible (singlet) pairing vertex $\Gamma_s^{PP'Q}$ which represents the effective attractive interaction between the electrons. Unfortunately, such an interplay is difficult to analyze because the objects in Eq.~(\ref{equ:bschisinglet}) are matrices in the fermionic (frequency and momentum) variables $P$ and $P'$. In order to get a better physical intuition we follow the ideas\footnote{Note that, differently from Ref.~\onlinecite{Yang2011,Chen2013} we have projected $[\chi_0^{PP'Q}]^{-1}$ instead of $\chi_0^{PP'Q}$ as this allows for a direct projection of the BS equation~(\ref{equ:bschisinglet}).} of Ref.~\onlinecite{Chen2013} and average all quantities over the fermionic indices $P\!=\!(\nu,\mathbf{k})$ and $P'\!=\!(\nu',\mathbf{k'})$ whereas for the momentum average we include the $d$-wave form factor $f_d(\mathbf{k})$:
\begin{subequations}
\label{equ:dwaveprojection}
\begin{align}
    \label{equ:dwaveprojectionbubble}
    &\chi_0^{-1}(\delta,T)=\sum_P[\chi_0^{PP'Q}]^{-1}f_d^2(\mathbf{k})\\
    \label{equ:dwaveprojectiongamma}
    &V_{\text{eff}}(\delta,T)=-\sum_{PP'}f_d(\mathbf{k})\Gamma_s^{PP'Q}f_d(\mathbf{k'}).
\end{align}
\end{subequations}
Applying this projection to Eq.~(\ref{equ:bschisinglet}) we obtain a Stoner-like criterion for the vanishing of the projected inverse susceptibility $[\chi_s^{PP'Q}]^{-1}$:
\begin{equation}
    \label{equ:stoner}
    \chi_0^{-1}(\delta,T)=V_{\text{eff}}(\delta,T),
\end{equation}
where $V_{\text{eff}}$ can be interpreted as effective pairing interaction. In Fig.~\ref{fig:Vd} we have plotted $\chi_0^{-1}(\delta,T)$ and $V_{\text{eff}}(\delta,T)$ at the maximal temperature $T_{\text{max}}$ of the superconducting dome as a function of doping. A touching of the two curves corresponds to the onset of superconducting order. One can clearly see that neither $V_d$ nor $\chi_0^{-1}$ exhibit a dome structure as a function of doping which, hence, originates from an interplay of these two quantities. Remarkably, $V_d$ (blue line) exhibits a monotonous increase of the effective pairing interaction upon lowering the doping and takes it maximum at half-filling where antiferromagnetic spin fluctuations dominate. In order to gain a better physical understanding of this interesting behavior of $V_d$, we use the parquet equation
\begin{align}
    \label{equ:effectivepairingparquet}
    \Gamma_s^{PP'Q}&=\Lambda_s^{PP'Q}\nonumber\\&+\frac{1}{2}\left[\Phi_d^{PP'(Q-P-P')}+\Phi_d^{P(Q-P)(P'-P)}\right]\nonumber\\ & -\frac{3}{2}\left[\Phi_m^{PP'(Q-P-P')}+\Phi_m^{P(Q-P)(P'-P)}\right]
\end{align}
and project it on the $d$-wave form factor as in Eqs.~(\ref{equ:dwaveprojection}). In this way, $V_d$ can be split into three contributions $V_\Lambda$, $V_{\text{CH}}$ and $V_{\text{SP}}$ originating from the fully irreducible vertex $\Lambda_s^{PP'Q}$, form the reducible charge vertex $\Phi_d^{PP'Q}$ and from the reducible spin vertex $\Phi_m^{PP'Q}$, respectively. Fig.~\ref{fig:Vdmagnetic} shows that the major contribution to the $d$-wave pairing glue $V_d$ can be traced back to spin fluctuations represented by $V_{\text{SP}}$ (i.e., the $d$-wave projection of the reducible spin vertex, red line). A further examination of the momentum sum for the calculation of $V_{\text{SP}}$ [see also Eqs.~(\ref{equ:dwaveprojection})] reveals that the by far most important contribution stems from a single grid point $\mathbf{k}\!-\!\mathbf{k'}\!=\!(\pi,\pi)$ (orange curve) which confirms that antiferromagnetic spin fluctuations can be identified as the main source for the pairing glue which leads to superconductivity in the model.

The same antiferromagnetic fluctuations which make $V_d$ large, on the other hand, open a gap in the spectral function and, hence, suppress the one particle Green's function $G_P$ and consequently the bubble $\chi_0$. This triggers a corresponding increase of $\chi_0^{-1}$ upon lowering the doping (black line), leading to the above mentioned competition between this quantity and $V_d$ which is responsible for the dome structure of the superconducting phase. 

Let us finally mention an interesting feature observed in $\chi_0^{-1}$ at optimal doping ($\delta\!=\!0.15)$, where the curves touch each other. There, $\chi_0^{-1}$ shows a non-monotonous behavior and exhibits a minimum as a function of doping. This feature can be seen as an increase of the effective density of states (DOS) near the Fermi level and appears, hence, to be responsible for the strong enhancement of superconducting fluctuations while $V_d$ is monotonous and rather featureless in this region of the phase diagram. In Fig.~\ref{fig:Vd}, we have also plotted the DMFT data for $\chi_0^{-1}$ (thin black line). It can be seen that the DMFT also shows a peculiarity in $\chi_0^{-1}$,  although it is less pronounced and slightly shifted to the smaller doping.

 The increase in the DOS can be attributed to the effect of a van Hove singularity, which crosses the Fermi level close to optimal doping. In correlated systems, the corresponding flattening of the dispersion law has to be attributed\cite{Dzyaloshinskii1996,Irkhin2001} to non-Fermi liquid effects. They enhance the contribution of van Hove singularities to the DOS with respect to the $\propto \ln^2 \beta$ expression known for the noninteracting Fermi gas in 2D. This enhancement is not seen on the DMFT level (see thin black line in Fig.~\ref{fig:Vd}) but arises in nonlocal extensions of DMFT\cite{Rubtsov2009}. Whether the observed behavior is related to the anomalous temperature behavior of the the bubble $\chi_0(T)\!\sim\!\frac{1}{\sqrt{T}}$, which has been reported in DCA studies\cite{Yang2011,Chen2012} and related to a quantum critical point below the superconducting dome, is a question for future research work.

\subsection{Spectral properties}
\label{sec:spectralproperties}
 
\begin{figure}[t!]
\begin{center}
\includegraphics[width=7.5 cm]{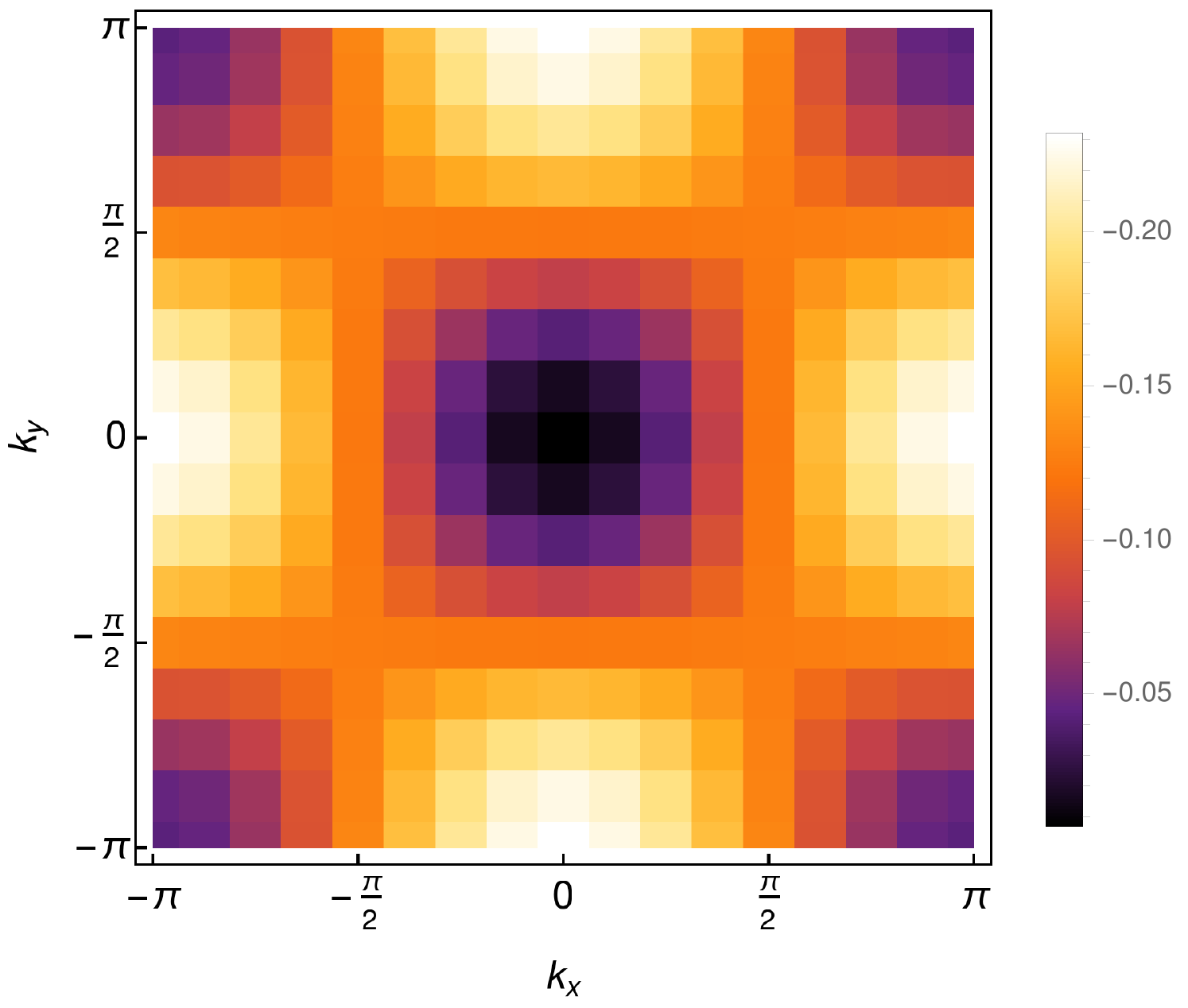}
\caption{Imaginary part of self-energy at the lowest Matsubara frequency at $\delta\!=\!0.09$ and $T\!=\!0.05t$}
\label{fig:afchi3}
\end{center}
\end{figure}

 Let us turn our attention to spectral properties obtained from our parquet equations. Since the self-energy and the Green's function are known only for the first fermionic Matsubara frequencies it is difficult to perform an analytic continuation to the real frequency axis in order to obtain a spectral function. However, interesting information can be already extracted from the self-energy at the lowest Matsubara frequency. Fig.~\ref{fig:afchi3} shows $\Sigma_K$ for $\nu\!=\!\pi/\beta$ as a function of $k_x$ and $k_y$. Our results confirm the  general wisdom that in the underdoped regime of the 2D Hubbard model above the superconducting dome a pseudogap region emerges with a momentum selective suppression of spectral weight\cite{Gull2009}. Our data indeed show a large value of the imaginary part of the self-energy (corresponding to a suppression of spectral weight) at $\mathbf{k}\!=\!(0,\pi)$ and considerable lower value of Im$\Sigma$ for $\mathbf{k}\!=\!(\pi/2,\pi/2)$. Such a behavior, we find only at low doping while for higher values of $\delta$, e.g., for optimal doping this feature disappears (not shown). 
 
\subsection{Comparison with other methods}
\label{sec:compare}

\begin{figure}[t!]
\begin{center}
\includegraphics[width=\linewidth]{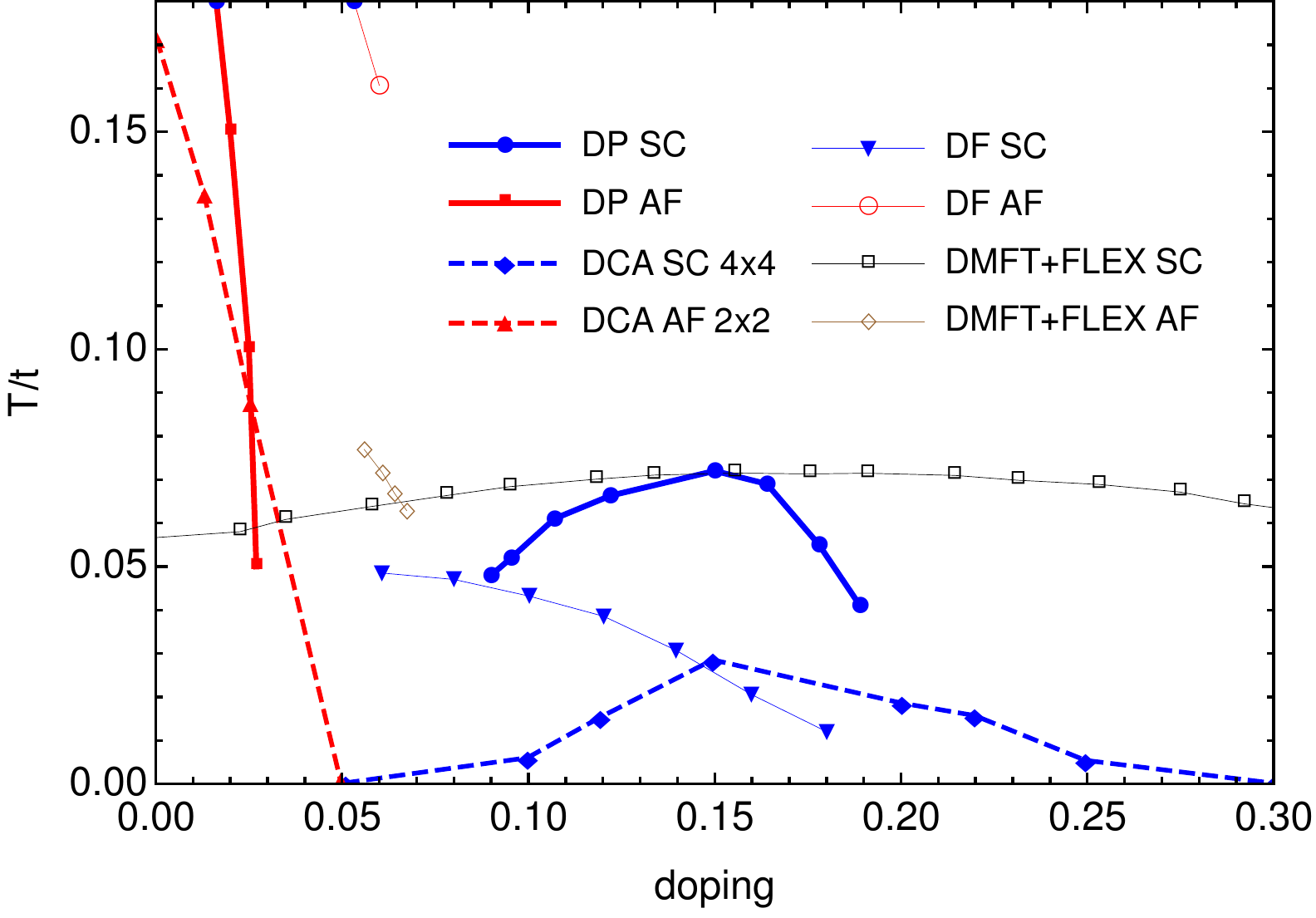}
\caption{Phase diagram for different methods: dual-fermion parquet for $U=8t$, $t'=-0.2t$, $t''=0.1t$, DCA\cite{Chen2013} for $U=6t$, $t'=-0.2t$, $t''=0$, dual-fermion\cite{Otsuki2014} for $U=8t$, $t'=t''=0$, DMFT+FLEX\cite{Kitatani2015} for $U=5t$, $t'=-0.2t$, $t''=0.16t$.}
\label{fig:pd2}
\end{center}
\end{figure}

In this section, we compare our phase diagram with corresponding results obtained by a number of different many body methods (see Fig.~\ref{fig:pd2}). Let us however stress that the purpose of this part of the paper is {\em not} a systematic analysis of our new methods w.r.t. other approaches but rather to demonstrate its applicability for obtaining reasonable results which are consistent with those of other approaches and potentially have the capability to improve them. A more complete study of similarities and differences between our new approach and existing techniques will be performed in future research work. Here, instead, we restrict our comparison to a small number of selected methods without any claim of completeness.

First, let us analyze various results for the superconducting phase in Fig.~\ref{fig:pd2}. We observe that our results (blue dots/line) feature a clear dome structure while DMFT combined with FLEX (thin black line) and dual fermions (thin blue line) exhibit only a rather weak or no superconducting dome. In fact, in the later approaches the superconducting phase seems to extend to half-filling, crossing also the AF phase which exists close to particle hole symmetry. On the contrary, a dome structure is obtained by dynamical cluster approximation (DCA) calculations (dashed blue line) at lower temperatures. The latter difference might be attributed to the lower value of the bare Hubbard interaction used in the DCA calculation ($U\!=\!6t$ in DCA vs. $U\!=\!8t$ for our results).

The absence of the dome structure in DF and FLEX based approximations might be attributed to the missing renormalization of the two-particle vertex functions which is provided by the parquet equations but not taken into account by FLEX and the corresponding FLEX-like diagram in DF. Moreover, for the later situation the feedback of the superconducting fluctuations onto the irreducible particle-hole vertices in the spin and charge channels is neglected. While, in general, it is difficult to trace how such differences in the methods might propagate to the final results, with a lack of mutual screening between competing channels it cannot be expected to fully capture the electronic properties of the system. On the other hand, our method has been considerably simplified due to reduction of the frequency space which allows to take into account much finer momentum grids w.r.t. DCA which can lead to quantitative improvements of the results.


\section{Conclusions and Outlook}
\label{sec:conclusions}

We have developed a new many-body approach for systems characterized by local interactions but essentially non-local correlations (which can be also referred to as ultra quantum matter). Physically our approach relies on the relation between the length and energy scales for correlations: the high-energy part of correlations is assumed to be local, whereas non-local correlations are associated with mutually interacting low-energy collective modes. Technically, the method can be described as a three-step procedure: (i) We map the original problem of interacting electrons onto a corresponding problem for dual particles which include all fully local correlation effects of DMFT already at the lowest (i.e., 0th) order of the perturbation theory. (ii) Within the path integral representation for the partition function we have integrated out all higher Matsubara frequencies $\lvert\nu\rvert\!>\!\pi/\beta$ by means of selected diagrams yielding a low-frequency effective theory including only the fermionic frequencies $\nu\!=\!\pm\pi/\beta$. This simplification has (iii) allowed as to apply one of the most complete theories, i.e., the parquet formalism, to this problem which -- in contrast to simple ladder approaches -- takes into account mutual screening effects between all different scattering channels. Moreover, due to the reduction of complexity in frequency space we were able to apply the parquet equations for much finer momentum grids w.r.t. to previous works. 


We have applied our new approach to the hole-doped two dimensional Hubbard model in parameter regimes which are relevant for the high-temperature superconducting cuprates. Consistent with earlier DCA studies on this problem, we have found a dome shaped superconducting region in the doping vs. temperature phase diagram and an antiferromagnetic region for small values of the doping. Furthermore, we could demonstrate that antiferromagnetic spin fluctuations represent the pairing glue for superconductivity which, consistently, exhibits a $d$-wave nature. Interestingly, AF fluctuations are sizable in the entire phase diagram while their superconducting counterparts are restricted to an area very close to the superconducting dome.
The same AF fluctuations are responsible for a suppression of the single-particle Green's function and the interplay between this effect and the enhancement of the effective pairing interaction could be shown to be responsible for the dome-like shape of the superconducting region. We have presented evidence that the effective DOS at the optimal doping is increased due to the presence of van Hove singuliarity increased by the non-Fermi-liquid effects. 
Finally, we have also observed enhanced charge fluctuations in the weak-to-intermediate doping regime which is consistent with the scenario of charge-fluctuation driven quantum critical point below the superconducting dome. Observed charge fluctuations also support the scenario for the breakdown of the AF order due to phase separation. Hence, overall it can be stated that our new method captures well the main features of the 2D Hubbard model in the considered parameter regime.

Nevertheless, our consideration left some open questions. The reduction of the effective model to only the first fermionic Matsubara frequency makes difficult the extraction of spectral properties or the consistent calculation of susceptibilities. 
This is one of the reasons why we left out of a detailed analysis of the single particle properties. 
Moreover, it is unclear how strongly the actual choice of the diagrams which are used for integrating out the higher Matsubara frequencies affects the final results. These questions require a further intense investigation by, for instance, including two or more fermionic frequencies in the effective model and/or comparing the effect of different Feynman diagrams when integrating out the high-frequency degrees of freedom. Furthermore, a better comparison of our findings with the results of other approaches requires a systematic analysis in the entire phase space spanned by the parameters $U$, $\mu$, $T$ and $t'$. The simplified frequency nature of our approach might also allow for an extension to more realistic models including, for instance, more orbitals. Finally, it might be possible to incorporate further simplifications into our scheme, such as an improved momentum grid \cite{Eckhardt2018} or alternatives to parquet equations which are also unbiased \cite{Krien2019}.

\paragraph*{Acknowledgements}
The authors thank A.Antipov, G.Cohen, K. Held, A. Kauch, A. Lichtenstein and A. Toschi for useful discussions. This research work was funded by the Russian Science Foundation through Grant 16-42-01057 (all authors) and the Deutsche Forschungsgemeinschaft (DFG, German Research Foundation) through Project No. 407372336 (G.R.). 

\newpage

\begin{widetext}

\appendix

\section{Definitions}
\label{app:def}

In this section we will introduce definitions of vertex functions, which are used throughout the paper. The two-particle imaginary time Green's function is defined as

\begin{equation}
G^{(2)}_{\mathbf{k}\mathbf{k}'\mathbf{q},\sigma\sigma'}(\tau_1,\tau_2,\tau_3)=\langle\text{T}_\tau c_{\mathbf{k},\sigma}^\dagger(\tau_1)c_{\mathbf{k}+\mathbf{q},\sigma}(\tau_2)c_{\mathbf{k}'+\mathbf{q},\sigma'}^\dagger(\tau_3)c_{\mathbf{k}',\sigma'}(0) \rangle.
\end{equation}
Performing Fourier transform for imaginary time arguments, we get Matsubara frequency Green's function in particle-hole notation\cite{Rohringer2012}:
\begin{equation}
G^{(2),\nu\nu'\omega}_{\mathbf{k}\mathbf{k}'\mathbf{q},\sigma\sigma'}=\int_0^\beta d\tau_1 d\tau_2 d\tau_3 e^{i\nu\tau_1}e^{i(\nu+\omega)\tau_2}e^{i(\nu'+\omega)\tau_3}G^{(2)}_{\mathbf{k}\mathbf{k}'\mathbf{q},\sigma\sigma'}(\tau_1,\tau_2,\tau_3).
\end{equation}
Since the parquet equations include all scattering channels on equal footing, we also introduce the particle-particle representation where the transfer momentum is defined as $\mathbf{q}_{pp}=\mathbf{q}+\mathbf{k}+\mathbf{k}'$, $\omega_{pp}=\omega+\nu+\nu'$, so $G^{pp(2),\nu\nu'\omega}_{\mathbf{k}\mathbf{k}'\mathbf{q},\sigma\sigma'}=G^{(2),\nu\nu'\omega+\nu+\nu'}_{\mathbf{k}\mathbf{k}'\mathbf{q}+\mathbf{k}+\mathbf{k}',\sigma\sigma'}$. Next, we introduce generalized susceptibilities:
\begin{equation}
\chi^{\nu\nu'\omega}_{\mathbf{k}\mathbf{k}'\mathbf{q},\sigma\sigma'}=G^{(2),\nu\nu'\omega}_{\mathbf{k}\mathbf{k}'\mathbf{q},\sigma\sigma'}-\beta G_{\nu, \mathbf{k}}G_{\nu',\mathbf{k}'}\delta_{\mathbf{q}0}\delta_{\omega0}.
\end{equation}

The full two-particle vertex function $F^{\nu\nu'\omega}_{r,\mathbf{k}\mathbf{k}'\mathbf{q}}$ is defined in the following way:

\begin{align}
\chi_{r,\mathbf{k}\mathbf{k}'\mathbf{q}}^{\nu\nu'\omega}&=\chi_{0,\mathbf{k}\mathbf{k}'\mathbf{q}}^{\nu\nu'\omega}-G_{\nu, \mathbf{k}}G_{\nu+\omega, \mathbf{k}+\mathbf{q}}F^{\nu\nu'\omega}_{r,\mathbf{k}\mathbf{k}'\mathbf{q}}G_{\nu', \mathbf{k}'}G_{\nu'+\omega, \mathbf{k}'+\mathbf{q}},\\
\chi_{0,\mathbf{k}\mathbf{k}'\mathbf{q}}^{\nu\nu'\omega}&=-\beta G_{\nu, \mathbf{k}}G_{\nu+\omega,\mathbf{k}+\mathbf{q}}\delta_{\nu\nu'}\delta_{\mathbf{k}\mathbf{k}'},
\end{align}
where $r$ is one of four spin channels: density (charge), magnetic (spin), singlet and triplet
\begin{align}
F_d&=F^{ph}_{\uparrow\uparrow;\uparrow\uparrow}+F^{ph}_{\uparrow\uparrow;\downarrow\downarrow},\\
F_m&=F^{ph}_{\uparrow\uparrow;\uparrow\uparrow}-F^{ph}_{\uparrow\uparrow;\downarrow\downarrow},\\
F_s&=F^{pp}_{\uparrow\downarrow;\uparrow\downarrow}-F^{pp}_{\uparrow\downarrow;\downarrow\uparrow},\\
F_t&=F^{pp}_{\uparrow\downarrow;\uparrow\downarrow}+F^{pp}_{\uparrow\downarrow;\downarrow\uparrow}.
\end{align}
We also introduce the vertex functions $\Gamma_r$ which are irreducible in channel $r$ and can be defined via the Bethe-Salpeter equations:
\begin{align}
F_{d/m}^{KK'Q} &= \Gamma_{d/m}^{KK'Q} + \frac{1}{\beta}\sum_{P}\Gamma_{d/m}^{KPQ}G_PG_{P+Q}F_{d/m}^{PK'Q},\\
F_{s/t}^{KK'Q} &= \Gamma_{s/t}^{KK'Q} + \frac{1}{2\beta}\sum_{P}\Gamma_{s/t}^{PK'Q}G_PG_{Q-P}F_{s/t}^{K(Q-P)Q},
\end{align}
where $K = (\mathbf{k},\nu)$. The corresponding complementary vertices $\Phi_r$ which are reducible in channel $r$ are defined as
\begin{equation}
\Phi_r^{KK'Q}+\Gamma_r^{KK'Q}=F_r^{KK'Q}.
\end{equation}

\section{Effective model}
\label{app:model}

In this section we present more detailed expressions for the propagator and the interaction of the effective low-frequency model. It has to be emphasized that the downfoalding shares some similarities with Wilson's renormalization group with the key difference that instead of high momenta high frequencies are integrated out. It is, hence, useful to introduce auxillary Green's functions for high and low frequencies:
\begin{eqnarray}
G^{<}_{\nu, \mathbf{k}} = \begin{cases}
\widetilde{G}_{0,\nu\mathbf{k}}, & \text{if $|\nu| \leqslant \nu_\text{max}$},\\
0, & \text{if $|\nu| > \nu_\text{max}$}.
\end{cases}\qquad
G^{>}_{\nu, \mathbf{k}} = \begin{cases}
0, & \text{if $|\nu| \leqslant \nu_\text{max}$},\\
\widetilde{G}_{0,\nu\mathbf{k}}, & \text{if $|\nu| > \nu_\text{max}$},
\end{cases}
\end{eqnarray}
where $\widetilde{G}_{0,\nu\mathbf{k}}$ is the bare dual Green's function.

As discussed in Sec.~\ref{sec:lowfreqmodel}, we use second-order perturbation theory in order to construct the effective action:
\begin{equation}
    S_\text{eff}[f_<]\approx S_<[f_<]-\frac{1}{2}\left(\langle S^2_>[f_<,f_>]\rangle_>-\langle S_>[f_<,f_>]\rangle_>^2 \right) = -\sum \mathcal{G}_{12}^{-1} f^\dagger_{1<} f_{2<} + \frac{1}{4}\sum \mathcal{V}_{1234} f^\dagger_{1<} f_{2<} f^\dagger_{3<} f_{4<},
\end{equation}
where $S_<[f_<]=-G^{<-1}_{12}f^\dagger_{1<} f_{2<}$, $S_>[f_<,f_>]=-\frac{1}{4}\sum \gamma^{(2)}_{1234}(f^\dagger_{1<}+f^\dagger_{1>}) (f_{2<}+f_{2>})(f^\dagger_{3<}+f^\dagger_{3>})(f_{4<}+f_{4>})$, $\mathcal{G}_{12}$ is the effective propagator and $\mathcal{V}_{1234}$ the effective interaction of our downfolded model. We can also rewrite the action in a three-index notation:
\begin{equation}
    S_\text{eff} = -\sum_{\nu \mathbf{k} \sigma}\mathcal{G}^{-1}_{\nu, \mathbf{k}}f_{\nu \mathbf{k} \sigma}^\dagger f_{\nu \mathbf{k} \sigma} - \sum_{\nu\nu'\omega \atop \mathbf{k} \mathbf{k}' \mathbf{q}}\sum_{\sigma\sigma'}\mathcal{V}^{\nu\nu'\omega}_{\mathbf{k}\mathbf{k}'\mathbf{q},\sigma\sigma'}f^\dagger_{\nu \mathbf{k} \sigma}f_{(\nu+\omega) (\mathbf{k}+\mathbf{q}) \sigma}f^\dagger_{(\nu'+\omega)(\mathbf{k}'+\mathbf{q})\sigma'}f_{\nu' \mathbf{k}'\sigma'},
\end{equation}

where the effective propagator and interaction are
\begin{equation}
    \mathcal{G}_{\nu, \mathbf{k}}^{-1} = G^{<-1}_{\nu, \mathbf{k}}-\frac{1}{4\beta^2}\sum_{\nu_{1}\omega \widetilde{\sigma} \atop \mathbf{k}_1\mathbf{q}}\gamma^{(2)}_{\nu\nu_1\omega,\sigma\widetilde{\sigma}}G^>_{\nu_1, \mathbf{k}_1}G^>_{\nu_1+\omega,\mathbf{k}_1+\mathbf{q}}G^>_{\nu+\omega,\mathbf{k}+\mathbf{q}}\gamma^{(2)}_{\nu_1\nu\omega,\widetilde{\sigma}\sigma},
\end{equation}

\begin{equation}
\begin{split}
    \mathcal{V}^{\nu\nu'\omega}_{\mathbf{k}\mathbf{k}'\mathbf{q},\sigma\sigma'}=\gamma^{(2)}_{\nu\nu'\omega,\sigma\sigma'} + \frac{2}{\beta}\sum_{\nu_1 \mathbf{k}_1 \widetilde{\sigma}} \gamma^{(2)}_{\nu\nu_1\omega,\sigma\widetilde{\sigma}} G^>_{\nu_1, \mathbf{k}_1} G^>_{\nu_1+\omega,\mathbf{k}_1+\mathbf{q}} \gamma^{(2)}_{\nu_1\nu'\omega,\widetilde{\sigma}\sigma'}-\\
    -\frac{1}{2\beta}\sum_{\omega_1 \mathbf{q}_1 \widetilde{\sigma}}\gamma^{(2)}_{\nu(\nu-\omega_1)(\omega+\omega_1),\sigma\widetilde{\sigma}} G^>_{\nu'-\omega_1, \mathbf{k}'-\mathbf{q}_1} G^>_{\nu+\omega_1+\omega,\mathbf{k}+\mathbf{q}_1+\mathbf{q}} \gamma^{(2)}_{(\nu'-\omega_1)\nu'(\nu-\nu'+\omega_1+\omega),\widetilde{\sigma}\sigma'}.
\end{split}
\end{equation}

\section{Parquet equations for the low-frequency model}
\label{app:details}

First, we consider the structure of vertex functions which depend on two fermionic frequencies, ignoring for the moment the momentum dependence. A full vertex function is usually written in a three index notation, $F^{\nu\nu'\omega}$, where $\nu$ and $\nu'$ are fermionic frequencies, which correspond to an ingoing and an outgoing electrons respectively, and $\omega$ is a bosonic transfer frequency. In case of a particle-hole vertex function the two ingoing electrons have frequencies $\nu'$ and $\nu'+\omega$, whereas the two outgoing electrons $\nu$ and $\nu+\omega$. In our downfolded model, the frequencies of these electrons are only allowed to take values $\pm\pi/\beta$ corresponding to the Matsubara indices $0$ and $-1$. Therefore, there are only 6 six frequency components of $ph$-vertex functions (where the numbers refer to fermionic Matsubara indices): $F^{000}$, $F^{0-10}$, $F^{00-1}$, $F^{-100}$, $F^{-1-10}$, and $F^{-1-11}$. In case of a particle-particle vertex functions outgoing and ingoing frequencies are $\nu$, $\omega-\nu$, and $\nu'$, $\omega-\nu'$ respectively. Therefore, there are other six types of vertex functions in $pp$-notation: $F_\text{pp}^{000}$, $F_\text{pp}^{00-1}$, $F_\text{pp}^{0-1-1}$, $F_\text{pp}^{-10-1}$, $F_\text{pp}^{-1-1-1}$, and $F_\text{pp}^{-1-1-2}$. In this section we will show that standard diagrammatic equations do not generate other types of frequency structures. The most straightforward way to demonstrate this is simply to write down explicitly the equations for all possible frequency combinations. Let us consider two examples: the case $(\nu,\nu',\omega)=(-1,-1,0)$ for the density channel, and $(\nu,\nu',\omega)=(-1,-1,-2)$ for the triplet channel:

\begin{equation}
    F_{d,\mathbf{k}\mathbf{k}'\mathbf{q}}^{-1-10} = \Gamma_{d,\mathbf{k}\mathbf{k}'\mathbf{q}}^{-1-10} + \frac{1}{\beta}\sum_{\mathbf{p}}\left(\Gamma_{d,\mathbf{k}\mathbf{p}\mathbf{q}}^{-1-10}G_{-1,\mathbf{p}}G_{-1,\mathbf{p}+\mathbf{q}}F_{d,\mathbf{p}\mathbf{k}'\mathbf{q}}^{-1-10}+\Gamma_{d,\mathbf{k}\mathbf{p}\mathbf{q}}^{-100}G_{0,\mathbf{p}}G_{0,\mathbf{p}+\mathbf{q}}F_{d,\mathbf{p}\mathbf{k}'\mathbf{q}}^{0-10}\right),
\end{equation}

\begin{equation}
    F_{t,\mathbf{k}\mathbf{k}'\mathbf{q}}^{-1-1-2} = \Gamma_{t,\mathbf{k}\mathbf{k}'\mathbf{q}}^{-1-1-2} + \frac{1}{2\beta}\sum_{\mathbf{p}}\Gamma_{t,\mathbf{p}\mathbf{k}'\mathbf{q}}^{-1-1-2}G_{-1,\mathbf{p}}G_{-1,\mathbf{q}-\mathbf{p}}F_{t,\mathbf{k}(\mathbf{q}-\mathbf{p})\mathbf{q}}^{-1-1-2}.
\end{equation}
It is easy to see that the vertices on the right-hand side are from the list of allowed vertex types.  

Next, we consider parquet equations:
\begin{subequations}
\begin{align}
   \Lambda_d^{KK'Q} &= \Gamma_d^{KK'Q} + \frac{1}{2}\Phi_d^{K(K+Q)(K'-K)}+\frac{3}{2}\Phi_m^{K(K+Q)(K'-K)}-\frac{1}{2}\Phi_s^{KK'(K+K'+Q)}-\frac{3}{2}\Phi_t^{KK'(K+K'+Q)},\\
   \Lambda_m^{KK'Q} &= \Gamma_m^{KK'Q} + \frac{1}{2}\Phi_d^{K(K+Q)(K'-K)}-\frac{1}{2}\Phi_m^{K(K+Q)(K'-K)}+\frac{1}{2}\Phi_s^{KK'(K+K'+Q)}-\frac{1}{2}\Phi_t^{KK'(K+K'+Q)},\\
   \Lambda_s^{KK'Q} &= \Gamma_s^{KK'Q} - \frac{1}{2}\Phi_d^{KK'(Q-K-K')}+\frac{3}{2}\Phi_m^{KK'(Q-K-K')}-\frac{1}{2}\Phi_d^{K(Q-K')(K'-K)}+\frac{3}{2}\Phi_m^{K(Q-K')(K'-K)},\\
   \Lambda_t^{KK'Q} &= \Gamma_t^{KK'Q} - \frac{1}{2}\Phi_d^{KK'(Q-K-K')}-\frac{1}{2}\Phi_m^{KK'(Q-K-K')}+\frac{1}{2}\Phi_d^{K(Q-K')(K'-K)}+\frac{1}{2}\Phi_m^{K(Q-K')(K'-K)}.
\end{align}
\end{subequations}
We can do the same analysis as before. For example, let us consider frequencies $(\nu,\nu',\omega)=(0,-1,0)$ for a magnetic channel, and $(\nu,\nu',\omega)=(-1,-1,-1)$ for a singlet channel:
\begin{subequations}
\begin{align}
    \Lambda_{m,\mathbf{k}\mathbf{k}'\mathbf{q}}^{0-10} &= \Gamma_{m,\mathbf{k}\mathbf{k}'\mathbf{q}}^{0-10} + \frac{1}{2}\Phi_{d,\mathbf{k}(\mathbf{k}+\mathbf{q})(\mathbf{k}'-\mathbf{k})}^{00-1}-\frac{1}{2}\Phi_{m,\mathbf{k}(\mathbf{k}+\mathbf{q})(\mathbf{k}'-\mathbf{k})}^{00-1}+\frac{1}{2}\Phi_{s,\mathbf{k}\mathbf{k}'(\mathbf{k}+\mathbf{k}'+\mathbf{q})}^{0-1-1}-\frac{1}{2}\Phi_{t,\mathbf{k}\mathbf{k}'(\mathbf{k}+\mathbf{k}'+\mathbf{q})}^{0-1-1},\\
    \Lambda_{s,\mathbf{k}\mathbf{k}'\mathbf{q}}^{-1-1-1} &= \Gamma_{s,\mathbf{k}\mathbf{k}'\mathbf{q}}^{-1-1-1} - \frac{1}{2}\Phi_{d,\mathbf{k}\mathbf{k}'(\mathbf{q}-\mathbf{k}-\mathbf{k}')}^{-1-11}+\frac{3}{2}\Phi_{m,\mathbf{k}\mathbf{k}'(\mathbf{q}-\mathbf{k}-\mathbf{k}')}^{-1-11}-\frac{1}{2}\Phi_{d,\mathbf{k}(\mathbf{q}-\mathbf{k}')(\mathbf{k}'-\mathbf{k})}^{-100}+\frac{3}{2}\Phi_{m,\mathbf{k}(\mathbf{q}-\mathbf{k}')(\mathbf{k}'-\mathbf{k})}^{-100}.
\end{align}
\end{subequations}

\end{widetext}

\bibliographystyle{apsrev4-1}

\begin{thebibliography}{96}%
	\makeatletter
	\providecommand \@ifxundefined [1]{%
		\@ifx{#1\undefined}
	}%
	\providecommand \@ifnum [1]{%
		\ifnum #1\expandafter \@firstoftwo
		\else \expandafter \@secondoftwo
		\fi
	}%
	\providecommand \@ifx [1]{%
		\ifx #1\expandafter \@firstoftwo
		\else \expandafter \@secondoftwo
		\fi
	}%
	\providecommand \natexlab [1]{#1}%
	\providecommand \enquote  [1]{``#1''}%
	\providecommand \bibnamefont  [1]{#1}%
	\providecommand \bibfnamefont [1]{#1}%
	\providecommand \citenamefont [1]{#1}%
	\providecommand \href@noop [0]{\@secondoftwo}%
	\providecommand \href [0]{\begingroup \@sanitize@url \@href}%
	\providecommand \@href[1]{\@@startlink{#1}\@@href}%
	\providecommand \@@href[1]{\endgroup#1\@@endlink}%
	\providecommand \@sanitize@url [0]{\catcode `\\12\catcode `\$12\catcode
		`\&12\catcode `\#12\catcode `\^12\catcode `\_12\catcode `\%12\relax}%
	\providecommand \@@startlink[1]{}%
	\providecommand \@@endlink[0]{}%
	\providecommand \url  [0]{\begingroup\@sanitize@url \@url }%
	\providecommand \@url [1]{\endgroup\@href {#1}{\urlprefix }}%
	\providecommand \urlprefix  [0]{URL }%
	\providecommand \Eprint [0]{\href }%
	\providecommand \doibase [0]{http://dx.doi.org/}%
	\providecommand \selectlanguage [0]{\@gobble}%
	\providecommand \bibinfo  [0]{\@secondoftwo}%
	\providecommand \bibfield  [0]{\@secondoftwo}%
	\providecommand \translation [1]{[#1]}%
	\providecommand \BibitemOpen [0]{}%
	\providecommand \bibitemStop [0]{}%
	\providecommand \bibitemNoStop [0]{.\EOS\space}%
	\providecommand \EOS [0]{\spacefactor3000\relax}%
	\providecommand \BibitemShut  [1]{\csname bibitem#1\endcsname}%
	\let\auto@bib@innerbib\@empty
	\bibitem [{\citenamefont {Damascelli}\ \emph {et~al.}(2003)\citenamefont
		{Damascelli}, \citenamefont {Hussain},\ and\ \citenamefont
		{Shen}}]{Damascelli2003}%
	\BibitemOpen
	\bibfield  {author} {\bibinfo {author} {\bibfnamefont {A.}~\bibnamefont
			{Damascelli}}, \bibinfo {author} {\bibfnamefont {Z.}~\bibnamefont {Hussain}},
		\ and\ \bibinfo {author} {\bibfnamefont {Z.-X.}\ \bibnamefont {Shen}},\
	}\href@noop {} {\bibfield  {journal} {\bibinfo  {journal} {Rev. Mod. Phys.}\
		}\textbf {\bibinfo {volume} {75}},\ \bibinfo {pages} {473} (\bibinfo {year}
		{2003})}\BibitemShut {NoStop}%
	\bibitem [{\citenamefont {Vishik}\ \emph {et~al.}(2010)\citenamefont {Vishik},
		\citenamefont {Lee}, \citenamefont {He}, \citenamefont {Hashimoto},
		\citenamefont {Hussain}, \citenamefont {Devereaux},\ and\ \citenamefont
		{Shen}}]{Vishik2010}%
	\BibitemOpen
	\bibfield  {author} {\bibinfo {author} {\bibfnamefont {I.~M.}\ \bibnamefont
			{Vishik}}, \bibinfo {author} {\bibfnamefont {W.~S.}\ \bibnamefont {Lee}},
		\bibinfo {author} {\bibfnamefont {R.~H.}\ \bibnamefont {He}}, \bibinfo
		{author} {\bibfnamefont {M.}~\bibnamefont {Hashimoto}}, \bibinfo {author}
		{\bibfnamefont {Z.}~\bibnamefont {Hussain}}, \bibinfo {author} {\bibfnamefont
			{T.~P.}\ \bibnamefont {Devereaux}}, \ and\ \bibinfo {author} {\bibfnamefont
			{Z.~X.}\ \bibnamefont {Shen}},\ }\href {\doibase
		10.1088/1367-2630/12/10/105008} {\bibfield  {journal} {\bibinfo  {journal}
			{New Journal of Physics}\ }\textbf {\bibinfo {volume} {12}},\ \bibinfo
		{pages} {1} (\bibinfo {year} {2010})},\ \Eprint
	{http://arxiv.org/abs/arXiv:1009.0274v1} {arXiv:1009.0274v1} \BibitemShut
	{NoStop}%
	\bibitem [{\citenamefont {Vishik}(2018)}]{Vishik2018}%
	\BibitemOpen
	\bibfield  {author} {\bibinfo {author} {\bibfnamefont {I.~M.}\ \bibnamefont
			{Vishik}},\ }\href {\doibase 10.1088/1361-6633/aaba96} {\bibfield  {journal}
		{\bibinfo  {journal} {Reports on Progress in Physics}\ }\textbf {\bibinfo
			{volume} {81}} (\bibinfo {year} {2018}),\
		10.1088/1361-6633/aaba96}\BibitemShut {NoStop}%
	\bibitem [{\citenamefont {Dagotto}(1994)}]{Dagotto94}%
	\BibitemOpen
	\bibfield  {author} {\bibinfo {author} {\bibfnamefont {E.}~\bibnamefont
			{Dagotto}},\ }\href {\doibase 10.1103/RevModPhys.66.763} {\bibfield
		{journal} {\bibinfo  {journal} {Rev. Mod. Phys.}\ }\textbf {\bibinfo {volume}
			{66}},\ \bibinfo {pages} {763} (\bibinfo {year} {1994})}\BibitemShut
	{NoStop}%
	\bibitem [{\citenamefont {Mermin}\ and\ \citenamefont
		{Wagner}(1966)}]{Mermin1966}%
	\BibitemOpen
	\bibfield  {author} {\bibinfo {author} {\bibfnamefont {N.~D.}\ \bibnamefont
			{Mermin}}\ and\ \bibinfo {author} {\bibfnamefont {H.}~\bibnamefont
			{Wagner}},\ }\href {\doibase 10.1103/PhysRevLett.17.1307} {\bibfield
		{journal} {\bibinfo  {journal} {Phys. Rev. Lett.}\ }\textbf {\bibinfo
			{volume} {17}},\ \bibinfo {pages} {1307} (\bibinfo {year}
		{1966})}\BibitemShut {NoStop}%
	\bibitem [{\citenamefont {Sch\"afer}\ \emph {et~al.}(2015)\citenamefont
		{Sch\"afer}, \citenamefont {Geles}, \citenamefont {Rost}, \citenamefont
		{Rohringer}, \citenamefont {Arrigoni}, \citenamefont {Held}, \citenamefont
		{Bl\"umer}, \citenamefont {Aichhorn},\ and\ \citenamefont
		{Toschi}}]{Schaefer2015-2}%
	\BibitemOpen
	\bibfield  {author} {\bibinfo {author} {\bibfnamefont {T.}~\bibnamefont
			{Sch\"afer}}, \bibinfo {author} {\bibfnamefont {F.}~\bibnamefont {Geles}},
		\bibinfo {author} {\bibfnamefont {D.}~\bibnamefont {Rost}}, \bibinfo {author}
		{\bibfnamefont {G.}~\bibnamefont {Rohringer}}, \bibinfo {author}
		{\bibfnamefont {E.}~\bibnamefont {Arrigoni}}, \bibinfo {author}
		{\bibfnamefont {K.}~\bibnamefont {Held}}, \bibinfo {author} {\bibfnamefont
			{N.}~\bibnamefont {Bl\"umer}}, \bibinfo {author} {\bibfnamefont
			{M.}~\bibnamefont {Aichhorn}}, \ and\ \bibinfo {author} {\bibfnamefont
			{A.}~\bibnamefont {Toschi}},\ }\href {\doibase 10.1103/PhysRevB.91.125109}
	{\bibfield  {journal} {\bibinfo  {journal} {Phys. Rev. B}\ }\textbf {\bibinfo
			{volume} {91}},\ \bibinfo {pages} {125109} (\bibinfo {year}
		{2015})}\BibitemShut {NoStop}%
	\bibitem [{\citenamefont {Vojta}(2009)}]{Vojta2009}%
	\BibitemOpen
	\bibfield  {author} {\bibinfo {author} {\bibfnamefont {M.}~\bibnamefont
			{Vojta}},\ }\href {\doibase 10.1080/00018730903122242} {\bibfield  {journal}
		{\bibinfo  {journal} {Advances in Physics}\ }\textbf {\bibinfo {volume}
			{58}},\ \bibinfo {pages} {699} (\bibinfo {year} {2009})},\ \Eprint
	{http://arxiv.org/abs/https://doi.org/10.1080/00018730903122242}
	{https://doi.org/10.1080/00018730903122242} \BibitemShut {NoStop}%
	\bibitem [{\citenamefont {Fujita}\ \emph {et~al.}(2012)\citenamefont {Fujita},
		\citenamefont {Hiraka}, \citenamefont {Matsuda}, \citenamefont {Matsuura},
		\citenamefont {M.~Tranquada}, \citenamefont {Wakimoto}, \citenamefont {Xu},\
		and\ \citenamefont {Yamada}}]{Fujita2012}%
	\BibitemOpen
	\bibfield  {author} {\bibinfo {author} {\bibfnamefont {M.}~\bibnamefont
			{Fujita}}, \bibinfo {author} {\bibfnamefont {H.}~\bibnamefont {Hiraka}},
		\bibinfo {author} {\bibfnamefont {M.}~\bibnamefont {Matsuda}}, \bibinfo
		{author} {\bibfnamefont {M.}~\bibnamefont {Matsuura}}, \bibinfo {author}
		{\bibfnamefont {J.}~\bibnamefont {M.~Tranquada}}, \bibinfo {author}
		{\bibfnamefont {S.}~\bibnamefont {Wakimoto}}, \bibinfo {author}
		{\bibfnamefont {G.}~\bibnamefont {Xu}}, \ and\ \bibinfo {author}
		{\bibfnamefont {K.}~\bibnamefont {Yamada}},\ }\href {\doibase
		10.1143/JPSJ.81.011007} {\bibfield  {journal} {\bibinfo  {journal} {Journal
				of the Physical Society of Japan}\ }\textbf {\bibinfo {volume} {81}},\
		\bibinfo {pages} {011007} (\bibinfo {year} {2012})},\ \Eprint
	{http://arxiv.org/abs/https://doi.org/10.1143/JPSJ.81.011007}
	{https://doi.org/10.1143/JPSJ.81.011007} \BibitemShut {NoStop}%
	\bibitem [{\citenamefont {Stepanov}\ \emph {et~al.}(2018)\citenamefont
		{Stepanov}, \citenamefont {Peters}, \citenamefont {Krivenko}, \citenamefont
		{Lichtenstein}, \citenamefont {Katsnelson},\ and\ \citenamefont
		{Rubtsov}}]{Stepanov2018}%
	\BibitemOpen
	\bibfield  {author} {\bibinfo {author} {\bibfnamefont {E.~A.}\ \bibnamefont
			{Stepanov}}, \bibinfo {author} {\bibfnamefont {L.}~\bibnamefont {Peters}},
		\bibinfo {author} {\bibfnamefont {I.~S.}\ \bibnamefont {Krivenko}}, \bibinfo
		{author} {\bibfnamefont {A.~I.}\ \bibnamefont {Lichtenstein}}, \bibinfo
		{author} {\bibfnamefont {M.~I.}\ \bibnamefont {Katsnelson}}, \ and\ \bibinfo
		{author} {\bibfnamefont {A.~N.}\ \bibnamefont {Rubtsov}},\ }\href {\doibase
		10.1038/s41535-018-0128-x} {\bibfield  {journal} {\bibinfo  {journal} {npj
				Quantum Mater.}\ }\textbf {\bibinfo {volume} {3}} (\bibinfo {year} {2018}),\
		10.1038/s41535-018-0128-x}\BibitemShut {NoStop}%
	\bibitem [{\citenamefont {Seibold}\ \emph {et~al.}(2014)\citenamefont
		{Seibold}, \citenamefont {Di~Castro}, \citenamefont {Grilli},\ and\
		\citenamefont {Lorenzana}}]{Seibold2014}%
	\BibitemOpen
	\bibfield  {author} {\bibinfo {author} {\bibfnamefont {G.}~\bibnamefont
			{Seibold}}, \bibinfo {author} {\bibfnamefont {C.}~\bibnamefont {Di~Castro}},
		\bibinfo {author} {\bibfnamefont {M.}~\bibnamefont {Grilli}}, \ and\ \bibinfo
		{author} {\bibfnamefont {J.}~\bibnamefont {Lorenzana}},\ }\href
	{https://doi.org/10.1038/srep05319} {\bibfield  {journal} {\bibinfo
			{journal} {Scientific Reports}\ }\textbf {\bibinfo {volume} {4}},\ \bibinfo
		{pages} {5319} (\bibinfo {year} {2014})}\BibitemShut {NoStop}%
	\bibitem [{\citenamefont {Yoshida}\ \emph {et~al.}(2006)\citenamefont
		{Yoshida}, \citenamefont {Zhou}, \citenamefont {Tanaka}, \citenamefont
		{Yang}, \citenamefont {Hussain}, \citenamefont {Shen}, \citenamefont
		{Fujimori}, \citenamefont {Sahrakorpi}, \citenamefont {Lindroos},
		\citenamefont {Markiewicz}, \citenamefont {Bansil}, \citenamefont {Komiya},
		\citenamefont {Ando}, \citenamefont {Eisaki}, \citenamefont {Kakeshita},\
		and\ \citenamefont {Uchida}}]{Yoshida2006}%
	\BibitemOpen
	\bibfield  {author} {\bibinfo {author} {\bibfnamefont {T.}~\bibnamefont
			{Yoshida}}, \bibinfo {author} {\bibfnamefont {X.~J.}\ \bibnamefont {Zhou}},
		\bibinfo {author} {\bibfnamefont {K.}~\bibnamefont {Tanaka}}, \bibinfo
		{author} {\bibfnamefont {W.~L.}\ \bibnamefont {Yang}}, \bibinfo {author}
		{\bibfnamefont {Z.}~\bibnamefont {Hussain}}, \bibinfo {author} {\bibfnamefont
			{Z.-X.}\ \bibnamefont {Shen}}, \bibinfo {author} {\bibfnamefont
			{A.}~\bibnamefont {Fujimori}}, \bibinfo {author} {\bibfnamefont
			{S.}~\bibnamefont {Sahrakorpi}}, \bibinfo {author} {\bibfnamefont
			{M.}~\bibnamefont {Lindroos}}, \bibinfo {author} {\bibfnamefont {R.~S.}\
			\bibnamefont {Markiewicz}}, \bibinfo {author} {\bibfnamefont
			{A.}~\bibnamefont {Bansil}}, \bibinfo {author} {\bibfnamefont
			{S.}~\bibnamefont {Komiya}}, \bibinfo {author} {\bibfnamefont
			{Y.}~\bibnamefont {Ando}}, \bibinfo {author} {\bibfnamefont {H.}~\bibnamefont
			{Eisaki}}, \bibinfo {author} {\bibfnamefont {T.}~\bibnamefont {Kakeshita}}, \
		and\ \bibinfo {author} {\bibfnamefont {S.}~\bibnamefont {Uchida}},\ }\href
	{\doibase 10.1103/PhysRevB.74.224510} {\bibfield  {journal} {\bibinfo
			{journal} {Phys. Rev. B}\ }\textbf {\bibinfo {volume} {74}},\ \bibinfo
		{pages} {224510} (\bibinfo {year} {2006})}\BibitemShut {NoStop}%
	\bibitem [{\citenamefont {Gunnarsson}\ \emph {et~al.}(2015)\citenamefont
		{Gunnarsson}, \citenamefont {Sch\"afer}, \citenamefont {LeBlanc},
		\citenamefont {Gull}, \citenamefont {Merino}, \citenamefont {Sangiovanni},
		\citenamefont {Rohringer},\ and\ \citenamefont {Toschi}}]{Gunnarsson2015}%
	\BibitemOpen
	\bibfield  {author} {\bibinfo {author} {\bibfnamefont {O.}~\bibnamefont
			{Gunnarsson}}, \bibinfo {author} {\bibfnamefont {T.}~\bibnamefont
			{Sch\"afer}}, \bibinfo {author} {\bibfnamefont {J.~P.~F.}\ \bibnamefont
			{LeBlanc}}, \bibinfo {author} {\bibfnamefont {E.}~\bibnamefont {Gull}},
		\bibinfo {author} {\bibfnamefont {J.}~\bibnamefont {Merino}}, \bibinfo
		{author} {\bibfnamefont {G.}~\bibnamefont {Sangiovanni}}, \bibinfo {author}
		{\bibfnamefont {G.}~\bibnamefont {Rohringer}}, \ and\ \bibinfo {author}
		{\bibfnamefont {A.}~\bibnamefont {Toschi}},\ }\href {\doibase
		10.1103/PhysRevLett.114.236402} {\bibfield  {journal} {\bibinfo  {journal}
			{Phys. Rev. Lett.}\ }\textbf {\bibinfo {volume} {114}},\ \bibinfo {pages}
		{236402} (\bibinfo {year} {2015})}\BibitemShut {NoStop}%
	\bibitem [{\citenamefont {Wollman}\ \emph {et~al.}(1993)\citenamefont
		{Wollman}, \citenamefont {Van~Harlingen}, \citenamefont {Lee}, \citenamefont
		{Ginsberg},\ and\ \citenamefont {Leggett}}]{Wollman1993}%
	\BibitemOpen
	\bibfield  {author} {\bibinfo {author} {\bibfnamefont {D.~A.}\ \bibnamefont
			{Wollman}}, \bibinfo {author} {\bibfnamefont {D.~J.}\ \bibnamefont
			{Van~Harlingen}}, \bibinfo {author} {\bibfnamefont {W.~C.}\ \bibnamefont
			{Lee}}, \bibinfo {author} {\bibfnamefont {D.~M.}\ \bibnamefont {Ginsberg}}, \
		and\ \bibinfo {author} {\bibfnamefont {A.~J.}\ \bibnamefont {Leggett}},\
	}\href {\doibase 10.1103/PhysRevLett.71.2134} {\bibfield  {journal} {\bibinfo
			{journal} {Phys. Rev. Lett.}\ }\textbf {\bibinfo {volume} {71}},\ \bibinfo
		{pages} {2134} (\bibinfo {year} {1993})}\BibitemShut {NoStop}%
	\bibitem [{\citenamefont {Shen}\ \emph {et~al.}(1993)\citenamefont {Shen},
		\citenamefont {Dessau}, \citenamefont {Wells}, \citenamefont {King},
		\citenamefont {Spicer}, \citenamefont {Arko}, \citenamefont {Marshall},
		\citenamefont {Lombardo}, \citenamefont {Kapitulnik}, \citenamefont
		{Dickinson}, \citenamefont {Doniach}, \citenamefont {DiCarlo}, \citenamefont
		{Loeser},\ and\ \citenamefont {Park}}]{Shen1993}%
	\BibitemOpen
	\bibfield  {author} {\bibinfo {author} {\bibfnamefont {Z.-X.}\ \bibnamefont
			{Shen}}, \bibinfo {author} {\bibfnamefont {D.~S.}\ \bibnamefont {Dessau}},
		\bibinfo {author} {\bibfnamefont {B.~O.}\ \bibnamefont {Wells}}, \bibinfo
		{author} {\bibfnamefont {D.~M.}\ \bibnamefont {King}}, \bibinfo {author}
		{\bibfnamefont {W.~E.}\ \bibnamefont {Spicer}}, \bibinfo {author}
		{\bibfnamefont {A.~J.}\ \bibnamefont {Arko}}, \bibinfo {author}
		{\bibfnamefont {D.}~\bibnamefont {Marshall}}, \bibinfo {author}
		{\bibfnamefont {L.~W.}\ \bibnamefont {Lombardo}}, \bibinfo {author}
		{\bibfnamefont {A.}~\bibnamefont {Kapitulnik}}, \bibinfo {author}
		{\bibfnamefont {P.}~\bibnamefont {Dickinson}}, \bibinfo {author}
		{\bibfnamefont {S.}~\bibnamefont {Doniach}}, \bibinfo {author} {\bibfnamefont
			{J.}~\bibnamefont {DiCarlo}}, \bibinfo {author} {\bibfnamefont
			{T.}~\bibnamefont {Loeser}}, \ and\ \bibinfo {author} {\bibfnamefont {C.~H.}\
			\bibnamefont {Park}},\ }\href {\doibase 10.1103/PhysRevLett.70.1553}
	{\bibfield  {journal} {\bibinfo  {journal} {Phys. Rev. Lett.}\ }\textbf
		{\bibinfo {volume} {70}},\ \bibinfo {pages} {1553} (\bibinfo {year}
		{1993})}\BibitemShut {NoStop}%
	\bibitem [{\citenamefont {Pringle}\ \emph {et~al.}(2000)\citenamefont
		{Pringle}, \citenamefont {Williams},\ and\ \citenamefont
		{Tallon}}]{Pringle2000}%
	\BibitemOpen
	\bibfield  {author} {\bibinfo {author} {\bibfnamefont {D.~J.}\ \bibnamefont
			{Pringle}}, \bibinfo {author} {\bibfnamefont {G.~V.~M.}\ \bibnamefont
			{Williams}}, \ and\ \bibinfo {author} {\bibfnamefont {J.~L.}\ \bibnamefont
			{Tallon}},\ }\href {\doibase 10.1103/PhysRevB.62.12527} {\bibfield  {journal}
		{\bibinfo  {journal} {Phys. Rev. B}\ }\textbf {\bibinfo {volume} {62}},\
		\bibinfo {pages} {12527} (\bibinfo {year} {2000})}\BibitemShut {NoStop}%
	\bibitem [{\citenamefont {Greco}\ and\ \citenamefont
		{Zeyher}(2015)}]{Greco2015}%
	\BibitemOpen
	\bibfield  {author} {\bibinfo {author} {\bibfnamefont {A.}~\bibnamefont
			{Greco}}\ and\ \bibinfo {author} {\bibfnamefont {R.}~\bibnamefont {Zeyher}},\
	}\href {\doibase 10.1088/0953-2048/29/1/015002} {\bibfield  {journal}
		{\bibinfo  {journal} {Superconductor Science and Technology}\ }\textbf
		{\bibinfo {volume} {29}},\ \bibinfo {pages} {015002} (\bibinfo {year}
		{2015})}\BibitemShut {NoStop}%
	\bibitem [{\citenamefont {Schmalian}\ \emph {et~al.}(1998)\citenamefont
		{Schmalian}, \citenamefont {Pines},\ and\ \citenamefont
		{Stojkovi{\'{c}}}}]{Schmalian1998}%
	\BibitemOpen
	\bibfield  {author} {\bibinfo {author} {\bibfnamefont {J.}~\bibnamefont
			{Schmalian}}, \bibinfo {author} {\bibfnamefont {D.}~\bibnamefont {Pines}}, \
		and\ \bibinfo {author} {\bibfnamefont {B.}~\bibnamefont {Stojkovi{\'{c}}}},\
	}\href {\doibase 10.1016/S0022-3697(98)00104-8} {\bibfield  {journal}
		{\bibinfo  {journal} {J. Phys. Chem. Solids}\ }\textbf {\bibinfo {volume}
			{59}},\ \bibinfo {pages} {1764} (\bibinfo {year} {1998})}\BibitemShut
	{NoStop}%
	\bibitem [{\citenamefont {Scalapino}(2012)}]{Scalapino2012}%
	\BibitemOpen
	\bibfield  {author} {\bibinfo {author} {\bibfnamefont {D.~J.}\ \bibnamefont
			{Scalapino}},\ }\href {\doibase 10.1103/RevModPhys.84.1383} {\bibfield
		{journal} {\bibinfo  {journal} {Rev. Mod. Phys.}\ }\textbf {\bibinfo {volume}
			{84}},\ \bibinfo {pages} {1383} (\bibinfo {year} {2012})}\BibitemShut
	{NoStop}%
	\bibitem [{\citenamefont {Markiewicz}(1997)}]{Markiewicz1997}%
	\BibitemOpen
	\bibfield  {author} {\bibinfo {author} {\bibfnamefont {R.~S.}\ \bibnamefont
			{Markiewicz}},\ }\href {\doibase 10.1016/S0022-3697(97)00025-5} {\bibfield
		{journal} {\bibinfo  {journal} {J. Phys. Chem. Solids}\ }\textbf {\bibinfo
			{volume} {58}},\ \bibinfo {pages} {1179} (\bibinfo {year}
		{1997})}\BibitemShut {NoStop}%
	\bibitem [{\citenamefont {Piriou}\ \emph {et~al.}(2011)\citenamefont {Piriou},
		\citenamefont {Jenkins}, \citenamefont {Berthod}, \citenamefont
		{Maggio-Aprile},\ and\ \citenamefont {Fischer}}]{Piriou2011}%
	\BibitemOpen
	\bibfield  {author} {\bibinfo {author} {\bibfnamefont {A.}~\bibnamefont
			{Piriou}}, \bibinfo {author} {\bibfnamefont {N.}~\bibnamefont {Jenkins}},
		\bibinfo {author} {\bibfnamefont {C.}~\bibnamefont {Berthod}}, \bibinfo
		{author} {\bibfnamefont {I.}~\bibnamefont {Maggio-Aprile}}, \ and\ \bibinfo
		{author} {\bibfnamefont {Ã.}~\bibnamefont {Fischer}},\ }\href
	{https://doi.org/10.1038/ncomms1229} {\bibfield  {journal} {\bibinfo
			{journal} {Nature Communications}\ }\textbf {\bibinfo {volume} {2}},\
		\bibinfo {pages} {221} (\bibinfo {year} {2011})}\BibitemShut {NoStop}%
	\bibitem [{\citenamefont {Legros}\ \emph {et~al.}(2019)\citenamefont {Legros},
		\citenamefont {Benhabib}, \citenamefont {Tabis}, \citenamefont {LalibertÃ©},
		\citenamefont {Dion}, \citenamefont {Lizaire}, \citenamefont {Vignolle},
		\citenamefont {Vignolles}, \citenamefont {Raffy}, \citenamefont {Li},
		\citenamefont {Auban-Senzier}, \citenamefont {Doiron-Leyraud}, \citenamefont
		{Fournier}, \citenamefont {Colson}, \citenamefont {Taillefer},\ and\
		\citenamefont {Proust}}]{Legros2019}%
	\BibitemOpen
	\bibfield  {author} {\bibinfo {author} {\bibfnamefont {A.}~\bibnamefont
			{Legros}}, \bibinfo {author} {\bibfnamefont {S.}~\bibnamefont {Benhabib}},
		\bibinfo {author} {\bibfnamefont {W.}~\bibnamefont {Tabis}}, \bibinfo
		{author} {\bibfnamefont {F.}~\bibnamefont {LalibertÃ©}}, \bibinfo {author}
		{\bibfnamefont {M.}~\bibnamefont {Dion}}, \bibinfo {author} {\bibfnamefont
			{M.}~\bibnamefont {Lizaire}}, \bibinfo {author} {\bibfnamefont
			{B.}~\bibnamefont {Vignolle}}, \bibinfo {author} {\bibfnamefont
			{D.}~\bibnamefont {Vignolles}}, \bibinfo {author} {\bibfnamefont
			{H.}~\bibnamefont {Raffy}}, \bibinfo {author} {\bibfnamefont {Z.~Z.}\
			\bibnamefont {Li}}, \bibinfo {author} {\bibfnamefont {P.}~\bibnamefont
			{Auban-Senzier}}, \bibinfo {author} {\bibfnamefont {N.}~\bibnamefont
			{Doiron-Leyraud}}, \bibinfo {author} {\bibfnamefont {P.}~\bibnamefont
			{Fournier}}, \bibinfo {author} {\bibfnamefont {D.}~\bibnamefont {Colson}},
		\bibinfo {author} {\bibfnamefont {L.}~\bibnamefont {Taillefer}}, \ and\
		\bibinfo {author} {\bibfnamefont {C.}~\bibnamefont {Proust}},\ }\href
	{https://doi.org/10.1038/s41567-018-0334-2} {\bibfield  {journal} {\bibinfo
			{journal} {Nature Physics}\ }\textbf {\bibinfo {volume} {15}},\ \bibinfo
		{pages} {142} (\bibinfo {year} {2019})}\BibitemShut {NoStop}%
	\bibitem [{\citenamefont {Jenkins}\ \emph {et~al.}(2010)\citenamefont
		{Jenkins}, \citenamefont {Schmadel}, \citenamefont {Bach}, \citenamefont
		{Greene}, \citenamefont {B\'echamp-Lagani\`ere}, \citenamefont {Roberge},
		\citenamefont {Fournier}, \citenamefont {Kontani},\ and\ \citenamefont
		{Drew}}]{Jenkins2010}%
	\BibitemOpen
	\bibfield  {author} {\bibinfo {author} {\bibfnamefont {G.~S.}\ \bibnamefont
			{Jenkins}}, \bibinfo {author} {\bibfnamefont {D.~C.}\ \bibnamefont
			{Schmadel}}, \bibinfo {author} {\bibfnamefont {P.~L.}\ \bibnamefont {Bach}},
		\bibinfo {author} {\bibfnamefont {R.~L.}\ \bibnamefont {Greene}}, \bibinfo
		{author} {\bibfnamefont {X.}~\bibnamefont {B\'echamp-Lagani\`ere}}, \bibinfo
		{author} {\bibfnamefont {G.}~\bibnamefont {Roberge}}, \bibinfo {author}
		{\bibfnamefont {P.}~\bibnamefont {Fournier}}, \bibinfo {author}
		{\bibfnamefont {H.}~\bibnamefont {Kontani}}, \ and\ \bibinfo {author}
		{\bibfnamefont {H.~D.}\ \bibnamefont {Drew}},\ }\href {\doibase
		10.1103/PhysRevB.81.024508} {\bibfield  {journal} {\bibinfo  {journal} {Phys.
				Rev. B}\ }\textbf {\bibinfo {volume} {81}},\ \bibinfo {pages} {024508}
		(\bibinfo {year} {2010})}\BibitemShut {NoStop}%
	\bibitem [{\citenamefont {Lee}\ \emph {et~al.}(2006)\citenamefont {Lee},
		\citenamefont {Nagaosa},\ and\ \citenamefont {Wen}}]{Lee2006}%
	\BibitemOpen
	\bibfield  {author} {\bibinfo {author} {\bibfnamefont {P.~A.}\ \bibnamefont
			{Lee}}, \bibinfo {author} {\bibfnamefont {N.}~\bibnamefont {Nagaosa}}, \ and\
		\bibinfo {author} {\bibfnamefont {X.~G.}\ \bibnamefont {Wen}},\ }\href
	{\doibase 10.1103/RevModPhys.78.17} {\bibfield  {journal} {\bibinfo
			{journal} {Rev. Mod. Phys.}\ }\textbf {\bibinfo {volume} {78}} (\bibinfo
		{year} {2006}),\ 10.1103/RevModPhys.78.17}\BibitemShut {NoStop}%
	\bibitem [{\citenamefont {Hubbard}(1964)}]{Hubbard64}%
	\BibitemOpen
	\bibfield  {author} {\bibinfo {author} {\bibfnamefont {J.}~\bibnamefont
			{Hubbard}},\ }\href {\doibase 10.1098/rspa.1964.0190} {\bibfield  {journal}
		{\bibinfo  {journal} {Proc R. Soc. London}\ }\textbf {\bibinfo {volume}
			{281}},\ \bibinfo {pages} {401} (\bibinfo {year} {1964})}\BibitemShut
	{NoStop}%
	\bibitem [{\citenamefont {Reznik}\ \emph {et~al.}(2006)\citenamefont {Reznik},
		\citenamefont {Pintschovius}, \citenamefont {Ito}, \citenamefont {Iikubo},
		\citenamefont {Sato}, \citenamefont {Goka}, \citenamefont {Fujita},
		\citenamefont {Yamada}, \citenamefont {Gu},\ and\ \citenamefont
		{Tranquada}}]{Reznik2006}%
	\BibitemOpen
	\bibfield  {author} {\bibinfo {author} {\bibfnamefont {D.}~\bibnamefont
			{Reznik}}, \bibinfo {author} {\bibfnamefont {L.}~\bibnamefont
			{Pintschovius}}, \bibinfo {author} {\bibfnamefont {M.}~\bibnamefont {Ito}},
		\bibinfo {author} {\bibfnamefont {S.}~\bibnamefont {Iikubo}}, \bibinfo
		{author} {\bibfnamefont {M.}~\bibnamefont {Sato}}, \bibinfo {author}
		{\bibfnamefont {H.}~\bibnamefont {Goka}}, \bibinfo {author} {\bibfnamefont
			{M.}~\bibnamefont {Fujita}}, \bibinfo {author} {\bibfnamefont
			{K.}~\bibnamefont {Yamada}}, \bibinfo {author} {\bibfnamefont {G.~D.}\
			\bibnamefont {Gu}}, \ and\ \bibinfo {author} {\bibfnamefont {J.~M.}\
			\bibnamefont {Tranquada}},\ }\href {https://doi.org/10.1038/nature04704}
	{\bibfield  {journal} {\bibinfo  {journal} {Nature}\ }\textbf {\bibinfo
			{volume} {440}},\ \bibinfo {pages} {1170} (\bibinfo {year}
		{2006})}\BibitemShut {NoStop}%
	\bibitem [{\citenamefont {Miao}\ \emph {et~al.}(2017)\citenamefont {Miao},
		\citenamefont {Lorenzana}, \citenamefont {Seibold}, \citenamefont {Peng},
		\citenamefont {Amorese}, \citenamefont {Yakhou-Harris}, \citenamefont
		{Kummer}, \citenamefont {Brookes}, \citenamefont {Konik}, \citenamefont
		{Thampy}, \citenamefont {Gu}, \citenamefont {Ghiringhelli}, \citenamefont
		{Braicovich},\ and\ \citenamefont {Dean}}]{Miao2017}%
	\BibitemOpen
	\bibfield  {author} {\bibinfo {author} {\bibfnamefont {H.}~\bibnamefont
			{Miao}}, \bibinfo {author} {\bibfnamefont {J.}~\bibnamefont {Lorenzana}},
		\bibinfo {author} {\bibfnamefont {G.}~\bibnamefont {Seibold}}, \bibinfo
		{author} {\bibfnamefont {Y.~Y.}\ \bibnamefont {Peng}}, \bibinfo {author}
		{\bibfnamefont {A.}~\bibnamefont {Amorese}}, \bibinfo {author} {\bibfnamefont
			{F.}~\bibnamefont {Yakhou-Harris}}, \bibinfo {author} {\bibfnamefont
			{K.}~\bibnamefont {Kummer}}, \bibinfo {author} {\bibfnamefont {N.~B.}\
			\bibnamefont {Brookes}}, \bibinfo {author} {\bibfnamefont {R.~M.}\
			\bibnamefont {Konik}}, \bibinfo {author} {\bibfnamefont {V.}~\bibnamefont
			{Thampy}}, \bibinfo {author} {\bibfnamefont {G.~D.}\ \bibnamefont {Gu}},
		\bibinfo {author} {\bibfnamefont {G.}~\bibnamefont {Ghiringhelli}}, \bibinfo
		{author} {\bibfnamefont {L.}~\bibnamefont {Braicovich}}, \ and\ \bibinfo
		{author} {\bibfnamefont {M.~P.~M.}\ \bibnamefont {Dean}},\ }\href {\doibase
		10.1073/pnas.1708549114} {\bibfield  {journal} {\bibinfo  {journal}
			{Proceedings of the National Academy of Sciences}\ }\textbf {\bibinfo
			{volume} {114}},\ \bibinfo {pages} {12430} (\bibinfo {year} {2017})},\
	\Eprint
	{http://arxiv.org/abs/https://www.pnas.org/content/114/47/12430.full.pdf}
	{https://www.pnas.org/content/114/47/12430.full.pdf} \BibitemShut {NoStop}%
	\bibitem [{\citenamefont {Avella}\ \emph {et~al.}(2013)\citenamefont {Avella},
		\citenamefont {Mancini}, \citenamefont {Mancini},\ and\ \citenamefont
		{Plekhanov}}]{Avella2013}%
	\BibitemOpen
	\bibfield  {author} {\bibinfo {author} {\bibfnamefont {A.}~\bibnamefont
			{Avella}}, \bibinfo {author} {\bibfnamefont {F.}~\bibnamefont {Mancini}},
		\bibinfo {author} {\bibfnamefont {F.~P.}\ \bibnamefont {Mancini}}, \ and\
		\bibinfo {author} {\bibfnamefont {E.}~\bibnamefont {Plekhanov}},\ }\href
	{\doibase 10.1140/epjb/e2013-40115-3} {\bibfield  {journal} {\bibinfo
			{journal} {The European Physical Journal B}\ }\textbf {\bibinfo {volume}
			{86}},\ \bibinfo {pages} {265} (\bibinfo {year} {2013})}\BibitemShut
	{NoStop}%
	\bibitem [{\citenamefont {Hansmann}\ \emph {et~al.}(2014)\citenamefont
		{Hansmann}, \citenamefont {Parragh}, \citenamefont {Toschi}, \citenamefont
		{Sangiovanni},\ and\ \citenamefont {Held}}]{Hansmann2014}%
	\BibitemOpen
	\bibfield  {author} {\bibinfo {author} {\bibfnamefont {P.}~\bibnamefont
			{Hansmann}}, \bibinfo {author} {\bibfnamefont {N.}~\bibnamefont {Parragh}},
		\bibinfo {author} {\bibfnamefont {A.}~\bibnamefont {Toschi}}, \bibinfo
		{author} {\bibfnamefont {G.}~\bibnamefont {Sangiovanni}}, \ and\ \bibinfo
		{author} {\bibfnamefont {K.}~\bibnamefont {Held}},\ }\href
	{http://stacks.iop.org/1367-2630/16/i=3/a=033009} {\bibfield  {journal}
		{\bibinfo  {journal} {New J. Phys.}\ }\textbf {\bibinfo {volume} {16}},\
		\bibinfo {pages} {033009} (\bibinfo {year} {2014})}\BibitemShut {NoStop}%
	\bibitem [{\citenamefont {Anderson}(2002)}]{Anderson2002}%
	\BibitemOpen
	\bibfield  {author} {\bibinfo {author} {\bibfnamefont {P.~W.}\ \bibnamefont
			{Anderson}},\ }\href {\doibase 10.1238/physica.topical.102a00010} {\bibfield
		{journal} {\bibinfo  {journal} {Physica Scripta}\ }\textbf {\bibinfo {volume}
			{T102}},\ \bibinfo {pages} {10} (\bibinfo {year} {2002})}\BibitemShut
	{NoStop}%
	\bibitem [{\citenamefont {Loh}\ \emph {et~al.}(1990)\citenamefont {Loh},
		\citenamefont {Gubernatis}, \citenamefont {Scalettar}, \citenamefont {White},
		\citenamefont {Scalapino},\ and\ \citenamefont {Sugar}}]{Loh1990}%
	\BibitemOpen
	\bibfield  {author} {\bibinfo {author} {\bibfnamefont {E.~Y.}\ \bibnamefont
			{Loh}}, \bibinfo {author} {\bibfnamefont {J.~E.}\ \bibnamefont {Gubernatis}},
		\bibinfo {author} {\bibfnamefont {R.~T.}\ \bibnamefont {Scalettar}}, \bibinfo
		{author} {\bibfnamefont {S.~R.}\ \bibnamefont {White}}, \bibinfo {author}
		{\bibfnamefont {D.~J.}\ \bibnamefont {Scalapino}}, \ and\ \bibinfo {author}
		{\bibfnamefont {R.~L.}\ \bibnamefont {Sugar}},\ }\href {\doibase
		10.1103/PhysRevB.41.9301} {\bibfield  {journal} {\bibinfo  {journal} {Phys.
				Rev. B}\ }\textbf {\bibinfo {volume} {41}},\ \bibinfo {pages} {9301}
		(\bibinfo {year} {1990})}\BibitemShut {NoStop}%
	\bibitem [{\citenamefont {Knizia}\ and\ \citenamefont
		{Chan}(2012)}]{Knizia2012}%
	\BibitemOpen
	\bibfield  {author} {\bibinfo {author} {\bibfnamefont {G.}~\bibnamefont
			{Knizia}}\ and\ \bibinfo {author} {\bibfnamefont {G.~K.-L.}\ \bibnamefont
			{Chan}},\ }\href {\doibase 10.1103/PhysRevLett.109.186404} {\bibfield
		{journal} {\bibinfo  {journal} {Phys. Rev. Lett.}\ }\textbf {\bibinfo
			{volume} {109}},\ \bibinfo {pages} {186404} (\bibinfo {year}
		{2012})}\BibitemShut {NoStop}%
	\bibitem [{\citenamefont {Senjean}(2019)}]{Senjean2019}%
	\BibitemOpen
	\bibfield  {author} {\bibinfo {author} {\bibfnamefont {B.}~\bibnamefont
			{Senjean}},\ }\href {\doibase 10.1103/PhysRevB.100.035136} {\bibfield
		{journal} {\bibinfo  {journal} {Phys. Rev. B}\ }\textbf {\bibinfo {volume}
			{100}},\ \bibinfo {pages} {035136} (\bibinfo {year} {2019})}\BibitemShut
	{NoStop}%
	\bibitem [{\citenamefont {Hallberg}(2006)}]{Hallberg2006}%
	\BibitemOpen
	\bibfield  {author} {\bibinfo {author} {\bibfnamefont {K.~A.}\ \bibnamefont
			{Hallberg}},\ }\href {\doibase 10.1080/00018730600766432} {\bibfield
		{journal} {\bibinfo  {journal} {Advances in Physics}\ }\textbf {\bibinfo
			{volume} {55}},\ \bibinfo {pages} {477} (\bibinfo {year} {2006})},\ \Eprint
	{http://arxiv.org/abs/https://doi.org/10.1080/00018730600766432}
	{https://doi.org/10.1080/00018730600766432} \BibitemShut {NoStop}%
	\bibitem [{\citenamefont {Huang}\ \emph {et~al.}(2018)\citenamefont {Huang},
		\citenamefont {Mendl}, \citenamefont {Jiang}, \citenamefont {Moritz},\ and\
		\citenamefont {Devereaux}}]{Huang2018}%
	\BibitemOpen
	\bibfield  {author} {\bibinfo {author} {\bibfnamefont {E.~W.}\ \bibnamefont
			{Huang}}, \bibinfo {author} {\bibfnamefont {C.~B.}\ \bibnamefont {Mendl}},
		\bibinfo {author} {\bibfnamefont {H.~C.}\ \bibnamefont {Jiang}}, \bibinfo
		{author} {\bibfnamefont {B.}~\bibnamefont {Moritz}}, \ and\ \bibinfo {author}
		{\bibfnamefont {T.~P.}\ \bibnamefont {Devereaux}},\ }\href {\doibase
		10.1038/s41535-018-0097-0} {\bibfield  {journal} {\bibinfo  {journal} {npj
				Quantum Materials}\ }\textbf {\bibinfo {volume} {3}},\ \bibinfo {pages} {1}
		(\bibinfo {year} {2018})}\BibitemShut {NoStop}%
	\bibitem [{\citenamefont {Halboth}\ and\ \citenamefont
		{Metzner}(2000)}]{Halboth2000}%
	\BibitemOpen
	\bibfield  {author} {\bibinfo {author} {\bibfnamefont {C.~J.}\ \bibnamefont
			{Halboth}}\ and\ \bibinfo {author} {\bibfnamefont {W.}~\bibnamefont
			{Metzner}},\ }\href {\doibase 10.1103/PhysRevB.61.7364} {\bibfield  {journal}
		{\bibinfo  {journal} {Phys. Rev. B}\ }\textbf {\bibinfo {volume} {61}},\
		\bibinfo {pages} {7364} (\bibinfo {year} {2000})}\BibitemShut {NoStop}%
	\bibitem [{\citenamefont {Metzner}\ \emph {et~al.}(2012)\citenamefont
		{Metzner}, \citenamefont {Salmhofer}, \citenamefont {Honerkamp},
		\citenamefont {Meden},\ and\ \citenamefont {Schoenhammer}}]{Metzner2012}%
	\BibitemOpen
	\bibfield  {author} {\bibinfo {author} {\bibfnamefont {W.}~\bibnamefont
			{Metzner}}, \bibinfo {author} {\bibfnamefont {M.}~\bibnamefont {Salmhofer}},
		\bibinfo {author} {\bibfnamefont {C.}~\bibnamefont {Honerkamp}}, \bibinfo
		{author} {\bibfnamefont {V.}~\bibnamefont {Meden}}, \ and\ \bibinfo {author}
		{\bibfnamefont {K.}~\bibnamefont {Schoenhammer}},\ }\href {\doibase
		10.1103/RevModPhys.84.299} {\bibfield  {journal} {\bibinfo  {journal} {Rev.
				Mod. Phys.}\ }\textbf {\bibinfo {volume} {84}},\ \bibinfo {pages} {299}
		(\bibinfo {year} {2012})},\ \Eprint {http://arxiv.org/abs/1105.5289}
	{arXiv:1105.5289} \BibitemShut {NoStop}%
	\bibitem [{\citenamefont {Metzner}\ and\ \citenamefont
		{Vollhardt}(1989)}]{Metzner1989}%
	\BibitemOpen
	\bibfield  {author} {\bibinfo {author} {\bibfnamefont {W.}~\bibnamefont
			{Metzner}}\ and\ \bibinfo {author} {\bibfnamefont {D.}~\bibnamefont
			{Vollhardt}},\ }\href {\doibase 10.1103/PhysRevLett.62.324} {\bibfield
		{journal} {\bibinfo  {journal} {Phys. Rev. Lett.}\ }\textbf {\bibinfo
			{volume} {62}},\ \bibinfo {pages} {324} (\bibinfo {year} {1989})}\BibitemShut
	{NoStop}%
	\bibitem [{\citenamefont {Georges}\ and\ \citenamefont
		{Krauth}(1992)}]{Georges1992}%
	\BibitemOpen
	\bibfield  {author} {\bibinfo {author} {\bibfnamefont {A.}~\bibnamefont
			{Georges}}\ and\ \bibinfo {author} {\bibfnamefont {W.}~\bibnamefont
			{Krauth}},\ }\href {\doibase 10.1103/PhysRevLett.69.1240} {\bibfield
		{journal} {\bibinfo  {journal} {Phys. Rev. Lett.}\ }\textbf {\bibinfo
			{volume} {69}},\ \bibinfo {pages} {1240} (\bibinfo {year}
		{1992})}\BibitemShut {NoStop}%
	\bibitem [{\citenamefont {Georges}\ \emph {et~al.}(1996)\citenamefont
		{Georges}, \citenamefont {Kotliar}, \citenamefont {Krauth},\ and\
		\citenamefont {Rozenberg}}]{Georges1996}%
	\BibitemOpen
	\bibfield  {author} {\bibinfo {author} {\bibfnamefont {A.}~\bibnamefont
			{Georges}}, \bibinfo {author} {\bibfnamefont {G.}~\bibnamefont {Kotliar}},
		\bibinfo {author} {\bibfnamefont {W.}~\bibnamefont {Krauth}}, \ and\ \bibinfo
		{author} {\bibfnamefont {M.~J.}\ \bibnamefont {Rozenberg}},\ }\href {\doibase
		10.1103/RevModPhys.68.13} {\bibfield  {journal} {\bibinfo  {journal} {Rev.
				Mod. Phys.}\ }\textbf {\bibinfo {volume} {68}},\ \bibinfo {pages} {13}
		(\bibinfo {year} {1996})}\BibitemShut {NoStop}%
	\bibitem [{\citenamefont {Lichtenstein}\ and\ \citenamefont
		{Katsnelson}(2000)}]{Lichtenstein2000}%
	\BibitemOpen
	\bibfield  {author} {\bibinfo {author} {\bibfnamefont {A.~I.}\ \bibnamefont
			{Lichtenstein}}\ and\ \bibinfo {author} {\bibfnamefont {M.~I.}\ \bibnamefont
			{Katsnelson}},\ }\href {\doibase 10.1103/PhysRevB.62.R9283} {\bibfield
		{journal} {\bibinfo  {journal} {Phys. Rev. B - Condens. Matter Mater. Phys.}\
		}\textbf {\bibinfo {volume} {62}},\ \bibinfo {pages} {9283} (\bibinfo {year}
		{2000})},\ \Eprint {http://arxiv.org/abs/9911320v1} {arXiv:9911320v1
		[arXiv:cond-mat]} \BibitemShut {NoStop}%
	\bibitem [{\citenamefont {Maier}\ \emph {et~al.}(2005)\citenamefont {Maier},
		\citenamefont {Jarrell}, \citenamefont {Pruschke},\ and\ \citenamefont
		{Hettler}}]{Maier2005}%
	\BibitemOpen
	\bibfield  {author} {\bibinfo {author} {\bibfnamefont {T.}~\bibnamefont
			{Maier}}, \bibinfo {author} {\bibfnamefont {M.}~\bibnamefont {Jarrell}},
		\bibinfo {author} {\bibfnamefont {T.}~\bibnamefont {Pruschke}}, \ and\
		\bibinfo {author} {\bibfnamefont {M.~H.}\ \bibnamefont {Hettler}},\ }\href
	{\doibase 10.1103/RevModPhys.77.1027} {\bibfield  {journal} {\bibinfo
			{journal} {Rev. Mod. Phys.}\ }\textbf {\bibinfo {volume} {77}},\ \bibinfo
		{pages} {1027} (\bibinfo {year} {2005})},\ \Eprint
	{http://arxiv.org/abs/0404055} {arXiv:0404055 [cond-mat]} \BibitemShut
	{NoStop}%
	\bibitem [{\citenamefont {Yang}\ \emph {et~al.}(2011)\citenamefont {Yang},
		\citenamefont {Fotso}, \citenamefont {Su}, \citenamefont {Galanakis},
		\citenamefont {Khatami}, \citenamefont {She}, \citenamefont {Moreno},
		\citenamefont {Zaanen},\ and\ \citenamefont {Jarrell}}]{Yang2011}%
	\BibitemOpen
	\bibfield  {author} {\bibinfo {author} {\bibfnamefont {S.~X.}\ \bibnamefont
			{Yang}}, \bibinfo {author} {\bibfnamefont {H.}~\bibnamefont {Fotso}},
		\bibinfo {author} {\bibfnamefont {S.~Q.}\ \bibnamefont {Su}}, \bibinfo
		{author} {\bibfnamefont {D.}~\bibnamefont {Galanakis}}, \bibinfo {author}
		{\bibfnamefont {E.}~\bibnamefont {Khatami}}, \bibinfo {author} {\bibfnamefont
			{J.~H.}\ \bibnamefont {She}}, \bibinfo {author} {\bibfnamefont
			{J.}~\bibnamefont {Moreno}}, \bibinfo {author} {\bibfnamefont
			{J.}~\bibnamefont {Zaanen}}, \ and\ \bibinfo {author} {\bibfnamefont
			{M.}~\bibnamefont {Jarrell}},\ }\href {\doibase
		10.1103/PhysRevLett.106.047004} {\bibfield  {journal} {\bibinfo  {journal}
			{Phys. Rev. Lett.}\ }\textbf {\bibinfo {volume} {106}},\ \bibinfo {pages} {1}
		(\bibinfo {year} {2011})}\BibitemShut {NoStop}%
	\bibitem [{\citenamefont {Chen}\ \emph {et~al.}(2013)\citenamefont {Chen},
		\citenamefont {Meng}, \citenamefont {Yang}, \citenamefont {Pruschke},
		\citenamefont {Moreno},\ and\ \citenamefont {Jarrell}}]{Chen2013}%
	\BibitemOpen
	\bibfield  {author} {\bibinfo {author} {\bibfnamefont {K.~S.}\ \bibnamefont
			{Chen}}, \bibinfo {author} {\bibfnamefont {Z.~Y.}\ \bibnamefont {Meng}},
		\bibinfo {author} {\bibfnamefont {S.~X.}\ \bibnamefont {Yang}}, \bibinfo
		{author} {\bibfnamefont {T.}~\bibnamefont {Pruschke}}, \bibinfo {author}
		{\bibfnamefont {J.}~\bibnamefont {Moreno}}, \ and\ \bibinfo {author}
		{\bibfnamefont {M.}~\bibnamefont {Jarrell}},\ }\href {\doibase
		10.1103/PhysRevB.88.245110} {\bibfield  {journal} {\bibinfo  {journal} {Phys.
				Rev. B - Condens. Matter Mater. Phys.}\ }\textbf {\bibinfo {volume} {88}},\
		\bibinfo {pages} {1} (\bibinfo {year} {2013})}\BibitemShut {NoStop}%
	\bibitem [{\citenamefont {Gull}\ \emph {et~al.}(2013)\citenamefont {Gull},
		\citenamefont {Parcollet},\ and\ \citenamefont {Millis}}]{Gull2013}%
	\BibitemOpen
	\bibfield  {author} {\bibinfo {author} {\bibfnamefont {E.}~\bibnamefont
			{Gull}}, \bibinfo {author} {\bibfnamefont {O.}~\bibnamefont {Parcollet}}, \
		and\ \bibinfo {author} {\bibfnamefont {A.~J.}\ \bibnamefont {Millis}},\
	}\href {\doibase 10.1103/PhysRevLett.110.216405} {\bibfield  {journal}
		{\bibinfo  {journal} {Phys. Rev. Lett.}\ }\textbf {\bibinfo {volume} {110}},\
		\bibinfo {pages} {216405} (\bibinfo {year} {2013})}\BibitemShut {NoStop}%
	\bibitem [{\citenamefont {Chen}\ \emph {et~al.}(2015)\citenamefont {Chen},
		\citenamefont {LeBlanc},\ and\ \citenamefont {Gull}}]{Chen2015}%
	\BibitemOpen
	\bibfield  {author} {\bibinfo {author} {\bibfnamefont {X.}~\bibnamefont
			{Chen}}, \bibinfo {author} {\bibfnamefont {J.~P.~F.}\ \bibnamefont
			{LeBlanc}}, \ and\ \bibinfo {author} {\bibfnamefont {E.}~\bibnamefont
			{Gull}},\ }\href {\doibase 10.1103/PhysRevLett.115.116402} {\bibfield
		{journal} {\bibinfo  {journal} {Phys. Rev. Lett.}\ }\textbf {\bibinfo
			{volume} {115}},\ \bibinfo {pages} {116402} (\bibinfo {year}
		{2015})}\BibitemShut {NoStop}%
	\bibitem [{\citenamefont {Khatami}\ \emph {et~al.}(2010)\citenamefont
		{Khatami}, \citenamefont {Mikelsons}, \citenamefont {Galanakis},
		\citenamefont {Macridin}, \citenamefont {Moreno}, \citenamefont {Scalettar},\
		and\ \citenamefont {Jarrell}}]{Khatami2010}%
	\BibitemOpen
	\bibfield  {author} {\bibinfo {author} {\bibfnamefont {E.}~\bibnamefont
			{Khatami}}, \bibinfo {author} {\bibfnamefont {K.}~\bibnamefont {Mikelsons}},
		\bibinfo {author} {\bibfnamefont {D.}~\bibnamefont {Galanakis}}, \bibinfo
		{author} {\bibfnamefont {A.}~\bibnamefont {Macridin}}, \bibinfo {author}
		{\bibfnamefont {J.}~\bibnamefont {Moreno}}, \bibinfo {author} {\bibfnamefont
			{R.~T.}\ \bibnamefont {Scalettar}}, \ and\ \bibinfo {author} {\bibfnamefont
			{M.}~\bibnamefont {Jarrell}},\ }\href {\doibase 10.1103/PhysRevB.81.201101}
	{\bibfield  {journal} {\bibinfo  {journal} {Phys. Rev. B}\ }\textbf {\bibinfo
			{volume} {81}},\ \bibinfo {pages} {201101} (\bibinfo {year}
		{2010})}\BibitemShut {NoStop}%
	\bibitem [{\citenamefont {Varma}(1999)}]{Varma1999}%
	\BibitemOpen
	\bibfield  {author} {\bibinfo {author} {\bibfnamefont {C.~M.}\ \bibnamefont
			{Varma}},\ }\href {\doibase 10.1103/PhysRevLett.83.3538} {\bibfield
		{journal} {\bibinfo  {journal} {Phys. Rev. Lett.}\ }\textbf {\bibinfo
			{volume} {83}},\ \bibinfo {pages} {3538} (\bibinfo {year}
		{1999})}\BibitemShut {NoStop}%
	\bibitem [{\citenamefont {Broun}(2008)}]{Broun2008}%
	\BibitemOpen
	\bibfield  {author} {\bibinfo {author} {\bibfnamefont {D.~M.}\ \bibnamefont
			{Broun}},\ }\href {\doibase 10.1038/nphys909} {\bibfield  {journal} {\bibinfo
			{journal} {Nat. Phys.}\ }\textbf {\bibinfo {volume} {4}},\ \bibinfo {pages}
		{170} (\bibinfo {year} {2008})}\BibitemShut {NoStop}%
	\bibitem [{\citenamefont {Sachdev}(2010)}]{Sachdev2010}%
	\BibitemOpen
	\bibfield  {author} {\bibinfo {author} {\bibfnamefont {S.}~\bibnamefont
			{Sachdev}},\ }\href {\doibase 10.1002/pssb.200983037} {\bibfield  {journal}
		{\bibinfo  {journal} {Phys. Status Solidi Basic Res.}\ }\textbf {\bibinfo
			{volume} {247}},\ \bibinfo {pages} {537} (\bibinfo {year}
		{2010})}\BibitemShut {NoStop}%
	\bibitem [{\citenamefont {Chen}\ \emph {et~al.}(2012)\citenamefont {Chen},
		\citenamefont {Meng}, \citenamefont {Pruschke}, \citenamefont {Moreno},\ and\
		\citenamefont {Jarrell}}]{Chen2012}%
	\BibitemOpen
	\bibfield  {author} {\bibinfo {author} {\bibfnamefont {K.-S.}\ \bibnamefont
			{Chen}}, \bibinfo {author} {\bibfnamefont {Z.~Y.}\ \bibnamefont {Meng}},
		\bibinfo {author} {\bibfnamefont {T.}~\bibnamefont {Pruschke}}, \bibinfo
		{author} {\bibfnamefont {J.}~\bibnamefont {Moreno}}, \ and\ \bibinfo {author}
		{\bibfnamefont {M.}~\bibnamefont {Jarrell}},\ }\href {\doibase
		10.1103/PhysRevB.86.165136} {\bibfield  {journal} {\bibinfo  {journal} {Phys.
				Rev. B}\ }\textbf {\bibinfo {volume} {86}},\ \bibinfo {pages} {165136}
		(\bibinfo {year} {2012})}\BibitemShut {NoStop}%
	\bibitem [{\citenamefont {Gull}\ \emph {et~al.}(2009)\citenamefont {Gull},
		\citenamefont {Parcollet}, \citenamefont {Werner},\ and\ \citenamefont
		{Millis}}]{Gull2009}%
	\BibitemOpen
	\bibfield  {author} {\bibinfo {author} {\bibfnamefont {E.}~\bibnamefont
			{Gull}}, \bibinfo {author} {\bibfnamefont {O.}~\bibnamefont {Parcollet}},
		\bibinfo {author} {\bibfnamefont {P.}~\bibnamefont {Werner}}, \ and\ \bibinfo
		{author} {\bibfnamefont {A.~J.}\ \bibnamefont {Millis}},\ }\href {\doibase
		10.1103/PhysRevB.80.245102} {\bibfield  {journal} {\bibinfo  {journal} {Phys.
				Rev. B}\ }\textbf {\bibinfo {volume} {80}},\ \bibinfo {pages} {245102}
		(\bibinfo {year} {2009})}\BibitemShut {NoStop}%
	\bibitem [{\citenamefont {Gunnarsson}\ \emph {et~al.}(2016)\citenamefont
		{Gunnarsson}, \citenamefont {Sch\"afer}, \citenamefont {LeBlanc},
		\citenamefont {Merino}, \citenamefont {Sangiovanni}, \citenamefont
		{Rohringer},\ and\ \citenamefont {Toschi}}]{Gunnarsson2016}%
	\BibitemOpen
	\bibfield  {author} {\bibinfo {author} {\bibfnamefont {O.}~\bibnamefont
			{Gunnarsson}}, \bibinfo {author} {\bibfnamefont {T.}~\bibnamefont
			{Sch\"afer}}, \bibinfo {author} {\bibfnamefont {J.~P.~F.}\ \bibnamefont
			{LeBlanc}}, \bibinfo {author} {\bibfnamefont {J.}~\bibnamefont {Merino}},
		\bibinfo {author} {\bibfnamefont {G.}~\bibnamefont {Sangiovanni}}, \bibinfo
		{author} {\bibfnamefont {G.}~\bibnamefont {Rohringer}}, \ and\ \bibinfo
		{author} {\bibfnamefont {A.}~\bibnamefont {Toschi}},\ }\href {\doibase
		10.1103/PhysRevB.93.245102} {\bibfield  {journal} {\bibinfo  {journal} {Phys.
				Rev. B}\ }\textbf {\bibinfo {volume} {93}},\ \bibinfo {pages} {245102}
		(\bibinfo {year} {2016})}\BibitemShut {NoStop}%
	\bibitem [{\citenamefont {Gunnarsson}\ \emph {et~al.}(2017)\citenamefont
		{Gunnarsson}, \citenamefont {Rohringer}, \citenamefont {Sch\"afer},
		\citenamefont {Sangiovanni},\ and\ \citenamefont {Toschi}}]{Gunnarsson2017}%
	\BibitemOpen
	\bibfield  {author} {\bibinfo {author} {\bibfnamefont {O.}~\bibnamefont
			{Gunnarsson}}, \bibinfo {author} {\bibfnamefont {G.}~\bibnamefont
			{Rohringer}}, \bibinfo {author} {\bibfnamefont {T.}~\bibnamefont
			{Sch\"afer}}, \bibinfo {author} {\bibfnamefont {G.}~\bibnamefont
			{Sangiovanni}}, \ and\ \bibinfo {author} {\bibfnamefont {A.}~\bibnamefont
			{Toschi}},\ }\href {\doibase 10.1103/PhysRevLett.119.056402} {\bibfield
		{journal} {\bibinfo  {journal} {Phys. Rev. Lett.}\ }\textbf {\bibinfo
			{volume} {119}},\ \bibinfo {pages} {056402} (\bibinfo {year}
		{2017})}\BibitemShut {NoStop}%
	\bibitem [{\citenamefont {Rohringer}\ \emph {et~al.}(2018)\citenamefont
		{Rohringer}, \citenamefont {Hafermann}, \citenamefont {Toschi}, \citenamefont
		{Katanin}, \citenamefont {Antipov}, \citenamefont {Katsnelson}, \citenamefont
		{Lichtenstein}, \citenamefont {Rubtsov},\ and\ \citenamefont
		{Held}}]{Rohringer2018RMP}%
	\BibitemOpen
	\bibfield  {author} {\bibinfo {author} {\bibfnamefont {G.}~\bibnamefont
			{Rohringer}}, \bibinfo {author} {\bibfnamefont {H.}~\bibnamefont
			{Hafermann}}, \bibinfo {author} {\bibfnamefont {A.}~\bibnamefont {Toschi}},
		\bibinfo {author} {\bibfnamefont {A.~A.}\ \bibnamefont {Katanin}}, \bibinfo
		{author} {\bibfnamefont {A.~E.}\ \bibnamefont {Antipov}}, \bibinfo {author}
		{\bibfnamefont {M.~I.}\ \bibnamefont {Katsnelson}}, \bibinfo {author}
		{\bibfnamefont {A.~I.}\ \bibnamefont {Lichtenstein}}, \bibinfo {author}
		{\bibfnamefont {A.~N.}\ \bibnamefont {Rubtsov}}, \ and\ \bibinfo {author}
		{\bibfnamefont {K.}~\bibnamefont {Held}},\ }\href {\doibase
		10.1103/RevModPhys.90.025003} {\bibfield  {journal} {\bibinfo  {journal}
			{Reviews of Modern Physics}\ }\textbf {\bibinfo {volume} {90}},\ \bibinfo
		{pages} {25003} (\bibinfo {year} {2018})}\BibitemShut {NoStop}%
	\bibitem [{\citenamefont {Toschi}\ \emph {et~al.}(2007)\citenamefont {Toschi},
		\citenamefont {Katanin},\ and\ \citenamefont {Held}}]{Toschi2007}%
	\BibitemOpen
	\bibfield  {author} {\bibinfo {author} {\bibfnamefont {A.}~\bibnamefont
			{Toschi}}, \bibinfo {author} {\bibfnamefont {A.~A.}\ \bibnamefont {Katanin}},
		\ and\ \bibinfo {author} {\bibfnamefont {K.}~\bibnamefont {Held}},\ }\href
	{\doibase 10.1103/PhysRevB.75.045118} {\bibfield  {journal} {\bibinfo
			{journal} {Phys Rev. B}\ }\textbf {\bibinfo {volume} {75}},\ \bibinfo {pages}
		{045118} (\bibinfo {year} {2007})}\BibitemShut {NoStop}%
	\bibitem [{\citenamefont {Rubtsov}\ \emph {et~al.}(2008)\citenamefont
		{Rubtsov}, \citenamefont {Katsnelson},\ and\ \citenamefont
		{Lichtenstein}}]{Rubtsov2008}%
	\BibitemOpen
	\bibfield  {author} {\bibinfo {author} {\bibfnamefont {A.~N.}\ \bibnamefont
			{Rubtsov}}, \bibinfo {author} {\bibfnamefont {M.~I.}\ \bibnamefont
			{Katsnelson}}, \ and\ \bibinfo {author} {\bibfnamefont {A.~I.}\ \bibnamefont
			{Lichtenstein}},\ }\href {\doibase 10.1103/PhysRevB.77.033101} {\bibfield
		{journal} {\bibinfo  {journal} {Phys. Rev. B}\ }\textbf {\bibinfo {volume}
			{77}},\ \bibinfo {pages} {033101} (\bibinfo {year} {2008})}\BibitemShut
	{NoStop}%
	\bibitem [{\citenamefont {Rubtsov}\ \emph {et~al.}(2012)\citenamefont
		{Rubtsov}, \citenamefont {Katsnelson},\ and\ \citenamefont
		{Lichtenstein}}]{Rubtsov12}%
	\BibitemOpen
	\bibfield  {author} {\bibinfo {author} {\bibfnamefont {A.~N.}\ \bibnamefont
			{Rubtsov}}, \bibinfo {author} {\bibfnamefont {M.~I.}\ \bibnamefont
			{Katsnelson}}, \ and\ \bibinfo {author} {\bibfnamefont {A.~I.}\ \bibnamefont
			{Lichtenstein}},\ }\href {\doibase 10.1016/j.aop.2012.01.002} {\bibfield
		{journal} {\bibinfo  {journal} {Ann. Phys.}\ }\textbf {\bibinfo {volume}
			{327}},\ \bibinfo {pages} {1320} (\bibinfo {year} {2012})}\BibitemShut
	{NoStop}%
	\bibitem [{\citenamefont {Rohringer}\ \emph {et~al.}(2013)\citenamefont
		{Rohringer}, \citenamefont {Toschi}, \citenamefont {Hafermann}, \citenamefont
		{Held}, \citenamefont {Anisimov},\ and\ \citenamefont
		{Katanin}}]{Rohringer2013}%
	\BibitemOpen
	\bibfield  {author} {\bibinfo {author} {\bibfnamefont {G.}~\bibnamefont
			{Rohringer}}, \bibinfo {author} {\bibfnamefont {A.}~\bibnamefont {Toschi}},
		\bibinfo {author} {\bibfnamefont {H.}~\bibnamefont {Hafermann}}, \bibinfo
		{author} {\bibfnamefont {K.}~\bibnamefont {Held}}, \bibinfo {author}
		{\bibfnamefont {V.~I.}\ \bibnamefont {Anisimov}}, \ and\ \bibinfo {author}
		{\bibfnamefont {A.~A.}\ \bibnamefont {Katanin}},\ }\href
	{http://link.aps.org/doi/10.1103/PhysRevB.88.115112} {\bibfield  {journal}
		{\bibinfo  {journal} {Phys. Rev. B}\ }\textbf {\bibinfo {volume} {88}},\
		\bibinfo {pages} {115112} (\bibinfo {year} {2013})}\BibitemShut {NoStop}%
	\bibitem [{\citenamefont {Ayral}\ and\ \citenamefont
		{Parcollet}(2015)}]{Ayral2015}%
	\BibitemOpen
	\bibfield  {author} {\bibinfo {author} {\bibfnamefont {T.}~\bibnamefont
			{Ayral}}\ and\ \bibinfo {author} {\bibfnamefont {O.}~\bibnamefont
			{Parcollet}},\ }\href {http://link.aps.org/doi/10.1103/PhysRevB.92.115109}
	{\bibfield  {journal} {\bibinfo  {journal} {Phys Rev. B}\ }\textbf {\bibinfo
			{volume} {92}},\ \bibinfo {pages} {115109} (\bibinfo {year}
		{2015})}\BibitemShut {NoStop}%
	\bibitem [{\citenamefont {Ayral}\ and\ \citenamefont
		{Parcollet}(2016{\natexlab{a}})}]{Ayral2016a}%
	\BibitemOpen
	\bibfield  {author} {\bibinfo {author} {\bibfnamefont {T.}~\bibnamefont
			{Ayral}}\ and\ \bibinfo {author} {\bibfnamefont {O.}~\bibnamefont
			{Parcollet}},\ }\href {\doibase 10.1103/PhysRevB.93.235124} {\bibfield
		{journal} {\bibinfo  {journal} {Phys. Rev. B}\ }\textbf {\bibinfo {volume}
			{93}},\ \bibinfo {pages} {235124} (\bibinfo {year}
		{2016}{\natexlab{a}})}\BibitemShut {NoStop}%
	\bibitem [{\citenamefont {Ayral}\ and\ \citenamefont
		{Parcollet}(2016{\natexlab{b}})}]{Ayral2016}%
	\BibitemOpen
	\bibfield  {author} {\bibinfo {author} {\bibfnamefont {T.}~\bibnamefont
			{Ayral}}\ and\ \bibinfo {author} {\bibfnamefont {O.}~\bibnamefont
			{Parcollet}},\ }\href {\doibase 10.1103/PhysRevB.94.075159} {\bibfield
		{journal} {\bibinfo  {journal} {Phys. Rev. B}\ }\textbf {\bibinfo {volume}
			{94}},\ \bibinfo {pages} {075159} (\bibinfo {year}
		{2016}{\natexlab{b}})}\BibitemShut {NoStop}%
	\bibitem [{\citenamefont {Taranto}\ \emph {et~al.}(2014)\citenamefont
		{Taranto}, \citenamefont {Andergassen}, \citenamefont {Bauer}, \citenamefont
		{Held}, \citenamefont {Katanin}, \citenamefont {Metzner}, \citenamefont
		{Rohringer},\ and\ \citenamefont {Toschi}}]{Taranto2014}%
	\BibitemOpen
	\bibfield  {author} {\bibinfo {author} {\bibfnamefont {C.}~\bibnamefont
			{Taranto}}, \bibinfo {author} {\bibfnamefont {S.}~\bibnamefont
			{Andergassen}}, \bibinfo {author} {\bibfnamefont {J.}~\bibnamefont {Bauer}},
		\bibinfo {author} {\bibfnamefont {K.}~\bibnamefont {Held}}, \bibinfo {author}
		{\bibfnamefont {A.}~\bibnamefont {Katanin}}, \bibinfo {author} {\bibfnamefont
			{W.}~\bibnamefont {Metzner}}, \bibinfo {author} {\bibfnamefont
			{G.}~\bibnamefont {Rohringer}}, \ and\ \bibinfo {author} {\bibfnamefont
			{A.}~\bibnamefont {Toschi}},\ }\href {\doibase
		10.1103/PhysRevLett.112.196402} {\bibfield  {journal} {\bibinfo  {journal}
			{Phys. Rev. Lett.}\ }\textbf {\bibinfo {volume} {112}},\ \bibinfo {pages}
		{196402} (\bibinfo {year} {2014})}\BibitemShut {NoStop}%
	\bibitem [{\citenamefont {Katanin}(2019)}]{Katanin2019}%
	\BibitemOpen
	\bibfield  {author} {\bibinfo {author} {\bibfnamefont {A.~A.}\ \bibnamefont
			{Katanin}},\ }\href {\doibase 10.1103/PhysRevB.99.115112} {\bibfield
		{journal} {\bibinfo  {journal} {Phys. Rev. B}\ }\textbf {\bibinfo {volume}
			{99}},\ \bibinfo {pages} {115112} (\bibinfo {year} {2019})}\BibitemShut
	{NoStop}%
	\bibitem [{\citenamefont {Mahan}(2000)}]{Mahan2000}%
	\BibitemOpen
	\bibfield  {author} {\bibinfo {author} {\bibfnamefont {G.~D.}\ \bibnamefont
			{Mahan}},\ }\href@noop {} {\emph {\bibinfo {title} {Many-Particle Physics}}}\
	(\bibinfo  {publisher} {Kluwer Academic/Plenum Publishers, New York},\
	\bibinfo {year} {2000})\BibitemShut {NoStop}%
	\bibitem [{\citenamefont {Bickers}\ and\ \citenamefont
		{Scalapino}(1989)}]{Bickers1989}%
	\BibitemOpen
	\bibfield  {author} {\bibinfo {author} {\bibfnamefont {N.~E.}\ \bibnamefont
			{Bickers}}\ and\ \bibinfo {author} {\bibfnamefont {D.~J.}\ \bibnamefont
			{Scalapino}},\ }\href@noop {} {\bibfield  {journal} {\bibinfo  {journal}
			{Ann. Phys. (N. Y.)}\ }\textbf {\bibinfo {volume} {193}},\ \bibinfo {pages}
		{206} (\bibinfo {year} {1989})}\BibitemShut {NoStop}%
	\bibitem [{\citenamefont {Bickers}(2004)}]{Bickers2004}%
	\BibitemOpen
	\bibfield  {author} {\bibinfo {author} {\bibfnamefont {N.~E.}\ \bibnamefont
			{Bickers}},\ }\enquote {\bibinfo {title} {Theoretical methods for strongly
			correlated electrons},}\ \ (\bibinfo  {publisher} {Springer-Verlag New York
		Berlin Heidelbert},\ \bibinfo {year} {2004})\ Chap.~\bibinfo {chapter} {6},
	pp.\ \bibinfo {pages} {237--296}\BibitemShut {NoStop}%
	\bibitem [{\citenamefont {Otsuki}\ \emph {et~al.}(2014)\citenamefont {Otsuki},
		\citenamefont {Hafermann},\ and\ \citenamefont {Lichtenstein}}]{Otsuki2014}%
	\BibitemOpen
	\bibfield  {author} {\bibinfo {author} {\bibfnamefont {J.}~\bibnamefont
			{Otsuki}}, \bibinfo {author} {\bibfnamefont {H.}~\bibnamefont {Hafermann}}, \
		and\ \bibinfo {author} {\bibfnamefont {A.~I.}\ \bibnamefont {Lichtenstein}},\
	}\href {\doibase 10.1103/PhysRevB.90.235132} {\bibfield  {journal} {\bibinfo
			{journal} {Phys. Rev. B - Condens. Matter Mater. Phys.}\ }\textbf {\bibinfo
			{volume} {90}},\ \bibinfo {pages} {1} (\bibinfo {year} {2014})},\ \Eprint
	{http://arxiv.org/abs/1410.1246} {arXiv:1410.1246} \BibitemShut {NoStop}%
	\bibitem [{\citenamefont {Kitatani}\ \emph {et~al.}(2015)\citenamefont
		{Kitatani}, \citenamefont {Tsuji},\ and\ \citenamefont
		{Aoki}}]{Kitatani2015}%
	\BibitemOpen
	\bibfield  {author} {\bibinfo {author} {\bibfnamefont {M.}~\bibnamefont
			{Kitatani}}, \bibinfo {author} {\bibfnamefont {N.}~\bibnamefont {Tsuji}}, \
		and\ \bibinfo {author} {\bibfnamefont {H.}~\bibnamefont {Aoki}},\ }\href
	{\doibase 10.1103/PhysRevB.92.085104} {\bibfield  {journal} {\bibinfo
			{journal} {Phys. Rev. B - Condens. Matter Mater. Phys.}\ }\textbf {\bibinfo
			{volume} {92}},\ \bibinfo {pages} {1} (\bibinfo {year} {2015})}\BibitemShut
	{NoStop}%
	\bibitem [{\citenamefont {Kitatani}\ \emph {et~al.}(2019)\citenamefont
		{Kitatani}, \citenamefont {Sch{\"{a}}fer}, \citenamefont {Aoki},\ and\
		\citenamefont {Held}}]{Kitatani2019}%
	\BibitemOpen
	\bibfield  {author} {\bibinfo {author} {\bibfnamefont {M.}~\bibnamefont
			{Kitatani}}, \bibinfo {author} {\bibfnamefont {T.}~\bibnamefont
			{Sch{\"{a}}fer}}, \bibinfo {author} {\bibfnamefont {H.}~\bibnamefont {Aoki}},
		\ and\ \bibinfo {author} {\bibfnamefont {K.}~\bibnamefont {Held}},\ }\href
	{\doibase 10.1103/PhysRevB.99.041115} {\bibfield  {journal} {\bibinfo
			{journal} {Phys. Rev. B}\ }\textbf {\bibinfo {volume} {99}},\ \bibinfo
		{pages} {1} (\bibinfo {year} {2019})}\BibitemShut {NoStop}%
	\bibitem [{\citenamefont {Diatlov}\ \emph {et~al.}(1957)\citenamefont
		{Diatlov}, \citenamefont {Sudakov},\ and\ \citenamefont
		{Ter-Martirosian}}]{Diatlov1957}%
	\BibitemOpen
	\bibfield  {author} {\bibinfo {author} {\bibfnamefont {I.~T.}\ \bibnamefont
			{Diatlov}}, \bibinfo {author} {\bibfnamefont {V.~V.}\ \bibnamefont
			{Sudakov}}, \ and\ \bibinfo {author} {\bibfnamefont {K.~A.}\ \bibnamefont
			{Ter-Martirosian}},\ }\href@noop {} {\bibfield  {journal} {\bibinfo
			{journal} {Sov. Phys. JETP}\ }\textbf {\bibinfo {volume} {5}},\ \bibinfo
		{pages} {631} (\bibinfo {year} {1957})}\BibitemShut {NoStop}%
	\bibitem [{\citenamefont {Rohringer}\ \emph {et~al.}(2012)\citenamefont
		{Rohringer}, \citenamefont {Valli},\ and\ \citenamefont
		{Toschi}}]{Rohringer2012}%
	\BibitemOpen
	\bibfield  {author} {\bibinfo {author} {\bibfnamefont {G.}~\bibnamefont
			{Rohringer}}, \bibinfo {author} {\bibfnamefont {A.}~\bibnamefont {Valli}}, \
		and\ \bibinfo {author} {\bibfnamefont {A.}~\bibnamefont {Toschi}},\ }\href
	{\doibase 10.1103/PhysRevB.86.125114} {\bibfield  {journal} {\bibinfo
			{journal} {Phys. Rev. B}\ }\textbf {\bibinfo {volume} {86}},\ \bibinfo
		{pages} {125114} (\bibinfo {year} {2012})}\BibitemShut {NoStop}%
	\bibitem [{\citenamefont {Yang}\ \emph {et~al.}(2009)\citenamefont {Yang},
		\citenamefont {Fotso}, \citenamefont {Liu}, \citenamefont {Maier},
		\citenamefont {Tomko}, \citenamefont {D'Azevedo}, \citenamefont {Scalettar},
		\citenamefont {Pruschke},\ and\ \citenamefont {Jarrell}}]{Yang2009}%
	\BibitemOpen
	\bibfield  {author} {\bibinfo {author} {\bibfnamefont {S.~X.}\ \bibnamefont
			{Yang}}, \bibinfo {author} {\bibfnamefont {H.}~\bibnamefont {Fotso}},
		\bibinfo {author} {\bibfnamefont {J.}~\bibnamefont {Liu}}, \bibinfo {author}
		{\bibfnamefont {T.~A.}\ \bibnamefont {Maier}}, \bibinfo {author}
		{\bibfnamefont {K.}~\bibnamefont {Tomko}}, \bibinfo {author} {\bibfnamefont
			{E.~F.}\ \bibnamefont {D'Azevedo}}, \bibinfo {author} {\bibfnamefont {R.~T.}\
			\bibnamefont {Scalettar}}, \bibinfo {author} {\bibfnamefont {T.}~\bibnamefont
			{Pruschke}}, \ and\ \bibinfo {author} {\bibfnamefont {M.}~\bibnamefont
			{Jarrell}},\ }\href {\doibase 10.1103/PhysRevE.80.046706} {\bibfield
		{journal} {\bibinfo  {journal} {Phys. Rev. E}\ }\textbf {\bibinfo {volume}
			{80}},\ \bibinfo {pages} {046706} (\bibinfo {year} {2009})}\BibitemShut
	{NoStop}%
	\bibitem [{\citenamefont {Tam}\ \emph {et~al.}(2013)\citenamefont {Tam},
		\citenamefont {Fotso}, \citenamefont {Yang}, \citenamefont {Lee},
		\citenamefont {Moreno}, \citenamefont {Ramanujam},\ and\ \citenamefont
		{Jarrell}}]{Tam2013}%
	\BibitemOpen
	\bibfield  {author} {\bibinfo {author} {\bibfnamefont {K.-M.}\ \bibnamefont
			{Tam}}, \bibinfo {author} {\bibfnamefont {H.}~\bibnamefont {Fotso}}, \bibinfo
		{author} {\bibfnamefont {S.-X.}\ \bibnamefont {Yang}}, \bibinfo {author}
		{\bibfnamefont {T.-W.}\ \bibnamefont {Lee}}, \bibinfo {author} {\bibfnamefont
			{J.}~\bibnamefont {Moreno}}, \bibinfo {author} {\bibfnamefont
			{J.}~\bibnamefont {Ramanujam}}, \ and\ \bibinfo {author} {\bibfnamefont
			{M.}~\bibnamefont {Jarrell}},\ }\href {\doibase 10.1103/PhysRevE.87.013311}
	{\bibfield  {journal} {\bibinfo  {journal} {Phys. Rev. E}\ }\textbf {\bibinfo
			{volume} {87}},\ \bibinfo {pages} {013311} (\bibinfo {year}
		{2013})}\BibitemShut {NoStop}%
	\bibitem [{\citenamefont {Li}\ \emph {et~al.}(2016)\citenamefont {Li},
		\citenamefont {Wentzell}, \citenamefont {Pudleiner}, \citenamefont
		{Thunstr\"om},\ and\ \citenamefont {Held}}]{Li2016}%
	\BibitemOpen
	\bibfield  {author} {\bibinfo {author} {\bibfnamefont {G.}~\bibnamefont
			{Li}}, \bibinfo {author} {\bibfnamefont {N.}~\bibnamefont {Wentzell}},
		\bibinfo {author} {\bibfnamefont {P.}~\bibnamefont {Pudleiner}}, \bibinfo
		{author} {\bibfnamefont {P.}~\bibnamefont {Thunstr\"om}}, \ and\ \bibinfo
		{author} {\bibfnamefont {K.}~\bibnamefont {Held}},\ }\href {\doibase
		10.1103/PhysRevB.93.165103} {\bibfield  {journal} {\bibinfo  {journal} {Phys.
				Rev. B}\ }\textbf {\bibinfo {volume} {93}},\ \bibinfo {pages} {165103}
		(\bibinfo {year} {2016})}\BibitemShut {NoStop}%
	\bibitem [{\citenamefont {Karrasch}\ \emph {et~al.}(2008)\citenamefont
		{Karrasch}, \citenamefont {Hedden}, \citenamefont {Peters}, \citenamefont
		{Pruschke}, \citenamefont {sch\"onhammer},\ and\ \citenamefont
		{Meden}}]{Karrasch2008}%
	\BibitemOpen
	\bibfield  {author} {\bibinfo {author} {\bibfnamefont {C.}~\bibnamefont
			{Karrasch}}, \bibinfo {author} {\bibfnamefont {R.}~\bibnamefont {Hedden}},
		\bibinfo {author} {\bibfnamefont {R.}~\bibnamefont {Peters}}, \bibinfo
		{author} {\bibfnamefont {T.}~\bibnamefont {Pruschke}}, \bibinfo {author}
		{\bibfnamefont {K.}~\bibnamefont {sch\"onhammer}}, \ and\ \bibinfo {author}
		{\bibfnamefont {V.}~\bibnamefont {Meden}},\ }\href@noop {} {\bibfield
		{journal} {\bibinfo  {journal} {J. Phys.: Condens. Matter}\ }\textbf
		{\bibinfo {volume} {20}},\ \bibinfo {pages} {345205} (\bibinfo {year}
		{2008})}\BibitemShut {NoStop}%
	\bibitem [{\citenamefont {Wentzell}\ \emph {et~al.}(2016)\citenamefont
		{Wentzell}, \citenamefont {Li}, \citenamefont {Tagliavini}, \citenamefont
		{Taranto}, \citenamefont {Rohringer}, \citenamefont {Held}, \citenamefont
		{Toschi},\ and\ \citenamefont {Andergassen}}]{Wentzell2016}%
	\BibitemOpen
	\bibfield  {author} {\bibinfo {author} {\bibfnamefont {N.}~\bibnamefont
			{Wentzell}}, \bibinfo {author} {\bibfnamefont {G.}~\bibnamefont {Li}},
		\bibinfo {author} {\bibfnamefont {A.}~\bibnamefont {Tagliavini}}, \bibinfo
		{author} {\bibfnamefont {C.}~\bibnamefont {Taranto}}, \bibinfo {author}
		{\bibfnamefont {G.}~\bibnamefont {Rohringer}}, \bibinfo {author}
		{\bibfnamefont {K.}~\bibnamefont {Held}}, \bibinfo {author} {\bibfnamefont
			{A.}~\bibnamefont {Toschi}}, \ and\ \bibinfo {author} {\bibfnamefont
			{S.}~\bibnamefont {Andergassen}},\ }\href@noop {} {\bibfield  {journal}
		{\bibinfo  {journal} {arXiv}\ } (\bibinfo {year} {2016})},\ \Eprint
	{http://arxiv.org/abs/1610.06520} {arXiv:1610.06520} \BibitemShut {NoStop}%
	\bibitem [{\citenamefont {Tagliavini}\ \emph {et~al.}(2017)\citenamefont
		{Tagliavini}, \citenamefont {Hummel}, \citenamefont {Wentzell}, \citenamefont
		{Andergassen}, \citenamefont {Toschi},\ and\ \citenamefont
		{Rohringer}}]{Tagliavini2017}%
	\BibitemOpen
	\bibfield  {author} {\bibinfo {author} {\bibfnamefont {A.}~\bibnamefont
			{Tagliavini}}, \bibinfo {author} {\bibfnamefont {S.}~\bibnamefont {Hummel}},
		\bibinfo {author} {\bibfnamefont {N.}~\bibnamefont {Wentzell}}, \bibinfo
		{author} {\bibfnamefont {S.}~\bibnamefont {Andergassen}}, \bibinfo {author}
		{\bibfnamefont {A.}~\bibnamefont {Toschi}}, \ and\ \bibinfo {author}
		{\bibfnamefont {G.}~\bibnamefont {Rohringer}},\ }\href@noop {} {\enquote
		{\bibinfo {title} {Efficient treatment of the bethe-salpeter equations
				inversion ind dynamical mean-field theory},}\ } (\bibinfo {year} {2017}),\
	\bibinfo {note} {(unpublished)}\BibitemShut {NoStop}%
	\bibitem [{\citenamefont {Eckhardt}\ \emph {et~al.}(2018)\citenamefont
		{Eckhardt}, \citenamefont {Schober}, \citenamefont {Ehrlich},\ and\
		\citenamefont {Honerkamp}}]{Eckhardt2018}%
	\BibitemOpen
	\bibfield  {author} {\bibinfo {author} {\bibfnamefont {C.~J.}\ \bibnamefont
			{Eckhardt}}, \bibinfo {author} {\bibfnamefont {G.~A.~H.}\ \bibnamefont
			{Schober}}, \bibinfo {author} {\bibfnamefont {J.}~\bibnamefont {Ehrlich}}, \
		and\ \bibinfo {author} {\bibfnamefont {C.}~\bibnamefont {Honerkamp}},\ }\href
	{\doibase 10.1103/PhysRevB.98.075143} {\bibfield  {journal} {\bibinfo
			{journal} {Phys. Rev. B}\ }\textbf {\bibinfo {volume} {98}},\ \bibinfo
		{pages} {075143} (\bibinfo {year} {2018})}\BibitemShut {NoStop}%
	\bibitem [{\citenamefont {Kugler}\ and\ \citenamefont {von
			Delft}(2018{\natexlab{a}})}]{Kugler2018}%
	\BibitemOpen
	\bibfield  {author} {\bibinfo {author} {\bibfnamefont {F.~B.}\ \bibnamefont
			{Kugler}}\ and\ \bibinfo {author} {\bibfnamefont {J.}~\bibnamefont {von
				Delft}},\ }\href {\doibase 10.1103/PhysRevLett.120.057403} {\bibfield
		{journal} {\bibinfo  {journal} {Phys. Rev. Lett.}\ }\textbf {\bibinfo
			{volume} {120}},\ \bibinfo {pages} {057403} (\bibinfo {year}
		{2018}{\natexlab{a}})}\BibitemShut {NoStop}%
	\bibitem [{\citenamefont {Kugler}\ and\ \citenamefont {von
			Delft}(2018{\natexlab{b}})}]{Kugler2018a}%
	\BibitemOpen
	\bibfield  {author} {\bibinfo {author} {\bibfnamefont {F.~B.}\ \bibnamefont
			{Kugler}}\ and\ \bibinfo {author} {\bibfnamefont {J.}~\bibnamefont {von
				Delft}},\ }\href {\doibase 10.1103/PhysRevB.97.035162} {\bibfield  {journal}
		{\bibinfo  {journal} {Phys. Rev. B}\ }\textbf {\bibinfo {volume} {97}},\
		\bibinfo {pages} {035162} (\bibinfo {year} {2018}{\natexlab{b}})}\BibitemShut
	{NoStop}%
	\bibitem [{\citenamefont {Mott}(1968)}]{Mott1968}%
	\BibitemOpen
	\bibfield  {author} {\bibinfo {author} {\bibfnamefont {N.~F.}\ \bibnamefont
			{Mott}},\ }\href@noop {} {\bibfield  {journal} {\bibinfo  {journal} {Rev.
				Mod. Phys.}\ }\textbf {\bibinfo {volume} {40}},\ \bibinfo {pages} {677}
		(\bibinfo {year} {1968})}\BibitemShut {NoStop}%
	\bibitem [{\citenamefont {Rubtsov}\ \emph {et~al.}(2009)\citenamefont
		{Rubtsov}, \citenamefont {Katsnelson}, \citenamefont {Lichtenstein},\ and\
		\citenamefont {Georges}}]{Rubtsov2009}%
	\BibitemOpen
	\bibfield  {author} {\bibinfo {author} {\bibfnamefont {A.~N.}\ \bibnamefont
			{Rubtsov}}, \bibinfo {author} {\bibfnamefont {M.~I.}\ \bibnamefont
			{Katsnelson}}, \bibinfo {author} {\bibfnamefont {A.~I.}\ \bibnamefont
			{Lichtenstein}}, \ and\ \bibinfo {author} {\bibfnamefont {A.}~\bibnamefont
			{Georges}},\ }\href {\doibase 10.1103/PhysRevB.79.045133} {\bibfield
		{journal} {\bibinfo  {journal} {Phys. Rev. B}\ }\textbf {\bibinfo {volume}
			{79}},\ \bibinfo {pages} {045133} (\bibinfo {year} {2009})}\BibitemShut
	{NoStop}%
	\bibitem [{\citenamefont {Hubbard}(1959)}]{Hubbard1959}%
	\BibitemOpen
	\bibfield  {author} {\bibinfo {author} {\bibfnamefont {J.}~\bibnamefont
			{Hubbard}},\ }\href {\doibase 10.1103/PhysRevLett.3.77} {\bibfield  {journal}
		{\bibinfo  {journal} {Phys. Rev. Lett.}\ }\textbf {\bibinfo {volume} {3}},\
		\bibinfo {pages} {77} (\bibinfo {year} {1959})}\BibitemShut {NoStop}%
	\bibitem [{\citenamefont {{Stratonovich}}(1957)}]{Stratonovich1957}%
	\BibitemOpen
	\bibfield  {author} {\bibinfo {author} {\bibfnamefont {R.~L.}\ \bibnamefont
			{{Stratonovich}}},\ }\href@noop {} {\bibfield  {journal} {\bibinfo  {journal}
			{Soviet Physics Doklady}\ }\textbf {\bibinfo {volume} {2}},\ \bibinfo {pages}
		{416} (\bibinfo {year} {1957})}\BibitemShut {NoStop}%
	\bibitem [{\citenamefont {Ribic}\ \emph {et~al.}(2017)\citenamefont {Ribic},
		\citenamefont {Gunacker}, \citenamefont {Iskakov}, \citenamefont
		{Wallerberger}, \citenamefont {Rohringer}, \citenamefont {Rubtsov},
		\citenamefont {Gull},\ and\ \citenamefont {Held}}]{Ribic2017b}%
	\BibitemOpen
	\bibfield  {author} {\bibinfo {author} {\bibfnamefont {T.}~\bibnamefont
			{Ribic}}, \bibinfo {author} {\bibfnamefont {P.}~\bibnamefont {Gunacker}},
		\bibinfo {author} {\bibfnamefont {S.}~\bibnamefont {Iskakov}}, \bibinfo
		{author} {\bibfnamefont {M.}~\bibnamefont {Wallerberger}}, \bibinfo {author}
		{\bibfnamefont {G.}~\bibnamefont {Rohringer}}, \bibinfo {author}
		{\bibfnamefont {A.~N.}\ \bibnamefont {Rubtsov}}, \bibinfo {author}
		{\bibfnamefont {E.}~\bibnamefont {Gull}}, \ and\ \bibinfo {author}
		{\bibfnamefont {K.}~\bibnamefont {Held}},\ }\href {\doibase
		10.1103/PhysRevB.96.235127} {\bibfield  {journal} {\bibinfo  {journal} {Phys.
				Rev. B}\ }\textbf {\bibinfo {volume} {96}},\ \bibinfo {pages} {235127}
		(\bibinfo {year} {2017})}\BibitemShut {NoStop}%
	\bibitem [{\citenamefont {Wilson}(1975)}]{Wilson1975}%
	\BibitemOpen
	\bibfield  {author} {\bibinfo {author} {\bibfnamefont {K.~G.}\ \bibnamefont
			{Wilson}},\ }\href {\doibase 10.1103/RevModPhys.47.773} {\bibfield  {journal}
		{\bibinfo  {journal} {Rev. Mod. Phys.}\ }\textbf {\bibinfo {volume} {47}},\
		\bibinfo {pages} {773} (\bibinfo {year} {1975})}\BibitemShut {NoStop}%
	\bibitem [{\citenamefont {Honerkamp}\ and\ \citenamefont
		{Salmhofer}(2003)}]{Honerkamp2003}%
	\BibitemOpen
	\bibfield  {author} {\bibinfo {author} {\bibfnamefont {C.}~\bibnamefont
			{Honerkamp}}\ and\ \bibinfo {author} {\bibfnamefont {M.}~\bibnamefont
			{Salmhofer}},\ }\href {\doibase 10.1103/PhysRevB.67.174504} {\bibfield
		{journal} {\bibinfo  {journal} {Phys. Rev. B}\ }\textbf {\bibinfo {volume}
			{67}},\ \bibinfo {pages} {174504} (\bibinfo {year} {2003})}\BibitemShut
	{NoStop}%
	\bibitem [{\citenamefont {{Li}}\ \emph {et~al.}(2017)\citenamefont {{Li}},
		\citenamefont {{Kauch}}, \citenamefont {{Pudleiner}},\ and\ \citenamefont
		{{Held}}}]{Li2017}%
	\BibitemOpen
	\bibfield  {author} {\bibinfo {author} {\bibfnamefont {G.}~\bibnamefont
			{{Li}}}, \bibinfo {author} {\bibfnamefont {A.}~\bibnamefont {{Kauch}}},
		\bibinfo {author} {\bibfnamefont {P.}~\bibnamefont {{Pudleiner}}}, \ and\
		\bibinfo {author} {\bibfnamefont {K.}~\bibnamefont {{Held}}},\ }\href
	{https://arxiv.org/abs/1708.07457} {\bibfield  {journal} {\bibinfo  {journal}
			{arXiv}\ } (\bibinfo {year} {2017})},\ \Eprint
	{http://arxiv.org/abs/1708.07457} {arXiv:1708.07457} \BibitemShut {NoStop}%
	\bibitem [{\citenamefont {Nicoletti}\ \emph {et~al.}(2010)\citenamefont
		{Nicoletti}, \citenamefont {Limaj}, \citenamefont {Calvani}, \citenamefont
		{Rohringer}, \citenamefont {Toschi}, \citenamefont {Sangiovanni},
		\citenamefont {Capone}, \citenamefont {Held}, \citenamefont {Ono},
		\citenamefont {Ando},\ and\ \citenamefont {Lupi}}]{Nicoletti2010}%
	\BibitemOpen
	\bibfield  {author} {\bibinfo {author} {\bibfnamefont {D.}~\bibnamefont
			{Nicoletti}}, \bibinfo {author} {\bibfnamefont {O.}~\bibnamefont {Limaj}},
		\bibinfo {author} {\bibfnamefont {P.}~\bibnamefont {Calvani}}, \bibinfo
		{author} {\bibfnamefont {G.}~\bibnamefont {Rohringer}}, \bibinfo {author}
		{\bibfnamefont {A.}~\bibnamefont {Toschi}}, \bibinfo {author} {\bibfnamefont
			{G.}~\bibnamefont {Sangiovanni}}, \bibinfo {author} {\bibfnamefont
			{M.}~\bibnamefont {Capone}}, \bibinfo {author} {\bibfnamefont
			{K.}~\bibnamefont {Held}}, \bibinfo {author} {\bibfnamefont {S.}~\bibnamefont
			{Ono}}, \bibinfo {author} {\bibfnamefont {Y.}~\bibnamefont {Ando}}, \ and\
		\bibinfo {author} {\bibfnamefont {S.}~\bibnamefont {Lupi}},\ }\href {\doibase
		10.1103/PhysRevLett.105.077002} {\bibfield  {journal} {\bibinfo  {journal}
			{Phys. Rev. Lett.}\ }\textbf {\bibinfo {volume} {105}},\ \bibinfo {pages}
		{077002} (\bibinfo {year} {2010})}\BibitemShut {NoStop}%
	\bibitem [{\citenamefont {Nourafkan}\ \emph {et~al.}(2019)\citenamefont
		{Nourafkan}, \citenamefont {C\^ot\'e},\ and\ \citenamefont
		{Tremblay}}]{Nourafkan2018}%
	\BibitemOpen
	\bibfield  {author} {\bibinfo {author} {\bibfnamefont {R.}~\bibnamefont
			{Nourafkan}}, \bibinfo {author} {\bibfnamefont {M.}~\bibnamefont {C\^ot\'e}},
		\ and\ \bibinfo {author} {\bibfnamefont {A.-M.~S.}\ \bibnamefont
			{Tremblay}},\ }\href {\doibase 10.1103/PhysRevB.99.035161} {\bibfield
		{journal} {\bibinfo  {journal} {Phys. Rev. B}\ }\textbf {\bibinfo {volume}
			{99}},\ \bibinfo {pages} {035161} (\bibinfo {year} {2019})}\BibitemShut
	{NoStop}%
	\bibitem [{\citenamefont {Kosterlitz}\ and\ \citenamefont
		{Thouless}(1973)}]{Kosterlitz1973}%
	\BibitemOpen
	\bibfield  {author} {\bibinfo {author} {\bibfnamefont {J.~M.}\ \bibnamefont
			{Kosterlitz}}\ and\ \bibinfo {author} {\bibfnamefont {D.~J.}\ \bibnamefont
			{Thouless}},\ }\href {\doibase 10.1088/0022-3719/6/7/010} {\bibfield
		{journal} {\bibinfo  {journal} {J. Phys.}\ }\textbf {\bibinfo {volume}
			{C6}},\ \bibinfo {pages} {1181} (\bibinfo {year} {1973})},\ \bibinfo {note}
	{[,349(1973)]}\BibitemShut {NoStop}%
	\bibitem [{\citenamefont {Tranquada}(2012)}]{Tranquada2012}%
	\BibitemOpen
	\bibfield  {author} {\bibinfo {author} {\bibfnamefont {J.~M.}\ \bibnamefont
			{Tranquada}},\ }\href {\doibase https://doi.org/10.1016/j.physb.2012.01.026}
	{\bibfield  {journal} {\bibinfo  {journal} {Physica B: Condensed Matter}\
		}\textbf {\bibinfo {volume} {407}},\ \bibinfo {pages} {1771 } (\bibinfo
		{year} {2012})},\ \bibinfo {note} {proceedings of the International Workshop
		on Electronic Crystals (ECRYS-2011)}\BibitemShut {NoStop}%
	\bibitem [{Note1()}]{Note1}%
	\BibitemOpen
	\bibinfo {note} {Note that, differently from Ref.~\protect \rev@citealpnum
		{Yang2011,Chen2013} we have projected $[\chi _0^{PP'Q}]^{-1}$ instead of
		$\chi _0^{PP'Q}$ as this allows for a direct projection of the BS
		equation~(\ref {equ:bschisinglet}).}\BibitemShut {Stop}%
	\bibitem [{\citenamefont {Dzyaloshinskii}(1996)}]{Dzyaloshinskii1996}%
	\BibitemOpen
	\bibfield  {author} {\bibinfo {author} {\bibfnamefont {I.}~\bibnamefont
			{Dzyaloshinskii}},\ }\href {\doibase 10.1051/jp1:1996127} {\bibfield
		{journal} {\bibinfo  {journal} {{Journal de Physique I}}\ }\textbf {\bibinfo
			{volume} {6}},\ \bibinfo {pages} {119} (\bibinfo {year} {1996})}\BibitemShut
	{NoStop}%
	\bibitem [{\citenamefont {Irkhin}\ \emph {et~al.}(2001)\citenamefont {Irkhin},
		\citenamefont {Katanin},\ and\ \citenamefont {Katsnelson}}]{Irkhin2001}%
	\BibitemOpen
	\bibfield  {author} {\bibinfo {author} {\bibfnamefont {V.~Y.}\ \bibnamefont
			{Irkhin}}, \bibinfo {author} {\bibfnamefont {A.~A.}\ \bibnamefont {Katanin}},
		\ and\ \bibinfo {author} {\bibfnamefont {M.~I.}\ \bibnamefont {Katsnelson}},\
	}\href {\doibase 10.1103/PhysRevB.64.165107} {\bibfield  {journal} {\bibinfo
			{journal} {Phys. Rev. B}\ }\textbf {\bibinfo {volume} {64}},\ \bibinfo
		{pages} {165107} (\bibinfo {year} {2001})}\BibitemShut {NoStop}%
	\bibitem [{\citenamefont {Krien}\ and\ \citenamefont
		{Valli}(2019)}]{Krien2019}%
	\BibitemOpen
	\bibfield  {author} {\bibinfo {author} {\bibfnamefont {F.}~\bibnamefont
			{Krien}}\ and\ \bibinfo {author} {\bibfnamefont {A.}~\bibnamefont {Valli}},\
	}\href {http://arxiv.org/abs/1909.02793} {\enquote {\bibinfo {title}
			{{Parquet-like equations for the Hedin three-leg vertex}},}\ } (\bibinfo
	{year} {2019}),\ \bibinfo {note} {(unpublished)},\ \Eprint
	{http://arxiv.org/abs/1909.02793} {arXiv:1909.02793} \BibitemShut {NoStop}%
\end{thebibliography}

\end{document}